\documentclass[11pt,a4paper]{article}
\usepackage{amsmath,amssymb,latexsym}
\usepackage[svgnames]{xcolor}
\usepackage{jheppub}
\usepackage{braket}
\usepackage{float}
\graphicspath{{figures/}}
\usepackage{bm}
\usepackage{setspace}
\usepackage{hepunits}
\usepackage{mciteplus}
\usepackage{booktabs,array}
\usepackage{enumitem}
\usepackage{placeins}
\usepackage{needspace}
\numberwithin{table}{section}

\newcommand{\nn}{\nonumber}
\newcommand{\nd}{\mathrm{d}}

\newcommand{\be}{\begin{equation}}
\newcommand{\ee}{\end{equation}}
\newcommand{\ba}{\begin{eqnarray}}
\newcommand{\ea}{\end{eqnarray}}
\newcommand{\dd}{\mathrm d}
\newcommand{\ii}{\mathrm i}
\newcommand{\Id}{\mathbb I}
\newcommand{\diag}{\operatorname{diag}}
\newcommand{\cV}{\mathcal V}
\newcommand{\cE}{\mathcal E}
\newcommand{\cH}{\mathcal H}
\newcommand{\cM}{\mathcal M}
\newcommand{\cB}{\mathcal B}
\newcommand{\cR}{\mathcal R}
\newcommand{\cN}{\mathcal N}
\newcommand{\Emb}{\operatorname{Emb}}
\def\ie{\begin{equation}\begin{aligned}}
\def\fe{\end{aligned}\end{equation}}

\title{MadStree: automated evaluation of tree-level de Sitter correlators} 
\author[a,b]{Jiaqi Chen}
\author[c,d]{Yi-Xiao Tao}
\author[e]{Xinyi Wang}
\author[f]{Dong Zhang}

\affiliation[a]{Beijing Key Laboratory of Optical Detection Technology for Oil and Gas, China University of Petroleum-Beijing, Beijing 102249, China}
\affiliation[b]{Basic Research Center for Energy Interdisciplinary, College of Science, China University of Petroleum-Beijing, Beijing 102249, China}
\affiliation[c]{Department of Mathematical Sciences, Tsinghua University, Beijing 100084, China}
\affiliation[d]{Institute for Theoretical and Mathematical Physics, Lomonosov Moscow State University, Moscow 119991, Russia}
\affiliation[e]{Nanophotonics and Biophotonics Key Laboratory of Jilin Province, School of Physics, Changchun University of Science and Technology, Changchun 130022, China}
\affiliation[f]{Department of Physics, School of Science, Tianjin University, Tianjin 300350, China}
\emailAdd{jiaqichen@cup.edu.cn}
\emailAdd{tyixiao@itmp.msu.ru}
\emailAdd{1437263818@qq.com}
\emailAdd{yanyandreamer@tju.edu.cn}

\abstract{We first derived the formulae of recursive reduction and d log-form differential equations for arbitrary tree-level and loop-integrand cosmological correlators, taking the external relations of massless bulk propagators into consideration. The differential equations could be systematically solved and provide the efficient evaluation of them. We have implemented this methodology as a Mathematica package \texttt{MadStree}. To demonstrate the efficiency of our method, we evaluated up to $10^6$ points for each master integral in our examples, and found that the average computational cost per numerical point per master integral is on the order of $\mu\mathrm{s}$.
}

\begin{document}
\maketitle

\section{Introduction}
\label{sec:introduction}

Late-time correlation functions in inflation connect quantum field dynamics to the observed
correlations of primordial fluctuations~\cite{Chen:2017ryl,Achucarro:2022qrl}. Their non-Gaussian structure~\cite{Maldacena:2002vr,Chen:2006nt} can
carry information about particles exchanged during inflation, including their masses and
spins~\cite{Arkani-Hamed:2018kmz,Arkani-Hamed:2015bza}. Trees with several massive scalar exchanges are relevant
in quasi-single-field models, where they can dominate the three-point
correlator~\cite{Xianyu:2023ytd,Chen:2009zp}. Evaluating such diagrams over kinematic ranges is therefore
useful for studying the momentum dependence of inflationary signals, which is also central to cosmological collider physics~\cite{Chen:2009we,Chen:2009zp,Baumann:2011nk,Noumi:2012vr,Arkani-Hamed:2015bza}.

In de~Sitter (dS) space, the in-in Feynman rules express scalar tree correlators as sums of
time integrals containing Hankel functions, external-energy exponentials and time-ordering
theta functions~\cite{Weinberg:2005vy,Chen:2017ryl}. Many analytic methods have been developed \cite{Arkani-Hamed:2018kmz,Baumann:2019oyu,Baumann:2020dch,Pajer:2020wnj,Pajer:2020wxk,Cabass:2021fnw,Baumann:2022jpr,Pimentel:2022fsc,Jazayeri:2022kjy,Wang:2022eop,Qin:2023ejc,Aoki:2024uyi,Liu:2024str,Xianyu:2025lbk,Chen:2023xlt,Tao:2022nqc,Chen:2023bji,Sleight:2019mgd,Sleight:2019hfp,Sleight:2020obc,Sleight:2021plv,Premkumar:2021mlz,Qin:2022lva,Qin:2022fbv,Xianyu:2023ytd,Qin:2024gtr,Marolf:2010zp,Xianyu:2022jwk,Loparco:2023rug,Werth:2024mjg,Qin:2025xct,Zhang:2025nzd,Altshuler:2025qmk,Liu:2024xyi,Das:2026vfv,Grafe:2026spectral,Belrhali:2026Laplace}, including the cosmological bootstrap, Mellin techniques and spectral decompositions. Strongly mixed cases have also been studied recently%
~\cite{Huenupi:2026ExactLinear,Huenupi:2026StrongMixing,
Wang:2026StrongMixing,Pinol:2026NewShapes,
Pinol:2026ExactBispectra,Huenupi:2026UnequalSpeeds}. See also \cite{Wang:2021qez,Werth:2023pfl,Pinol:2023oux,Werth:2024aui,Chen:2023iix,Chen:2024glu} for some recent progress in numerical methods.
Also see \cite{Albayrak:2026auu} for a review.

We use a different approach, which is widely used in precision calculations in
flat-spacetime QFT: Integration-By-Parts (IBP) relations \cite{Chetyrkin:1981qh} and the
differential equations (DE) that they induce \cite{Kotikov:1990kg,Kotikov:1991pm,Bern:1993kr,Gehrmann:1999as}.
An IBP relation is a linear relation within an integral family. A sufficient set of these
relations reduces the family to master integrals; differentiation and reduction then give a
closed system of first-order DE. A d log-form connection~\cite{Henn:2013pwa} exposes its
singularities and supports numerical transport from boundary data. In flat space, FIRE
\cite{Smirnov:2019qkx,Smirnov:2025prc} and Kira
\cite{Maierhofer:2017gsa,Klappert:2020nbg,Lange:2025fba} automate IBP reduction, while generalized
power series expansions~\cite{Moriello:2019yhu}, DiffExp
\cite{Hidding:2020ytt} and the auxiliary mass flow method implemented in AMFlow
\cite{Liu:2017jxz,Liu:2020kpc,Liu:2022chg,Huang:2026rjb} provide numerical evaluation.

The IBP/DE framework in dS spacetime was developed in \cite{Chen:2023iix,Chen:2024glu,Chen:2026dqp}. In \cite{Chen:2023iix}, the authors give general iterative reduction formulas and d log-form DE
for Hankel vertex integral families, including the contact terms needed for tree correlators. This method gives a systematic approach to calculate the dS tree correlators.
In~\cite{Chen:2024glu}, the authors applied the method proposed in \cite{Chen:2023iix} to some specific examples. They construct power-series solutions, determine boundary data
for two-vertex dS correlators through a blow-up of the multi-energy limit, and describe how the
boundary factorization extends to general trees. These results provide the analytic basis for a systematic numerical implementation. It is worth mentioning that the systematic treatment of general trees is also discussed in \cite{Liu:2024str,Xianyu:2025lbk,Grafe:2026spectral} by different frameworks.

We develop this framework in two directions. First, we incorporate the extra algebraic
relations of massless internal propagators into the general reduction and DE formulas.
The generic Hankel formulas already apply to these integrals, but retain redundant mode-function
states. We derive closed recurrences and a d log-form connection directly on this reduced space
(Secs.~\ref{sec:ibp} and \ref{sec:dlog}). This reduces the size of the system to be transported
without requiring a separate IBP solve for each diagram. Second, we automate the reduction and the numerical evaluation for arbitrary tree families.
From the tree specification, the package derives the closed reduction system of the family:
every convergent integral --- any contact sector and any integer shifts of the time powers ---
reduces to the master basis through closed recurrence formulas; the divergent configurations
are handled by the analytic regulator described below. Loop integrands --- the time integrals of loop diagrams, with the spatial loop-momentum
integrations left unperformed --- form families of the same kind and reduce the same way,
extended by the common-theta contacts that loops require (Sec.~\ref{sec:common-theta}).
Following the hierarchical energy limit of \cite{Chen:2024glu}, we construct the
boundary data for general tree topology: at leading order, the energy hierarchy fixes the
theta functions and the resulting factorized integrals supply Gamma-function coefficients.
Auxiliary energies include vertices without external legs in the same construction
(Sec.~\ref{sec:boundary}). Boundary data and local solution series can then be shared among
kinematic points. The benchmarks in Sec.~\ref{sec:app-massive-tree} demonstrate the resulting
multipoint efficiency.

The proposed framework is implemented as a software package \texttt{MadStree}, whose name stands for \emph{Mad calculations of dS tree diagrams}. It implements both the
reduced formulas and the numerical workflow in Mathematica. Its numerical backend is the
built-in \texttt{FlintNDE} module. For each specified Schwinger--Keldysh contour assignment, the
input gives the tree topology, propagators, vertex energies and time powers. The package
constructs the master system, the differential equation and boundary data, and returns values
along the requested path;
forming a physical correlator requires the contour sum and normalization factors appropriate
to the model. The construction applies to the scalar Hankel and exponential integral families
defined in Sec.~\ref{sec:notation}.

Numerically, the same workflow covers loop integrands; the bubble example in
Sec.~\ref{sec:app-bubble} illustrates the extension. For time integrals that require regularization, the package can
introduce a common analytic regulator on the vertex powers and determine the lowest Laurent power
from the boundary and DE. It then reconstructs the coefficients numerically following the procedure
used by AMFlow~\cite{Liu:2022chg,Huang:2026rjb}. The massless-chain
example demonstrates the extraction of a pole coefficient and a finite part.

Sec.~\ref{sec:method} presents the method, including the boundary construction, with the
common-theta derivation in App.~\ref{app:theta}. Sec.~\ref{sec:usage} describes package usage
and Laurent reconstruction. Sec.~\ref{sec:examples-section} presents examples, and
Sec.~\ref{sec:verification} compares the results with independent references. We conclude
in Sec.~\ref{sec:summary}.

\section{IBP reduction and differential equations for tree-level dS integrals}
\label{sec:method}

This section constructs the reduced master space, its IBP recurrences and d log-form DE,
and the boundary data needed for numerical evaluation. We first define the integral family
and conventions, then derive the local blocks and assemble them over the diagram. The massless
relations are imposed before assembly, so the formulas act directly on the reduced space.
Quantities inherited from the generic mode-function basis of \cite{Chen:2023iix} carry the
superscript ${\rm unreduced}$ of \eqref{eq:unreduced-space}. The boundary construction follows
in Sec.~\ref{sec:boundary}. Finally, Sec.~\ref{sec:common-theta} extends the contact rules to
several propagators sharing one theta function, as needed for loop diagrams.

\subsection{Notation: propagators, the integral family and index conventions}
\label{sec:notation}
\label{sec:conventions}

\paragraph{In-in Feynman rules and asymptotic behavior.}
Cosmological correlators are evaluated in the Schwinger--Keldysh (SK) formalism: the operator
product is ordered along a closed time path, and each vertex is labeled by a contour branch,
written \texttt{"+"} or \texttt{"-"}, with one time integration $\int_{-\infty}^0\dd\tau_v$ per
vertex. With the positive-frequency mode function
\begin{equation}
  u(\tau;k)=-\ii\frac{\sqrt{\pi}}{2}\,
  e^{\ii\pi(\nu/2+1/4)}H_{\rm Hubble}^{(d-1)/2}
  (-\tau)^{d/2}H^{(1)}_\nu(-k\tau),
  \label{eq:mode-function}
\end{equation}
where $k=|\bm k|$ and $\nu^2=d^2/4-m^2/H_{\rm Hubble}^2$ for a minimally coupled scalar in $d$
spatial dimensions ($\nu$ is real or imaginary for $d=3$), the bulk-to-bulk propagators are
\begin{equation}
  G_{>}(k;\tau_1,\tau_2)\equiv u(\tau_1,k)u^*(\tau_2,k),\qquad
  G_{<}(k;\tau_1,\tau_2)\equiv u^*(\tau_1,k)u(\tau_2,k),
  \label{eq:bulk-bulk-propagators}
\end{equation}
and the four Schwinger--Keldysh propagators used by the in-in rules are
\begin{align}
  G_{++}&=G_{>}\theta(\tau_1-\tau_2)+G_{<}\theta(\tau_2-\tau_1),
  &G_{--}&=G_{<}\theta(\tau_1-\tau_2)+G_{>}\theta(\tau_2-\tau_1),
  \notag\\
  G_{+-}&=G_{<},
  &G_{-+}&=G_{>}.
  \label{eq:sk-propagators}
\end{align}
Here $G_{++}$ is the time-ordered (Feynman) propagator and $G_{--}$ is
the anti-time-ordered (anti-Feynman) propagator. We refer to this pair
collectively as the time-ordered and anti-time-ordered propagators. The mixed-contour
components $G_{+-}$ and $G_{-+}$ are the Wightman propagators.
Throughout, $\theta_{ij}\equiv\theta(\tau_i-\tau_j)$ denotes the theta function of the time
difference of vertices $i$ and $j$. Bulk-to-boundary propagators of massless fields,
$G_\pm\propto(1\mp\ii k\tau)e^{\pm\ii k\tau}$, enter only as pure-exponential blocks.
In $d=3$, a minimally coupled massless field has $\nu=3/2$ and a conformally coupled one has
$\nu=1/2$; the two form one function family under the reduction, because the mode functions of
every half-integer order are finite sums of exponentials multiplied by powers of $\tau$, and the
recurrences of Sec.~\ref{sec:iterative} absorb the polynomial factors into shifted vertex time
powers. The master integrals therefore fix the massless blocks at $\nu=1/2$: every massless
building block below carries this order. The Hankel
functions obey
\begin{equation}
  \partial_\tau^2H^{(1,2)}_\nu(-k\tau)
  +\frac1\tau\,\partial_\tau H^{(1,2)}_\nu(-k\tau)
  +\left(k^2-\frac{\nu^2}{\tau^2}\right)H^{(1,2)}_\nu(-k\tau)=0,
  \label{eq:hankel-eom}
\end{equation}
with the two asymptotic regimes
\ie
  &H^{(1)}_\nu(-k\tau)=c_1(-k\tau)^\nu(1+O(\tau^2))+c_2(-k\tau)^{-\nu}(1+O(\tau^2)),
  \quad \tau\to0^-,\\
&c_1=e^{-\ii\pi\nu}c[\nu]\,, \quad c_2=c[-\nu]\,, \quad
c[\nu]\equiv \frac{ 2^{-\nu } \Gamma (-\nu )}{\ii \pi }  \, ,
  \label{eq:hankel-tau0}
\fe
and
\begin{align}
  &H^{(1)}_\nu(-k\tau)\sim\sqrt{\frac2\pi}(-k\tau)^{-1/2}
  e^{-\ii k\tau-\ii\pi(\nu/2+1/4)},\notag\\
  &H^{(2)}_\nu(-k\tau)\sim\sqrt{\frac2\pi}(-k\tau)^{-1/2}
  e^{\ii k\tau+\ii\pi(\nu/2+1/4)},
  \quad -k\tau\to\infty.
  \label{eq:hankel-infinity}
\end{align}
The first regime fixes the boundary conditions of the time integrals; the second justifies the
Wick rotation used in the numerical evaluation of Sec.~\ref{sec:numerics}.

\paragraph{The mode-function two-state $h$ block.}
A tree-level correlator with $c_V$ vertices contains one time integration per vertex, and each
integrand is a product of the elementary factors
\begin{equation}
  \left\{e^{\ii k\tau},\;H^{(1,2)}_\nu(-k\tau),\;
  \partial_\tau H^{(1,2)}_\nu(-k\tau),\;\theta(\tau_j-\tau_k)\right\}.
  \label{eq:integrand-factors}
\end{equation}
As in \cite{Chen:2024glu}, the Hankel prefactor and the time derivative are absorbed into two
mode-function states of one function system. We use the convention of the package, in which the prefactor is the
\emph{positive} power of $z$,
\begin{equation}
  \begin{aligned}
  h^{(\alpha)}(\nu,0;z)&\equiv z^{+\nu}H^{(\alpha)}_\nu(z),\\
  h^{(\alpha)}(\nu,1;z)&\equiv-\frac1k\,\partial_\tau h^{(\alpha)}(\nu,0;z)=\partial_z h^{(\alpha)}(\nu,0;z),
  \qquad z=-k\tau,
  \end{aligned}
  \label{eq:h-definition}
\end{equation}
where $\alpha\in\{1,2\}$ selects the Hankel branch and the second argument $a\in\{0,1\}$ selects the
undifferentiated function or its first derivative. A superscript on $h$ always denotes the
branch; when it is omitted, the statement applies to each branch separately. A massive
propagator carries two independent mode-function state bits, one at each vertex. The relations of
Sec.~\ref{sec:massless} reduce a massless (anti-)time-ordered propagator ($G_{++}$ or $G_{--}$) to one
shared two-state slot. A theta-free massless propagator ($G_{+-}$ or $G_{-+}$) contributes
no state slot. The package works directly in these reduced spaces.

The functions in \cite{Chen:2023iix,Chen:2024glu} differ in both their prefactor and the order
of the second Hankel branch. Denoting their order by $\nu_{\rm ref}$ and their functions by
$h_{\rm ref}^{(\alpha)}$, their definition is
\begin{equation}
  \begin{aligned}
  h_{\rm ref}^{(1)}(\nu_{\rm ref},a;z)
    &=\partial_z^a\!\left[z^{-\nu_{\rm ref}}H^{(1)}_{\nu_{\rm ref}}(z)\right],\\
  h_{\rm ref}^{(2)}(\nu_{\rm ref},a;z)
    &=\partial_z^a\!\left[z^{-\nu_{\rm ref}}H^{(2)}_{\nu_{\rm ref}^*}(z)\right],
    \qquad a\in\{0,1\}.
  \end{aligned}
  \label{eq:h-reference-definition}
\end{equation}
For real or purely imaginary order, the reflection identity gives
\begin{equation}
  H^{(2)}_{\nu_{\rm ref}^*}(z)
  =e^{\pi\operatorname{Im}\nu_{\rm ref}}H^{(2)}_{\nu_{\rm ref}}(z).
  \label{eq:h-conjugate-order}
\end{equation}
This factor is therefore present in the Wronskian and the pinched-propagator normalization
of those references. Setting $\nu_{\rm ref}=-\nu$ aligns their time powers and differential
systems with ours. The two branch reflection identities then give
\begin{equation}
  \begin{aligned}
  &h_{\rm ref}^{(1)}(-\nu,a;z_1)h_{\rm ref}^{(2)}(-\nu,b;z_2)\\
  &\qquad=e^{-\pi\operatorname{Im}\nu}
    h^{(1)}(\nu,a;z_1)h^{(2)}(\nu,b;z_2),
    \qquad a,b\in\{0,1\}.
  \end{aligned}
  \label{eq:h-reference-conversion}
\end{equation}
The same factor applies with the two Hankel branches interchanged. Thus the change of sign of
the order must be accompanied by this constant when translating a propagator kernel or its
contact from the reference convention.

The physical propagators also contain the mode prefactor in \eqref{eq:mode-function}. For
real times and momenta, its absolute square supplies $e^{-\pi\operatorname{Im}\nu}$, which
cancels the factor from replacing $H^{(2)}_{\nu^*}$ by $H^{(2)}_\nu$. In terms of the
common-order functions \eqref{eq:h-definition}, the result is
\begin{equation}
  \begin{aligned}
  G_>(k;\tau_1,\tau_2)
    ={}&\frac{\pi}{4}H_{\rm Hubble}^{d-1}k^{-2\nu}
    \bigl[(-\tau_1)(-\tau_2)\bigr]^{d/2-\nu}\\
    &\times h^{(1)}(\nu,0;-k\tau_1)h^{(2)}(\nu,0;-k\tau_2).
  \end{aligned}
  \label{eq:physical-common-order}
\end{equation}
For $G_<$ the branches are interchanged. The remaining powers and constants are restored
when forming a physical correlator; they are distinct from the sector normalization below.
These physical identities assume real or purely imaginary $\nu$, whereas the common-order
integral family itself admits analytic continuation in $\nu$.

Since the Hankel equation involves $\nu^2$ only, the sign change also relates the two
prefactor conventions in the equation of motion. In the convention
\eqref{eq:h-definition}, with \mbox{$\bm h=(h(\nu,0;z),h(\nu,1;z))^T$} and
$P_1\equiv\tfrac12(\Id-\sigma_3)=\diag(0,1)$,
\begin{equation}
  \partial_\tau\bm h
  =\left[-\ii k\sigma_2-\frac{1-2\nu}{\tau}\,P_1\right]\bm h,
  \label{eq:h-first-order-system}
\end{equation}
where $\sigma_{1,2,3}$ are the Pauli matrices and $\Id$ the $2\times2$ identity; the equivalent
dimensionless equation of motion is
\begin{equation}
  \partial_z^2h(\nu,0;z)+\frac{1-2\nu}{z}\,\partial_zh(\nu,0;z)+h(\nu,0;z)=0,
  \label{eq:h-dimensionless-eom}
\end{equation}
whose constant term is $1$; the physical $k^2$ enters only through $\partial_\tau=-k\partial_z$.

\paragraph{The vertex integral family and its index conventions.}
Every integral in this paper is written with one symbol, $\text{I}$, whose arguments are the sector,
the integer time-power shifts and the mode-function state bits. For a vertex carrying $n$ massive legs the
one-vertex family (an $n$-fold Hankel vertex integral family) reads
\begin{equation}
  \begin{aligned}
  \text{I}^{(a_0)}_{\tilde{\bm a};\alpha_1\cdots\alpha_n}
  &=\int_{-\infty}^0\dd\tau\,(-\tau)^{\nu_0+a_0}e^{\ii p_0\tau}
    \prod_{i=1}^{n}h^{(\alpha_i)}(\nu_i,a_i;-k_i\tau),\\
  &\hspace{2em} a_0\in\mathbb Z,\quad a_i\in\{0,1\},\quad \alpha_i\in\{1,2\},
  \end{aligned}
  \label{eq:vertex-integral-family}
\end{equation}
with the following index conventions, used throughout the diagram:
\begin{itemize}
  \item $\nu_0$ is the power of the vertex time, $\alpha_i\in\{1,2\}$ is the Hankel branch chosen at
        leg $i$ by the SK type of the propagator that ends there, $a_i\in\{0,1\}$ is the \emph{mode-function state bit}
        --- the index of the two components of \eqref{eq:h-definition} --- and $a_0$ is the integer
        \emph{shift} of the vertex time power;
  \item $p_0=\varsigma k_0$ is the signed energy of the vertex (Sec.~\ref{sec:ibp}), and
        $k_i=|\bm k_i|$ the momentum magnitude of leg $i$;
  \item the mode-function state bits are packed into one integer by the binary convention of \cite{Chen:2023iix},
  \begin{equation}
    \tilde{\bm a}=1+\sum_{i=1}^{n}a_i\,2^{\,n-i},
    \label{eq:binary-label}
  \end{equation}
  so that the first bit is the most significant and the $2^n$ entries
  $\text{I}_{\tilde{\bm a}}=\text{I}_{\{\bm a\}}$ of the master vector are ordered exactly as the
  Kronecker products of the leg spaces; in the package this is
  \texttt{IntegerDigits[...,2,n]} and the same order is used for the whole diagram.
\end{itemize}
For a diagram the same three kinds of index are kept, only made global. Fix a sector $s$ on a diagram
with $c_V$ vertices. It has $c_s$ time components, counted below, and $n^{\rm slot}_s$ mode-function two-state
slots: one for each massive
propagator at each of the two vertices it joins, and one shared by the two vertices of a massless
(anti-)time-ordered propagator. We write a master as
\begin{equation}
  \text{I}_{\tilde{\bm r}}\equiv \text{I}_{s}(\bm n;\bm a),
  \qquad \bm n\in\mathbb Z^{c_s},\quad \bm a\in\{0,1\}^{n^{\rm slot}_s},
  \label{eq:graph-index}
\end{equation}
where $\bm n$ assigns one integer time-power shift to each component (a component is a maximal
set of vertices merged by the contacts; a single vertex is one component), written $n_v$ for the
component $v$, with unit vector $\bm e_v$ ($\bm n\pm\bm e_v$ shifts $n_v$ by one and leaves the
other components unchanged), and $\bm a$ is the ordered
string of mode-function state bits of all slots, packed into $\tilde{\bm r}$ by \eqref{eq:binary-label}, and the
sector $s$ itself is recorded by a \emph{sector key}: with the root propagators listed in input
order $E=(e_1,\dots,e_{N_E})$ and $L_s^{\rm ctr}$ the set of pinched propagators,
\begin{equation}
  \operatorname{key}(s)=\xi_1\xi_2\cdots\xi_{N_E},
  \qquad \xi_j=\begin{cases}0,&e_j\in L_s^{\rm ctr},\\1,&e_j\notin L_s^{\rm ctr},\end{cases}
  \label{eq:sector-key}
\end{equation}
a fixed-length bit \emph{string}, not a binary integer, so leading zeros are kept and the top
sector is the all-$1$ string; every other sector contains at least one pinched propagator and
is called \emph{pinched}. Each pinched propagator merges two components and removes one time
integration, and the diagram is a tree, so
\begin{equation}
  c_s=c_V-\big|L_s^{\rm ctr}\big|,
  \label{eq:component-count}
\end{equation}
and only the top sector has as many shifts as the diagram has vertices. Throughout, we write $V_s$
for the set of components of $s$ and call its members the \emph{vertices} of $s$: a merged
component is integrated over a single time exactly as an unmerged vertex is, so it is one vertex
of $V_s$, and in the top sector $V_s$ is the vertex set of the diagram. All vertex-labeled objects
below --- $\tau_v$, $n_v$, $\bm e_v$, the energies, the generators and the letters --- are indexed
by these $v\in V_s$. Sectors are partially ordered
by $s_1\prec s_0$ when $s_1$ is reachable from $s_0$ by at least one contact, i.e.\ when
$L_{s_1}^{\rm ctr}\supsetneq L_{s_0}^{\rm ctr}$; we then call $s_1$ a \emph{subsector} of $s_0$ and
$s_0$ a \emph{supersector} of $s_1$. The subscripts distinguish sectors; they do not count
pinches. This ordering is the tree-level counterpart of the subsector relation of
flat-space IBP, where $S$ is a subsector of $S'$ if every propagator
power of $S$ is less than or equal to the corresponding power of $S'$ \cite{Klappert:2020nbg}: the
integer propagator powers that order flat-space sectors are here replaced by the contact indicator
bits $\xi_j$, since a propagator of a tree is either kept ($\xi_j=1$) or pinched out
($\xi_j=0$).

These three indices also label the package object
\texttt{MSIntegral[sectorKey, aShifts, stateBits]}, which represents the normalized integral
$J_s=\cN_s\widehat{\text I}_s$ defined below. Here
\texttt{sectorKey} is the string \eqref{eq:sector-key}, \texttt{aShifts} the vector $\bm n$ and
\texttt{stateBits} the string $\bm a$; the same slots are used by the returned records
\texttt{"sectorKey", "stateBits", "integral", "normalization"}.

\paragraph{Suppressed data and normalization.}
The indices in \eqref{eq:graph-index} label the integral within a fixed context. Its full
definition also includes the following data:
\begin{enumerate}
  \item the powers in \eqref{eq:physical-common-order} are split between the explicit vertex
        power $\nu_0$ and the definition \eqref{eq:h-definition}. The remaining physical mode
        constants and momentum powers multiply the correlator externally; they are not part
        of the package's sector normalization;
  \item the choice of the small-$\tau$ branch and of the vanishing powers at $\tau\to0$: both are
        contained in the superscript $(1,2)$ and the order $\nu$ of $h^{(\alpha)}$, which select between
        the two terms of \eqref{eq:hankel-tau0};
  \item the sector normalization $\cN_s$: the object the package propagates is
        $J_s(\bm n;\bm a)=\cN_s\,\widehat{\text I}_s(\bm n;\bm a)$, where
        $\widehat{\bm{\text I}}_s=Q_s\bm{\text I}_s$ includes the massless basis change
        defined in \eqref{eq:package-massless-basis}, with
        \begin{equation}
          \cN_s=\cN_s^{(0)}\!\!\prod_{\substack{e\in L_s^{\rm ctr}\\[1pt]e\ \text{massive (anti-)time-ordered}}}\!\!
          \left(-\sigma_e\,\frac{4\ii}{\pi}\,q_e^{\,2\nu_e-1}\right),
          \label{eq:sector-normalization}
        \end{equation}
        one factor $F_e$ of \eqref{eq:contact-factor} per pinched massive (anti-)time-ordered propagator;
        $\nu_e$ is the order and $q_e$ the momentum magnitude of that propagator, and $\sigma_e$ its contour sign,
        $+1$ for $G_{++}$ and $-1$ for $G_{--}$; a
        pinched massless (anti-)time-ordered propagator contributes the factor $1$, and $\cN_s^{(0)}$ is the
        normalization of the sector seed, equal to $1$ for every example of
        Sec.~\ref{sec:examples-section};
  \item the SK contour signs: the vertex phase sign $\varsigma_v$ of \eqref{eq:phase-block}, the incidence
        $\epsilon_{ve}$ of \eqref{eq:incidence} and the propagator sign $\sigma_e$ of
        \eqref{eq:contact-weight} are fixed by the
        vertex types and by the vertex order of each propagator. They determine the signed
        energies, the energy letters and the contact coefficients;
  \item the theta chamber: at a boundary chart each theta is frozen to $0$ or $1$ by the strict
        energy order (Sec.~\ref{sec:boundary}), and inside the bulk the ordering of a bundle is
        carried by the single $\delta$ of Sec.~\ref{sec:common-theta};
  \item the regulator: when a time power sits on a pole, a common analytic shift
        $\nu_0\to\nu_0+\varepsilon$ is understood (Sec.~\ref{sec:regulator}).
\end{enumerate}
These data determine the coefficients and boundary conditions even though they are suppressed
in the integral label. In particular, the momentum dependence of $\cN_s$ contributes to the
normalized connection, as described in Sec.~\ref{sec:remaining}.

\subsection{IBP relations from the three building blocks}
\label{sec:ibp}
\label{sec:global}

The building blocks assembled in this subsection drive the reduction: from them, the one-step
relations of Sec.~\ref{sec:iterative} express an integral whose
time power at one vertex is shifted by one unit in terms of integrals one step closer to the
master value, and their iteration reduces shifted integrals of the family to the zero-shift
masters. Three building blocks contribute --- the theta-free vertex
exponential, the massive vertex $h$ block and the massless shared block; their relative
normalization is fixed first.

\paragraph{The relative normalization of the recurrences.}
The recurrences below use the $+\nu$ convention and the $(-\tau)^{\nu_0}$ time power
of \eqref{eq:vertex-integral-family}.
With $(-\tau)^{A}$ instead of $\tau^{\nu_0}$, the derivative of the explicit
power picks up a sign, $\partial_\tau(-\tau)^A=-A(-\tau)^{A-1}$, so the one-step relation carries
$-\text{M}_1$ where the reference carries $+\text{M}_1$; the diagonal part of the DE is unaffected
but the contact block of the DE changes sign, which is why our \eqref{eq:contact-block-full} differs
from the printed \cite{Chen:2024glu} form by that overall sign.

\paragraph{Block (i): the theta-free exponential of a vertex.}
Each vertex $v$ carries exactly \emph{one} theta-free exponential factor,
\begin{equation}
  E_v(\tau_v)=\exp\!\left(\ii\varsigma_v k_{0,v}\tau_v\right),
  \qquad
  \varsigma_v=\begin{cases}-1,&v\ \text{is a $+$ vertex},\\[2pt]
  +1,&v\ \text{is a $-$ vertex},\end{cases}
  \qquad p_{0,v}=\varsigma_vk_{0,v},
  \label{eq:phase-block}
\end{equation}
so that a \texttt{"+"} vertex gives $e^{-\ii k_{0,v}\tau_v}$ and a \texttt{"-"} vertex gives
$e^{+\ii k_{0,v}\tau_v}$. Here $k_{0,v}$ is the
\emph{single summed} external-energy parameter of the vertex, the sum of the magnitudes of the theta-free
external legs attached to $v$; in the package it is the input field \texttt{externalLegEnergy}.
Eq.~(3.1) of \cite{Chen:2023iix}
likewise writes one $k_0$ per vertex, with no separate exponential parameters for individual legs or blocks. A theta-free propagator branch is not folded into
this symbol: it keeps its own magnitude, and that magnitude is added to the energy of whichever
component holds the vertex it ends on. After a contact has merged a set $C$ of vertices
into one component, the same scalar becomes the component energy
\begin{equation}
  p_{0,C}=\sum_{u\in C}\varsigma_uk_{0,u}
  +\!\!\sum_{\substack{\text{theta-free propagators }e\\[-1pt]
  \text{with exactly one vertex in }C}}\!\!(-\varsigma_{v_e})\,q_e,
  \label{eq:component-energy}
\end{equation}
where $q_e=|\bm q_e|$ is the momentum magnitude of the propagator and $v_e$ is the vertex of $e$ that
lies in $C$; a branch whose two vertices both lie in $C$ contributes $+q_e-q_e=0$ and is therefore
already absent from this sum. The
second sum collects the massless Wightman and external branches, which create no slot: they add only
their magnitudes to the same scalar, each with the contour sign of the vertex where the branch ends. The
first sum alone is $p_{0,u}$, so for a one-vertex component \eqref{eq:component-energy} reduces to
$p_{0,u}$ exactly when that vertex carries no theta-free propagator branch.

Writing $\cE_v\simeq\mathbb C$ for the one-dimensional space of this exponential, the
block creates no mode-function state bit and no slot: it contributes nothing to
$\text{M}_1$ and contributes the scalar $\ii p_{0,v}\Id$ to $\text{M}_0$.

\paragraph{Block (ii): the $h$ system of a massive propagator.}
A massive propagator of order $\nu$ contributes, at each of the two vertices it joins, one mode-function two-state
system
\eqref{eq:h-definition}, with the two-dimensional space $\cH_i\simeq\mathbb C^2$ --- one factor per
propagator end, indexed by $i$ in the diagram below --- and the generators
\eqref{eq:h-first-order-system}. This is the block of \cite{Chen:2023iix} unchanged, with $1-2\nu$
replacing the printed $2\nu+1$. Only the branch $\alpha\in\{1,2\}$, fixed by the SK type of the
propagator, differs between the two vertices it joins, and it does not change the generator
$-\ii k\sigma_2$.

\paragraph{Block (iii): the massless (anti-)time-ordered propagator.}
\label{sec:massless}
Before the extra massless relations are imposed, a massless (anti-)time-ordered propagator $e=(u_e,v_e)$ is exactly
like a massive one: the vertex $u_e$ carries two mode-function states $a\in\{0,1\}$, the vertex $v_e$ carries two
mode-function states $b\in\{0,1\}$, and the local space of this propagator is
$\cH_{e,u}\otimes\cH_{e,v}\simeq\mathbb C^4$ with, in the natural order,
\begin{equation}
  \bm{\text I}_e^{(4)}=(\text{I}_{00},\text{I}_{01},\text{I}_{10},\text{I}_{11})^T ,
  \label{eq:four-state}
\end{equation}
where the $\text{I}_{ab}$ are the integrals defined by \eqref{eq:massless-relations} below, in the
convention
\eqref{eq:graph-index}.
The diagram-level matrices follow from the Leibniz rule. The integrand is a product of
independent factors, one exponential per vertex and one first-order mode-function system per
propagator end, so $\partial_{\tau_v}$ differentiates one factor at a time and every other
factor is a spectator. We organize the states as a tensor product and define
$\Emb_i(A)=\Id\otimes\cdots\otimes A\otimes\cdots\otimes\Id$ to act on the slot $i$ alone, each
vertex matrix is the sum of the embedded generators of the factors that depend on $\tau_v$:
\ie
  \cV_s^{\,\rm unreduced}
  &=\bigotimes_{v\in V_s}\cE_v\otimes\!\!
    \bigotimes_{\substack{\text{active two-state}\\\text{factors }i}}\!\!\cH_i,
  \qquad
  \dim\cV_s^{\,\rm unreduced}=2^{N_s},
  \label{eq:unreduced-space}\\
  \text{M}_{1;v}^{\,\rm unreduced}
  &=\nu_{0,v}\Id-\sum_{i\in H_v^{\,\rm unreduced}}(1-2\nu_i)\Emb_i(P_1),\\
  \text{M}_{0;v}^{\,\rm unreduced}
  &=\ii p_{0,v}\Id-\ii\sum_{i\in H_v^{\,\rm unreduced}}k_i\Emb_i(\sigma_2).
  \fe
Here $V_s$ is the set of vertices of $s$ defined after \eqref{eq:component-count}, with
$p_{0,v}$ the component energy \eqref{eq:component-energy} for a merged component and the vertex
energy \eqref{eq:phase-block} otherwise, $H_v^{\,\rm unreduced}$ the mode-function
two-state factors of the propagators of $s$ ending on $v$ --- massive and massless alike, before the
reduction below --- and $N_s$ their total number in $s$; the product carries one $\cE_v$ per
vertex of $V_s$ and one $\cH_i$ per two-state factor. These are the matrices of
\cite{Chen:2023iix}, with the $+\nu$ prefactor \eqref{eq:h-definition} and
the $(-\tau)^{\nu_0}$ time power.

Imposing the extra massless relations reduces these four mode-function states at the two vertices to one
shared mode-function two-state system.
At the conformal value $\nu=1/2$ the positive-prefactor block degenerates: the two branches are pure
exponentials, $h^{(1)}(\tfrac12,0;z)=-\ii\sqrt{2/\pi}\,e^{\ii z}$ and
$h^{(2)}(\tfrac12,0;z)=+\ii\sqrt{2/\pi}\,e^{-\ii z}$, \eqref{eq:h-dimensionless-eom} becomes
$h''+h=0$, and the derivative mode-function state is a branch eigenstate of the zeroth-order mode-function state. Carrying the
derivative out explicitly,
\begin{equation}
  \begin{aligned}
  h^{(1)}(\tfrac12,1;z)
  &=\partial_z\!\left(-\ii\sqrt{\tfrac2\pi}\,e^{\ii z}\right)
  =\sqrt{\tfrac2\pi}\,e^{\ii z}
  =+\ii\,h^{(1)}(\tfrac12,0;z),\\[2pt]
  h^{(2)}(\tfrac12,1;z)
  &=\partial_z\!\left(+\ii\sqrt{\tfrac2\pi}\,e^{-\ii z}\right)
  =\sqrt{\tfrac2\pi}\,e^{-\ii z}
  =-\ii\,h^{(2)}(\tfrac12,0;z).
  \end{aligned}
  \label{eq:h-branch-eigenvalues}
\end{equation}
Consistently with the fixed massless order of Sec.~\ref{sec:conventions}, the quotient derived
here acts on the pure exponential blocks: an input propagator of another half-integer order must
first be expanded into $\nu=1/2$ blocks with shifted vertex powers.
Its defining property is that the two branches carry \emph{opposite} signs in
\eqref{eq:h-branch-eigenvalues}, while a single \emph{(anti-)time-ordered} propagator puts the two branches on its
two vertices in both possible orders, one per theta term. Write $\sigma_e=+1$ for $G_{++}$ and
$\sigma_e=-1$ for $G_{--}$. For a massless (anti-)time-ordered propagator $e=(u_e,v_e)$ of momentum magnitude $q_e$,
with $\tau_u\equiv\tau_{u_e}$ and $\tau_v\equiv\tau_{v_e}$, the four entries of
\eqref{eq:four-state} are, for both $G_{++}$ and $G_{--}$,
\begin{equation}
  \text{I}_{00}=\text{I}_{11}=A_e+B_e,
  \qquad
  \text{I}_{10}=-\text{I}_{01}=\ii\sigma_e\left(A_e-B_e\right),
  \label{eq:massless-relations}
\end{equation}
in terms of the two theta chambers of the propagator, taken undifferentiated,
\begin{equation}
  \begin{aligned}
  A_e={}&\int_{-\infty}^{0}\!\!\dd\tau_u\!\int_{-\infty}^{0}\!\!\dd\tau_v\;\Phi_e\,\theta_{uv}\,
  h^{(1)}\!\left(\tfrac12,0;-q_e\tau_u\right)h^{(2)}\!\left(\tfrac12,0;-q_e\tau_v\right),\\
  B_e={}&\int_{-\infty}^{0}\!\!\dd\tau_u\!\int_{-\infty}^{0}\!\!\dd\tau_v\;\Phi_e\,\theta_{vu}\,
  h^{(2)}\!\left(\tfrac12,0;-q_e\tau_u\right)h^{(1)}\!\left(\tfrac12,0;-q_e\tau_v\right),
  \end{aligned}
  \label{eq:chamber-integrals}
\end{equation}
which is \eqref{eq:sk-propagators} written out in the $h$ mode-function state basis: the branch pair
of each chamber is the one carried by $G_{\sigma_e}$, so \eqref{eq:chamber-integrals} displays
$G_{++}$, and $G_{--}$ exchanges the two Hankel branches inside each term. Here $\Phi_e$ collects
every other factor of the integrand: the vertex exponentials of $u$ and $v$, their time powers, and
the factors of all the remaining propagators. The relations \eqref{eq:massless-relations} follow
because raising the
mode-function state at $u_e$ from $0$ to $1$ multiplies $A_e$ by $+\ii\sigma_e$ and $B_e$ by
$-\ii\sigma_e$, and raising it at $v_e$ multiplies them by $-\ii\sigma_e$ and $+\ii\sigma_e$: the sign
of the eigenvalue \eqref{eq:h-branch-eigenvalues} is fixed by the branch, and the two chambers hold
the two branches in opposite order. The undifferentiated chambers themselves give the
$\text{I}_{00}$ entry, a derivative with respect to $\tau_u$ or $\tau_v$ is the corresponding
replacement times $-q_e$ by \eqref{eq:h-definition}, and differentiating both ends returns the
$\text{I}_{11}$ entry because the two eigenvalues at one propagator multiply to $+1$.
Only two of the four mode-function states are therefore independent, but exactly
two: all four are built from the \emph{sum} and the \emph{difference} of the two chambers, and the two
chambers are different objects, with complementary regions of time ordering and the two Hankel
branches interchanged between $u$ and $v$. No constant $\lambda$ can relate
$\text{I}_{10}=\lambda\,\text{I}_{00}$, since $A_e$ and $B_e$ coincide at the exchange-symmetric
point $\Phi_e(\tau_u,\tau_v)=\Phi_e(\tau_v,\tau_u)$, where $\text{I}_{10}$ vanishes while
$\text{I}_{00}$ does not, and are unequal at a generic point, where $\text{I}_{10}$ does not vanish.
Differentiating therefore does not return the same integral up to a phase, as it would for a block
made of a \emph{single} theta term: the theta-free massless branches of
\eqref{eq:component-energy} are of that kind, which is why they collapse to one dimension and create
no slot, while an (anti-)time-ordered propagator keeps a genuine mode-function two-state factor. We choose the shared mode-function two-state
$\bm g_e=(\text{I}_{00},\text{I}_{10})^T$ and write $\cM_e\simeq\mathbb C^2$ for the
two-dimensional space it spans. This space is the single slot that the massless (anti-)time-ordered propagator $e$
contributes to the diagram. The constant embedding and
projection are
\begin{equation}
  \bm{\text I}^{(4)}_e=S_e\bm g_e,\qquad
  S_e=\begin{pmatrix}1&0\\0&-1\\0&1\\1&0\end{pmatrix},\qquad
  P_e=\begin{pmatrix}1&0&0&0\\0&0&1&0\end{pmatrix},\qquad
  P_eS_e=\Id_2,
  \label{eq:SP-massless}
\end{equation}
and every four-dimensional operator that preserves the relation subspace \eqref{eq:massless-relations}
reduces to $P_eOS_e$. In particular
\begin{equation}
  P_e(\sigma_2\otimes\Id_2)S_e=\sigma_2,
  \qquad
  P_e(\Id_2\otimes\sigma_2)S_e=-\sigma_2,
  \label{eq:endpoint-projections}
\end{equation}
so the two vertices joined by a massless (anti-)time-ordered propagator act with opposite signs of $\sigma_2$
on one shared two-dimensional slot. Its bit $r_e$ enters the mode-function state string of both
vertices joined by $e$; the slot cannot be duplicated into two independent mode-function two-state factors. Consequently the full-graph tensor structure survives, but
the vertex-by-vertex factorization
$\cV_s\stackrel{?}{=}\bigotimes_{v\in V_s}\cV_v$ fails, because $\cM_e$ cannot belong to both $u_e$
and $v_e$ at once. The reduced space of a sector is
\begin{equation}
  \cV_s=\bigotimes_{v\in V_s}\cE_v
  \otimes\bigotimes_{i\in H(s)}\cH_i
  \otimes\bigotimes_{e\in E_m(s)}\cM_e,
  \qquad \dim\cV_s=2^{|H(s)|+|E_m(s)|},
  \label{eq:reduced-space}
\end{equation}
where $H(s)$ is the set of mode-function two-state factors that the massive propagators of $s$ still
contribute, one space $\cH_i$ each, and $E_m(s)$ the set of its unpinched massless (anti-)time-ordered
propagators, one shared space $\cM_e$ each.

The theta-derivative contact vector $\bm\rho_e^{(4)}=(0,-1,+1,0)^T=S_e|1\rangle_e$
fixes the shared-slot selector to $|1\rangle_e=(0,1)^T$ by \eqref{eq:SP-massless}.
With the propagator sign $\sigma_e$ of \eqref{eq:massless-relations}, the vertices
$u_e$ and $v_e$, in that order, carry the \emph{contact weights}
\begin{equation}
  w_{e}^{(u_e)}=+2\ii\sigma_e,\qquad w_{e}^{(v_e)}=-2\ii\sigma_e,
  \label{eq:contact-weight}
\end{equation}
in the branch pair $-\ii e^{\ii z}$ and $+\ii e^{-\ii z}$ that \eqref{eq:h-branch-eigenvalues} leaves
at $\nu=\tfrac12$ once the common factor $\sqrt{2/\pi}$ is divided out; in the $h$ mode-function
state basis of \eqref{eq:h-definition} the same pair reads $\pm4\ii\sigma_e/\pi$. A massive
propagator stays in $h$ normalization at both ends, and its contact keeps the $4\ii/\pi$ explicit.
In Sec.~\ref{sec:remaining} we separate this scalar Wronskian factor $F_e$ from a constant
selector $C^{(u_ev_e)}$: the local column of $C^{(u_ev_e)}$ is
$-\bm\rho_e^{(4)}$, so that at $\nu_e=\tfrac12$ the product $F_eC^{(u_ev_e)}$ carries precisely
these raw contact weights. The product, rather than $C^{(u_ev_e)}$ alone, enters
\eqref{eq:R-two-vertex}.

The package uses unit-amplitude sum and difference states. For one shared slot this is the
constant change of basis
\begin{equation}
  \widehat{\bm g}_e=\frac{\pi}{2}D_{\sigma_e}\bm g_e,
  \qquad D_{\sigma_e}=\diag(1,\ii\sigma_e),
  \qquad
  \widehat{\bm g}_e=\frac{\pi}{2}(A_e+B_e,\,B_e-A_e)^{\mathsf T}.
  \label{eq:package-massless-basis}
\end{equation}
The factor $\pi/2$ removes the two Hankel amplitudes. The transformed contact weights are
$(-2,+2)$ at $(u_e,v_e)$, while the single-vertex generator is proportional to
$\ii\sigma_e q_e\sigma_1$ and is diagonalized by the Hadamard matrix
$\tfrac1{\sqrt2}\left(\begin{smallmatrix}1&1\\1&-1\end{smallmatrix}\right)$.
On a sector, write $Q_s$ for the tensor product of these changes on its massless shared
slots and the identity on its massive slots, and set
$\widehat{\bm{\text I}}_s=Q_s\bm{\text I}_s$. The transformation preserves the energy letters
and is distinct from the sector normalization $\cN_s$.

\paragraph{Local generators and the full-graph $\text{M}_1$, $\text{M}_0$.}
For each massless (anti-)time-ordered propagator, whose two vertices we order as $(u_e,v_e)$, define the incidence
\begin{equation}
  \epsilon_{ve}=\begin{cases}+1,&v=u_e,\\-1,&v=v_e,\\0,&v\notin e,\end{cases}
  \label{eq:incidence}
\end{equation}
and let $H_v(s)$ be the set of mode-function two-state factors that the massive propagators of $s$ contribute at
$v$. Each building block then contributes to the vertex $v$ the pair of atoms
\begin{align}
  \Delta\text{M}_{1}^{(E_v)}&=0, & \Delta\text{M}_{0}^{(E_v)}&=\ii p_{0,v}\Id,
  &&\text{the exponential of the vertex }v,
  \label{eq:phase-block-M10}\\
  \Delta\text{M}_{1}^{(h_i)}&=-(1-2\nu_i)P_1, & \Delta\text{M}_{0}^{(h_i)}&=-\ii k_i\sigma_2,
  &&\text{a massive propagator of order }\nu_i\text{ at }v,
  \label{eq:massive-h-block-M10}\\
  \Delta\text{M}_{1}^{(\cM_e)}&=0, &
  \Delta\text{M}_{0}^{(\cM_e)}&=-\ii\epsilon_{ve}q_e\sigma_2,
  &&\text{a massless propagator }(\nu=\tfrac12)\text{ at }v.
  \label{eq:massless-shared-block-M10}
\end{align}
The massive atom acts on the two-state space of a massive propagator, the massless atom on the
shared slot, once per vertex with the sign $\epsilon_{ve}$; a massless propagator does not
enter $\text{M}_1$ at all, its $\nu=1/2$ making $1-2\nu$ vanish identically. The complete
matrices of a vertex $v$ in a sector $s$ are the sums of these atoms, each embedded in the slot
it acts on, together with the $\nu_{0,v}\Id_s$ of the vertex time power,
\begin{align}
  \text{M}_{1;v,s}(\nu_{0,v})
  &=\nu_{0,v}\Id_s+\Emb_{\cE_v}\!\left(\Delta\text{M}_{1}^{(E_v)}\right)
  +\sum_{i\in H_v(s)}\Emb_i\!\left(\Delta\text{M}_{1}^{(h_i)}\right)\nn\\
  &\quad+\sum_{e\in E_m(s)}\Emb_e\!\left(\Delta\text{M}_{1}^{(\cM_e)}\right)\nn\\
  &=\nu_{0,v}\Id_s-\sum_{i\in H_v(s)}(1-2\nu_i)\Emb_i(P_1),
  \label{eq:global-M1}\\
  \text{M}_{0;v,s}
  &=\Emb_{\cE_v}\!\left(\Delta\text{M}_{0}^{(E_v)}\right)
  +\sum_{i\in H_v(s)}\Emb_i\!\left(\Delta\text{M}_{0}^{(h_i)}\right)
  +\sum_{e\in E_m(s)}\Emb_e\!\left(\Delta\text{M}_{0}^{(\cM_e)}\right)\nn\\
  &=\ii p_{0,v}\Id_s+\sum_{i\in H_v(s)}\Emb_i(-\ii k_i\sigma_2)
  +\sum_{e\in E_m(s)}\Emb_e(-\ii\epsilon_{ve}q_e\sigma_2).
  \label{eq:global-M0}
\end{align}
The reduction only replaces the two mode-function two-state factors of a massless propagator by
one shared factor: $\partial_{\tau_v}$ still acts only on the blocks belonging to $v$, so the
vertex generators remain sums of embeddings on the common space \eqref{eq:reduced-space}, and
different vertices meet only on a shared massless slot, where they carry the same $\sigma_2$
and opposite $\epsilon_{ve}$.

\subsection{d log-form differential equations}
\label{sec:dlog}

This subsection derives the differential equations for the reduced master
integrals from the vertex IBP relations assembled in Sec.~\ref{sec:ibp}.
For a given sector, differentiation with respect to the kinematic variables produces master integrals with shifted time powers, which can be reduced to the master integrals of this sector and its subsectors via IBP relations. We give the diagonal sector blocks in d $\log$ form, identify their energy letters, and present the iterative reduction formulas.
The contact terms and off-diagonal blocks are constructed in
Sec.~\ref{sec:remaining}.

\paragraph{Master choice and the single-vertex block.}
The masters of a sector $s$ are the integrals \eqref{eq:graph-index} at vanishing time-power shift,
$\text{I}_s(\bm 0;\bm a)$. Collected in the bit order \eqref{eq:binary-label}, they form the ordered
vector
\begin{equation}
  \bm{\text I}_s(\bm n)=\big(\text{I}_s(\bm n;\bm a)\big)_{\bm a\in\{0,1\}^{n^{\rm slot}_s}},
  \qquad
  \bm{\text I}=\big(\bm{\text I}_s(\bm 0)\big)_{{\rm all\ sectors}\ s},
  \label{eq:master-vector}
\end{equation}
where $\bm{\text I}$ is the stack of the $\bm{\text I}_s(\bm 0)$ over all sectors. Writing its connection as $\mathcal A$, the single-vertex diagonal blocks admit logarithmic
potentials,
\begin{align}
&\nd \bm{\text I} = \mathcal A\bm{\text I},\qquad
\mathcal A=\sum_x\mathcal A_x\nd x,\qquad
\mathcal A_{ss}=\nd\Omega_{ss},\nn\\
&\Omega_{ss}
 =  \Omega_{{\rm ex},s} - \ii \sum_{v\in V_s}U_s^{-1}\cdot  \widetilde{\Omega}_{0;v,s} \cdot  U_s \cdot \text{M}_{1;v,s}(\nu_{0,v}+1) \, ,\label{eq:ndlogDE}
\end{align}
where $x$ runs over the independent kinematic variables. The full d log-form system is obtained
in the normalized basis of \eqref{eq:normalized-vector}.
The derivation is that of \cite{Chen:2023iix}: differentiate the master vector, reduce the
differentiated integrals back onto the masters by applying the one-step recurrence
\eqref{eq:one-step-ibp} iteratively, and read off the connection. The massless reduction changes
nothing in that derivation except the two matrices through which the answer is written:
$\text{M}_{1;v,s}$ and $\text{M}_{0;v,s}$ are now built on the reduced space
\eqref{eq:reduced-space}, and the diagonalizer of \cite{Chen:2023iix} is replaced by the $U_s$ of
\eqref{eq:global-M0-diagonal}, which carries an extra factor on every shared slot. Because the
reduced generators still commute slot by slot, that one $U_s$ diagonalizes every vertex matrix at
once, so \eqref{eq:ndlogDE} follows from the formula given in \cite{Chen:2023iix} by this single
replacement.
We use $\bm a,\bm b$ for the original state strings and $\bm d,\bm d'$ for the strings in the
simultaneous diagonal basis defined below; each string has $n_s^{\rm slot}$ bits.
The two ingredients are diagonal in different bases: $\widetilde\Omega_{0;v,s}$ in the
$\bm d$ basis and $\Omega_{{\rm ex},s}$ in the $\bm a$ basis,
\ie \label{eq:omega-form}
&\left( \widetilde{\Omega}_{0;v,s} \right)_{\bm d'\bm d}  \equiv \begin{cases} - \ii \log \kappa_{v,s;\bm d} , & \bm d'=\bm d \\
0, & \bm d'\neq\bm d
\end{cases} \, , \\
&\left( \Omega_{\rm ex,s} \right)_{\bm{b}\bm{a}}  \equiv \begin{cases} - \sum_{i\in H(s)} a_i(1-2\nu_i) \log k_i , & \bm{b}=\bm{a} \\
0, & \bm{b}\neq \bm{a}
\end{cases} \, ,
\fe
with the energy letters
\begin{equation}
  \kappa_{v,s;\bm d}=p_{0,v}
  +\sum_{i\in H_v(s)}(2d_i-1)k_i
  +\sum_{e\in E_m(s)}\epsilon_{ve}(2d_e-1)q_e,
  \label{eq:global-energy-letter}
\end{equation}
where $\bm d$ contains an eigenstate bit $d_i$ for each massive slot and $d_e$ for each shared slot.
The same $d_e$ enters the letters of $u_e$ and $v_e$ with opposite signs, so the letters
remain linear forms in the momenta even for diagrams with shared slots. Only the mode-function states of massive propagators contribute to
$\Omega_{\rm ex}$, because a massless slot has $1-2\nu=0$.

$\text{M}_{1;v,s}$ is diagonal in the original mode-function state basis. Each
$\text{M}_{0;v,s}$ is a sum of $\sigma_2$ matrices (or $\sigma_1$ matrices in the package basis)
acting on individual slots. The matrices $\text{M}_{0;v,s}$ for different vertices therefore
commute with one another and are simultaneously diagonalized by one constant matrix, the
Kronecker product of the massive diagonalizer $\text{T}$ with the diagonalizer of the shared slots:
\begin{align}
  &\left(\text{M}_{1;v,s}[\nu_{0,v},\bm{\nu}]\right)_{\bm{b}\bm{a}}= \begin{cases}
  \nu_{0,v}-\sum_{i\in H_v(s)} a_i(1-2\nu_i) , & \bm{b}=\bm{a} \\
  0, & \bm{b}\neq \bm{a}\end{cases}\, ,
  \label{eq:M-form}\\
  &\left(\text{T}_n\right)_{\bm d\bm a}=\prod_{i=1}^n \text{T}_{d_i a_i},
  \qquad
  \text{T}=\frac{1}{\sqrt{2}}\begin{pmatrix} 1 & -\ii \\ -\ii & 1 \end{pmatrix},
  \nn\\
  &U_s=\bigotimes_{i\in H(s)}\text{T}_i\otimes\bigotimes_{e\in E_m(s)}\text{T}_e,
  \qquad
  \widetilde{\text{M}}_{0;v,s}=U_s\text{M}_{0;v,s}U_s^{-1}=\diag(\ii\kappa_{v,s;\bm d}).
  \label{eq:global-M0-diagonal}
\end{align}
Here $\text{T}$ diagonalizes $\sigma_2$ ($\text{T}\sigma_2\text{T}^{-1}=\sigma_3$); on the shared
massless slots of the package quotient the same factor is instead the Hadamard matrix, which
diagonalizes $\sigma_1$, and the two choices give the same set of letters
\eqref{eq:global-energy-letter} because $D_{\sigma_e}$ only relabels the shared bit. The condition
for one constant $U_s$ to diagonalize every vertex matrix slot by slot is that all the Pauli
directions arriving on a given slot mutually commute, which holds here because every massive slot
receives only $\sigma_2$ and every shared slot only $\sigma_2$ (or, in the quotient, only
$\sigma_1$). For a
single vertex all of whose legs are massive the sums reduce to one term with $U=\text{T}_n$ and
\eqref{eq:ndlogDE} is the uniform formula of \cite{Chen:2024glu}.

Inverting $\text{M}_{0;v,s}$ gives
$U_s^{-1}\diag(1/(\ii\kappa_{v,s;\bm d}))U_s$, i.e.\ the letter-by-letter reciprocal of the scalar
letters \eqref{eq:global-M0-diagonal}, and $\text{M}_{1}^{-1}$ is the reciprocal of its diagonal
entries. User or Wick coordinates are implemented by pulling back the one-form,
and changing basis to the package quotient is the constant similarity
transformation generated by the $D_{\sigma_e}$.

\paragraph{What the extra relations change.}
Three statements isolate the effect of the massless reduction on the DE. (1) A massless shared
slot contributes nothing to $\Omega_{\rm ex}$ or to $\text{M}_1$, so the diagonal logarithmic
structure in the orders $\nu_i$ is exactly the massive one. (2) It contributes one
eigenstate bit $d_e$ to the letters of both vertices with opposite incidence
\eqref{eq:global-energy-letter}, so the two vertices share the bit while their recurrences
remain separate and in closed form. (3) The theta derivative of an (anti-)time-ordered propagator, massive or
massless, does not act within the sector: its delta function sets the two vertex times equal,
so the contact terms are integrals of strictly lower sectors and form the off-diagonal blocks
\eqref{eq:contact-block-full}. The vertices joined by the pinched propagators become one vertex;
its \emph{base power of time} is
\begin{equation}
  \nu_{0,C}=\sum_{u\in C}\nu_{0,u}+\sum_{e\in E_s(C)}\varrho_e+|C|-1,
  \label{eq:base-power}
\end{equation}
where $C$ is the set of vertices joined into one component by the pinched propagators, and
$E_s(C)$ the set of those propagators with both vertices in $C$. If propagators join $v_1$ with
$v_2$ and $v_2$ with $v_3$, then $C$ is the whole set $\{v_1,v_2,v_3\}$. Here $\varrho_e$ is
the raw contact power of the propagator --- $2\nu_e-1$ for a massive (anti-)time-ordered propagator, $0$ for a
massless one --- as will follow from \eqref{eq:contact-factor} and \eqref{eq:massless-relations}.
In a sector $s_1$ in which $C$ is one component, $C\in V_{s_1}$: for this
vertex $v=C$, the symbols $\tau_v$, $p_{0,v}$ and $\nu_{0,v}$ are $\tau_C$, the component energy
\eqref{eq:component-energy} and the base power \eqref{eq:base-power}, and the matrices
$\text{M}_{0;v,s_1}$, $\text{M}_{1;v,s_1}$ and the letters \eqref{eq:global-energy-letter} at $v$ are
built from the legs of the whole component. A later contact may merge $C$ with another component.
In the top sector every vertex is a single component, and all of these are the original vertex quantities.

\paragraph{Iterative reduction.}
\label{sec:iterative}
One step of the recurrence is taken at one vertex $v\in V_s$: in $\bm{\text I}_s(\bm n)$ the time
of $v$ enters at the power $\nu_{0,v}+n_v$, and the step changes $n_v$ by one. The complete
one-step IBP at $v$ is
\begin{equation}
  -\text{M}_{1;v,s}(\nu_{0,v}+n_v)\,\bm{\text I}_s(\bm n-\bm e_v)
  +\text{M}_{0;v,s}\,\bm{\text I}_s(\bm n)
  +\cR_{v,s}(\bm n)=0,
  \label{eq:one-step-ibp}
\end{equation}
where the minus in front of $\text{M}_1$ is the $(-\tau)^A$ derivative and $\cR_{v,s}$ collects the
theta contact terms of the propagators incident to $v$ (Sec.~\ref{sec:remaining}). Solving
\eqref{eq:one-step-ibp} for the vector at $\bm n-\bm e_v$, and the same relation taken one step
higher for the one at $\bm n+\bm e_v$, gives the two directions
\begin{align}
  A_{-,v,s}(\bm n)&=\text{M}_{1;v,s}(\nu_{0,v}+n_v)^{-1}\text{M}_{0;v,s},
  \label{eq:global-Aminus}\\
  A_{+,v,s}(\bm n)&=U_s^{-1}\widetilde{\text{M}}_{0;v,s}^{-1}U_s\,
  \text{M}_{1;v,s}(\nu_{0,v}+n_v+1),
  \label{eq:global-Aplus}
\end{align}
so $\bm{\text I}_s(\bm n\pm\bm e_v)$ is expressed in terms of $\bm{\text I}_s(\bm n)$ with only two
diagonal inverses, up to $\cR$: the lowering step carries $+\text{M}_{1;v,s}^{-1}\cR_{v,s}(\bm n)$
and the raising step
$-\text{M}_{0;v,s}^{-1}\cR_{v,s}(\bm n+\bm e_v)$. The lowering step is singular when an eigenvalue of $\text{M}_1$ vanishes,
$\nu_{0,v}+n_v-\sum_ia_i(1-2\nu_i)=0$; this is a property of the mode-function states and the time
powers, not of the kinematics, and the regulator of Sec.~\ref{sec:regulator} moves the solution away
from this singular layer. The raising
step is singular on the kinematic hypersurface $\kappa_{v,s;\bm d}=0$, i.e.\ when the momenta
become collinear and split into two opposite groups; there the DE has its poles, and the letter
count is exactly the number of such hypersurfaces.

\subsection{Vertex IBP, pinches and off-diagonal DE blocks}
\label{sec:remaining}

A contact joins two adjacent vertices and removes one time integration. We first describe one
such term, then sum over the neighbours of the vertex at which IBP is performed. This determines
the block of the DE between a sector and the subsector obtained by pinching the propagator joining that pair.
We start with massive propagators on a tree; the extra massless relations will subsequently act
on the rows and columns of the same block. Several propagators sharing one theta are treated in
Sec.~\ref{sec:common-theta}.

\paragraph{The contact term for one adjacent pair.}
Let $s_0$ be a sector and let an (anti-)time-ordered propagator $e_i$ join two of its vertices, denoted by
$v_0$ and $v_i$. These are current vertices of $s_0$, each carrying one time integration, and may
already represent merged components of the original diagram. For this calculation we differentiate
with respect to $\tau_{v_0}$. The ordered pair $(v_0,v_i)$ will record this choice: its first vertex
is the differentiated one. Write $q_{e_i}$ for the momentum magnitude, $\nu_{e_i}$ for the order,
and $\sigma_{e_i}=+1$ or $-1$ for a $G_{++}$ or $G_{--}$ propagator, respectively.

The derivative of its theta produces $\delta(\tau_{v_0}-\tau_{v_i})$ and the difference of the two
Hankel-branch arrangements. Write $a,b\in\{0,1\}$ for the state bits at $v_0,v_i$, respectively.

\begin{samepage}
The Wronskian relation gives
\begin{equation}
  \begin{aligned}
  &h_{\rm ref}^{(1)}(-\nu,a;z)h_{\rm ref}^{(2)}(-\nu,b;z)
   -h_{\rm ref}^{(2)}(-\nu,a;z)h_{\rm ref}^{(1)}(-\nu,b;z)\\
  &\qquad=-\frac{4\ii}{\pi}(b-a)e^{-\pi\operatorname{Im}\nu}z^{2\nu-1},
  \end{aligned}
  \label{eq:wronskian-h}
\end{equation}
for real or purely imaginary $\nu$.
\end{samepage}
The bit difference is zero for $00,11$, positive for $01$
and negative for $10$. By \eqref{eq:h-reference-conversion}, each product on the left is
$e^{-\pi\operatorname{Im}\nu}$ times the corresponding product of the common-order functions
\eqref{eq:h-definition}. Dividing \eqref{eq:wronskian-h} by this common factor gives the contact
term in our basis, with merged-time power and scalar coefficient
\begin{equation}
  \varrho_e=2\nu_e-1,
  \qquad F_e=-\sigma_e\frac{4\ii}{\pi}q_e^{\,\varrho_e}.
  \label{eq:contact-factor}
\end{equation}
Here the sign is chosen so that $F_e$ is exactly the factor contributed by a pinched massive
propagator to \eqref{eq:sector-normalization}. The two endpoint derivatives have opposite signs;
we keep this distinction in the ordered-pair selector below.

Denote the resulting subsector by $s_i=s_0/e_i$ and its merged vertex by $C_i$. All other vertices
and their time powers are unchanged. The base power \eqref{eq:base-power} and the actual power
left by the contact term give
\begin{align}
  \nu_{0,C_i}&=\nu_{0,v_0}+\nu_{0,v_i}+\varrho_{e_i}+1,\nn\\
  \nu_{0,v_0}+n_{v_0}+\nu_{0,v_i}+n_{v_i}+\varrho_{e_i}
  &=\nu_{0,C_i}+n^{(0i)}_{C_i},\nn\\
  n^{(0i)}_{C_i}&=n_{v_0}+n_{v_i}-1,\nn\\
  n^{(0i)}_w&=n_w\quad(w\in V_{s_0}\setminus\{v_0,v_i\}).
  \label{eq:contact-shift}
\end{align}
Thus $\bm n^{(0i)}$ is a shift vector for $s_i$, with one fewer entry than $\bm n$ for $s_0$.
In particular, $\bm n=\bm0$ gives shift $-1$ at $C_i$, whereas $\bm n=\bm e_{v_0}$ or
$\bm e_{v_i}$ gives $\bm n^{(0i)}=\bm0$. The raw contact at zero shift is one power of
$-\tau_{C_i}$ below the subsector master; raising either endpoint before taking the contact
supplies that power.

We next specify the state indices. Let $\bm a$ be the full bit string of $s_0$, and write
$a_{e_i,v_0}$ and $a_{e_i,v_i}$ for the two bits on the pinched propagator. The remaining bits
retain their slot labels and their relative order in $s_i$; call their string $\bm c$.
The constant matrix $C^{(v_0v_i)}_{s_0s_i}$ has rows labelled by $\bm a$ and columns by $\bm c$:
\begin{equation}
  \bigl(C^{(v_0v_i)}_{s_0s_i}\bigr)_{\bm a\bm c}
  =\bigl(a_{e_i,v_i}-a_{e_i,v_0}\bigr)
    \prod_{\ell\in H(s_i)}\delta_{a_\ell c_\ell}.
  \label{eq:contact-atom-massive}
\end{equation}
The product runs over all surviving massive endpoint slots and preserves every spectator bit.
In a temporary ordering that puts the two removed
slots first, this matrix is the tensor product
\begin{equation}
  C^{(v_0v_i)}_{s_0s_i}
  =\bigl(|0\rangle\otimes|1\rangle-|1\rangle\otimes|0\rangle\bigr)
     \otimes\bigotimes_{\ell\in H(s_i)}\Id_2,
  \qquad
  |0\rangle=\begin{pmatrix}1\\0\end{pmatrix},\quad
  |1\rangle=\begin{pmatrix}0\\1\end{pmatrix}.
  \label{eq:contact-tensor}
\end{equation}
The local column is $(0,1,-1,0)^T$ in the order $00,01,10,11$; every spectator contributes a
$2\times2$ identity. Restoring the global bit order only permutes rows and columns, as specified
unambiguously by \eqref{eq:contact-atom-massive}. If $s_0$ has $N$ endpoint slots, then $s_i$ has
$N-2$, so $C^{(v_0v_i)}_{s_0s_i}$ is $2^N\times2^{N-2}$. Its local factor is a $4\times1$
column because the contact removes two bits, rather than an operator within a fixed state space.

Define $\cR^{(v_0v_i)}_{s_0}(\bm n)$ to be just the term generated when the derivative at $v_0$
hits the theta of $e_i$. The coefficient and the pinched integral now have matching indices:
\begin{equation}
  \bigl(\cR^{(v_0v_i)}_{s_0}(\bm n)\bigr)_{\bm a}
  =F_{e_i}\sum_{\bm c\in\{0,1\}^{N-2}}
    \bigl(C^{(v_0v_i)}_{s_0s_i}\bigr)_{\bm a\bm c}\,
    \text{I}_{s_i}(\bm n^{(0i)};\bm c).
  \label{eq:R-two-vertex}
\end{equation}
For any fixed spectator string, the $00$ and $11$ rows vanish, and the $01$ and $10$ rows are
$+F_{e_i}\text{I}_{s_i}$ and $-F_{e_i}\text{I}_{s_i}$. Interchanging the differentiated endpoint
reverses the sign: in the same global bit order,
$C^{(v_iv_0)}_{s_0s_i}=-C^{(v_0v_i)}_{s_0s_i}$ and
$\cR^{(v_iv_0)}_{s_0}(\bm n)=-\cR^{(v_0v_i)}_{s_0}(\bm n)$.
The matrix maps the subsector vector into the supersector equation; the direction of this
matrix multiplication is opposite to the graphical operation of deleting the propagator.

\paragraph{IBP at a chosen vertex.}
Choose any vertex of $s_0$ and call it $v_0$ for this step. Let
$\operatorname{Adj}_{s_0}(v_0)$ be the vertices joined to it by an (anti-)time-ordered propagator carrying a theta.
We label these neighbours $v_i$, their propagators $e_i$, and their subsectors $s_i=s_0/e_i$.
Theta-free propagators still enter the local generators, but do not enter this neighbour sum.
The total derivative of the integrand at $v_0$ gives
\begin{equation}
  -\text{M}_{1;v_0,s_0}(\nu_{0,v_0}+n_{v_0})\,
       \bm{\text I}_{s_0}(\bm n-\bm e_{v_0})
  +\text{M}_{0;v_0,s_0}\,\bm{\text I}_{s_0}(\bm n)
  +\sum_{v_i\in\operatorname{Adj}_{s_0}(v_0)}\cR^{(v_0v_i)}_{s_0}(\bm n)=0.
  \label{eq:vertex-neighbour-ibp}
\end{equation}
The endpoints of the time integration give no contribution in the convergence domain used to
derive IBP; the contacts displayed here come from differentiating the internal theta factors.
The same relations then continue analytically with the regulator of Sec.~\ref{sec:regulator}.
Equation~\eqref{eq:vertex-neighbour-ibp} is \eqref{eq:one-step-ibp} with its remainder written
as the sum over the actual neighbours of the differentiated vertex.

For the DE we need the row at $\bm n=\bm e_{v_0}$. By \eqref{eq:contact-shift}, every term in its
remainder is already a subsector master:
\begin{equation}
  \bigl(\cR_{v_0,s_0}(\bm e_{v_0})\bigr)_{\bm a}
  =\sum_{v_i\in\operatorname{Adj}_{s_0}(v_0)}F_{e_i}
    \sum_{\bm c}
    \bigl(C^{(v_0v_i)}_{s_0s_i}\bigr)_{\bm a\bm c}\,
    \text{I}_{s_i}(\bm0;\bm c).
  \label{eq:contact-injection}
\end{equation}
Here $\cR_{v_0,s_0}$ denotes the full remainder of the chosen vertex, while the ordered-pair
superscript on $\cR^{(v_0v_i)}_{s_0}$ denotes one summand. The outer sum selects a pinched
propagator and hence its subsector; the inner sum runs over that subsector's state bits. No
additional raising label is needed: the argument $\bm e_{v_0}$ states which time power was raised.

\begin{samepage}
For the massive sectors considered here, $Q_s=\Id$ in the normalized basis
\eqref{eq:normalized-vector}. With unit seed normalizations $\cN_s^{(0)}=1$, the definition
\eqref{eq:sector-normalization} gives
\begin{equation}
  \bm{\text J}_s=\cN_s\bm{\text I}_s(\bm0),
  \qquad \cN_{s_i}=\cN_{s_0}F_{e_i}.
  \label{eq:massive-contact-normalization}
\end{equation}\par
\end{samepage}
Multiplying \eqref{eq:contact-injection} by $\cN_{s_0}$ therefore gives the normalized
contact source $\sum_i C^{(v_0v_i)}_{s_0s_i}\bm{\text J}_{s_i}$: each $F_{e_i}$ is absorbed
into the subsector normalization.
Other constant seed normalizations multiply each pair block below by
$\cN_{s_0}^{(0)}/\cN_{s_i}^{(0)}$.

\paragraph{The DE block for a fixed adjacent pair.}
\begin{samepage}
Keep the chosen vertex $v_0$ and fix its neighbour $v_1$, with propagator $e=e_1=(v_0,v_1)$
and subsector $s_1=s_0/e$. The off-diagonal block for this pinch multiplies
$\bm{\text J}_{s_1}$ in the differential equation for $\bm{\text J}_{s_0}$;
its rows belong to $s_0$ and its columns to $s_1$. It receives two contributions, from the
time IBPs at $v_0$ and $v_1$. Raising the time power at either endpoint makes the contact
integral the same zero-shift master $\bm{\text I}_{s_1}(\bm0)$, by \eqref{eq:contact-shift}.
The two contributions have opposite contact signs because the derivatives act on opposite
arguments of the same theta.\par
\end{samepage}

To obtain these contributions, let $v$ be either $v_0$ or $v_1$ and differentiate with respect
to its signed energy $p_{0,v}$. Since
$\partial_{p_{0,v}}e^{\ii p_{0,v}\tau_v}=\ii\tau_v e^{\ii p_{0,v}\tau_v}$, this gives
$\partial_{p_{0,v}}\bm{\text I}_{s_0}(\bm0)=-\ii\bm{\text I}_{s_0}(\bm e_v)$.
The time IBP at the same vertex reduces this raised integral through
\begin{equation}
  \text{M}_{0;v,s_0}\bm{\text I}_{s_0}(\bm e_v)
  =\text{M}_{1;v,s_0}(\nu_{0,v}+1)\bm{\text I}_{s_0}(\bm0)
   -\cR_{v,s_0}(\bm e_v),
  \label{eq:raising-row}
\end{equation}
and hence
\begin{equation}
  \partial_{p_{0,v}}\bm{\text I}_{s_0}(\bm0)
  =-\ii\text{M}_{0;v,s_0}^{-1}\text{M}_{1;v,s_0}(\nu_{0,v}+1)\bm{\text I}_{s_0}(\bm0)
   +\ii\text{M}_{0;v,s_0}^{-1}\cR_{v,s_0}(\bm e_v).
  \label{eq:derivative-raising}
\end{equation}
The first term belongs to the diagonal sector block already given in \eqref{eq:ndlogDE}; the
second supplies the subsector blocks. To read the latter as logarithmic derivatives, define the
vertex kernel
\begin{equation}
  \mathcal K_{v,s_0}=U_{s_0}^{-1}\widetilde\Omega_{0;v,s_0}U_{s_0},
  \qquad
  \partial_{p_{0,v}}\mathcal K_{v,s_0}
  =U_{s_0}^{-1}\diag\!\left(\frac{1}{\ii\kappa_{v,s_0;\bm d}}\right)U_{s_0}
  =\text{M}_{0;v,s_0}^{-1}.
  \label{eq:M0-inverse-derivative}
\end{equation}
Here $U_{s_0}$, $\widetilde\Omega_{0;v,s_0}$ and $\kappa_{v,s_0;\bm d}$ are the tensor
diagonalizer, diagonal logarithmic matrix and energy letters of Sec.~\ref{sec:dlog}.
The equality uses $\partial_{p_{0,v}}\kappa_{v,s_0;\bm d}=1$.
Extract the $s_1$ term of \eqref{eq:contact-injection} and substitute it into the remainder
in \eqref{eq:derivative-raising}. Multiplying by $\cN_{s_0}$ and using
\eqref{eq:massive-contact-normalization} absorbs $F_e$ into $\bm{\text J}_{s_1}$.
The normalization depends only on propagator momenta, so it commutes with the signed-energy
derivatives. With \eqref{eq:M0-inverse-derivative}, the two endpoint equations become
\begin{equation}
  \begin{aligned}
  \partial_{p_{0,v_0}}\bm{\text J}_{s_0}
  &=\cdots+\ii\bigl(\partial_{p_{0,v_0}}\mathcal K_{v_0,s_0}\bigr)
       C^{(v_0v_1)}_{s_0s_1}\bm{\text J}_{s_1},\\
  \partial_{p_{0,v_1}}\bm{\text J}_{s_0}
  &=\cdots-\ii\bigl(\partial_{p_{0,v_1}}\mathcal K_{v_1,s_0}\bigr)
       C^{(v_0v_1)}_{s_0s_1}\bm{\text J}_{s_1}.
  \end{aligned}
  \label{eq:energy-contact-normalized}
\end{equation}
The omitted terms act within $s_0$ or come from other adjacent pinches. The first displayed
source comes from the time IBP at $v_0$, the second from that at $v_1$; its minus sign uses
$C^{(v_1v_0)}_{s_0s_1}=-C^{(v_0v_1)}_{s_0s_1}$.

\begin{samepage}
The same reduction determines the derivatives with respect to the propagator momenta. Keeping
the orders fixed, differentiating the $h$ systems gives the full differential before reduction,
\begin{equation}
  \nd\bm{\text I}_{s_0}(\bm0)
  =(\nd\Omega_{{\rm ex},s_0})\bm{\text I}_{s_0}(\bm0)
   -\ii\sum_{v\in V_{s_0}}
    \left[\nd p_{0,v}\Id-\sum_{\ell\in H_v(s_0)}
      \Emb_\ell(\sigma_2)\nd k_\ell\right]\bm{\text I}_{s_0}(\bm e_v).
  \label{eq:full-master-differential}
\end{equation}\par
\end{samepage}
\begin{samepage}
A momentum shared by two endpoint slots occurs at both vertices in this expression.
In the constant tensor basis $U_{s_0}$ the bracket is diagonal, with entries
$\nd\kappa_{v,s_0;\bm d}$. Multiplying it by $\text{M}_{0;v,s_0}^{-1}$ therefore gives
$\nd\mathcal K_{v,s_0}$. Substituting \eqref{eq:raising-row} and applying
\eqref{eq:massive-contact-normalization} gives the full differential of the normalized masters.
Only $v_0$ and $v_1$ supply the selected $s_1$ source. Denoting the coefficient of
$\bm{\text J}_{s_1}$ in $\nd\bm{\text J}_{s_0}$ by $(\nd\Omega^{\rm J})_{s_0s_1}$, we obtain
\begin{equation}
  (\nd\Omega^{\rm J})_{s_0s_1}
  =\ii\bigl(\nd\mathcal K_{v_0,s_0}-\nd\mathcal K_{v_1,s_0}\bigr)
       C^{(v_0v_1)}_{s_0s_1}.
  \label{eq:contact-block-full}
\end{equation}\par
\end{samepage}
The kernel difference combines the two time IBPs into one off-diagonal block, including
all propagator-momentum derivatives.

To display the $2\times2$ factors in this block, let $d\in\{0,1\}$ label an eigenstate of
$\sigma_2$ and define its projector in the original mode-function basis,
\begin{equation}
  \Pi_d=\text{T}^{-1}|d\rangle\langle d|\text{T}
  =\frac12\begin{pmatrix}1&-\ii(-1)^d\\ \ii(-1)^d&1\end{pmatrix},
  \qquad
  \Pi_{\bm d}^{(s_0)}=\bigotimes_{\ell\in H(s_0)}\Pi_{d_\ell}.
  \label{eq:letter-projectors}
\end{equation}
The two-state matrix $\text{T}$ is the one in \eqref{eq:global-M0-diagonal}. The string
$\bm d$ labels the diagonal tensor basis, whereas $\bm a$ and $\bm b$ below label the original
mode-function basis. In particular,
\begin{equation}
  \nd\mathcal K_{v,s_0}
  =-\ii\sum_{\bm d\in\{0,1\}^{N}}
       \Pi_{\bm d}^{(s_0)}\,\nd\log\kappa_{v,s_0;\bm d}.
  \label{eq:kernel-dlog-tensor}
\end{equation}
For $N$ endpoint slots in $s_0$, the matrix element of the selected DE block is therefore
\begin{equation}
  \begin{aligned}
  \bigl((\nd\Omega^{\rm J})_{s_0s_1}\bigr)_{\bm a\bm c}
  ={}&\sum_{\bm d,\bm b\in\{0,1\}^{N}}
    \left[\prod_{\ell\in H(s_0)}(\Pi_{d_\ell})_{a_\ell b_\ell}\right]
    \bigl(C^{(v_0v_1)}_{s_0s_1}\bigr)_{\bm b\bm c}\\
   &\times\left(\nd\log\kappa_{v_0,s_0;\bm d}
                -\nd\log\kappa_{v_1,s_0;\bm d}\right).
  \end{aligned}
  \label{eq:massive-contact-bits}
\end{equation}
Here $\bm a$ is a row of the supersector, $\bm c\in\{0,1\}^{N-2}$ a column of the subsector,
and $\bm b$ the intermediate supersector string summed in the matrix product.
The projector entries and the selector \eqref{eq:contact-atom-massive} are constant. Every
kinematic dependence in this block is consequently a $\nd\log$ of an energy letter.
Together with the diagonal block of Sec.~\ref{sec:dlog} and the diagonal contribution
$\nd\log\cN_s\,\Id$, this gives the complete massive d log-form DE.

\paragraph{Massless reduction on the two sides of a block.}
For a massless (anti-)time-ordered propagator, the unreduced time-IBP relations follow by setting
$\nu_e=1/2$ while retaining its two endpoint bits. The additional relations
\eqref{eq:massless-relations} then replace those four states by one shared bit.
To describe this reduction, use the connection $\mathcal A$ of
the original integrals in \eqref{eq:ndlogDE}, before the basis change $Q_s$ and normalization
$\cN_s$. Write $\bm{\text I}^{\rm u}_s$ and $\mathcal A^{\rm u}$ for the vector and connection
before imposing the extra massless relations, and $\bm{\text I}_s$ and $\mathcal A$ for their
reduced counterparts. Both retain the $h$ amplitudes; the superscript ${\rm u}$ distinguishes
only the unreduced state space.

The local maps of \eqref{eq:SP-massless} can also be written in tensor notation,
\begin{equation}
  S_e=(|00\rangle+|11\rangle)\langle0|
       +( |10\rangle-|01\rangle)\langle1|,
  \qquad P_e=\Id_2\otimes\langle0|,
  \qquad P_eS_e=\Id_2.
  \label{eq:massless-tensor-maps}
\end{equation}
Here $|ab\rangle=|a\rangle\otimes|b\rangle$ is ordered by the endpoints $(u_e,v_e)$;
$S_e$ is $4\times2$ and restores the four endpoint states from the shared bit, while $P_e$ is
$2\times4$ and selects the $00$ and $10$ entries. Grouping the endpoint slots of each massless
propagator, define the diagram maps
\begin{align}
  S_s&=\bigotimes_{\ell\in H(s)}\Id_2\otimes
        \bigotimes_{e\in E_m(s)}S_e,
  &P_s&=\bigotimes_{\ell\in H(s)}\Id_2\otimes
        \bigotimes_{e\in E_m(s)}P_e,\nn\\
  \bm{\text I}^{\rm u}_s&=S_s\bm{\text I}_s,
  &\bm{\text I}_s&=P_s\bm{\text I}^{\rm u}_s,
  \qquad P_sS_s=\Id.
  \label{eq:sector-massless-maps}
\end{align}
The fixed global slot order is restored by the corresponding row and column permutations.
With $N_s=|H(s)|+2|E_m(s)|$ unreduced bits and
$n_s^{\rm slot}=|H(s)|+|E_m(s)|$ reduced bits, $S_s$ has size
$2^{N_s}\times2^{n_s^{\rm slot}}$ and $P_s$ the reverse size.
Every factor is displayed in \eqref{eq:massless-tensor-maps} or is a $2\times2$ identity.

Differentiation preserves the massless relations. Applying $P_{s_0}$ to the equation for the
supersector and substituting $\bm{\text I}^{\rm u}_{s_1}=S_{s_1}\bm{\text I}_{s_1}$ gives
\begin{equation}
  \mathcal A_{s_0s_1}=P_{s_0}\mathcal A^{\rm u}_{s_0s_1}S_{s_1},
  \qquad
  (\mathcal A_{s_0s_1})_{\bm a\bm c}
  =\sum_{\bm b,\bm f}(P_{s_0})_{\bm a\bm b}
       (\mathcal A^{\rm u}_{s_0s_1})_{\bm b\bm f}(S_{s_1})_{\bm f\bm c}.
  \label{eq:massless-block-projection}
\end{equation}
The row strings $\bm a,\bm b$ belong to the reduced and unreduced spaces of $s_0$,
respectively; the column strings $\bm c,\bm f$ belong to the reduced and unreduced spaces of
$s_1$. Thus the left map reduces the supersector equations, and the right map expresses their
subsector sources in the reduced basis. A massless propagator that survives the contact is
reduced on both sides. If it is the pinched propagator, its shared bit is present only on
the supersector side.

After this reduction, $Q_s$ of \eqref{eq:package-massless-basis} changes each shared slot to the
unit-amplitude sum and difference basis. Unlike $P_s$ and $S_s$, it is square and does not change
the number of masters. Explicitly, the normalized package vector is
\begin{equation}
  Q_s=\bigotimes_{\ell\in H(s)}\Id_2\otimes
      \bigotimes_{e\in E_m(s)}\left[\frac{\pi}{2}\diag(1,\ii\sigma_e)\right],
  \qquad
  \bm{\text J}_s=\cN_s Q_s\bm{\text I}_s(\bm0).
  \label{eq:normalized-vector}
\end{equation}
\begin{samepage}
For every sector, including the top sector, the normalized diagonal and off-diagonal blocks are
\begin{align}
  (\nd\Omega^{\rm J})_{ss}
  &=Q_sP_s\mathcal A^{\rm u}_{ss}S_sQ_s^{-1}
       +\nd\log\cN_s\,\Id,\nn\\
  (\nd\Omega^{\rm J})_{s_0s_1}
  &=\frac{\cN_{s_0}}{\cN_{s_1}}
       Q_{s_0}P_{s_0}\mathcal A^{\rm u}_{s_0s_1}S_{s_1}Q_{s_1}^{-1},
       \qquad s_1\prec s_0.
  \label{eq:normalized-de}
\end{align}\par
\end{samepage}
This displays separately the reduction of dimensions, the shared-slot basis change, and the
sector normalization. In particular, specializing the massive formula to $\nu_e=1/2$ is followed
by the rectangular maps; specialization alone still leaves redundant endpoint states.

For the same pair $(v_0,v_1)$, write $C^{\rm u}$ for the unreduced selector of
\eqref{eq:contact-atom-massive}, and set
\begin{align}
  \widehat C^{(v_0v_1)}_{s_0s_1}
  &=\frac{\cN_{s_0}F_e}{\cN_{s_1}}
       Q_{s_0}P_{s_0}C^{{\rm u},(v_0v_1)}_{s_0s_1}S_{s_1}Q_{s_1}^{-1},\nn\\
  \widehat{\mathcal K}_{v,s_0}
  &=Q_{s_0}P_{s_0}\mathcal K^{\rm u}_{v,s_0}S_{s_0}Q_{s_0}^{-1}.
  \label{eq:normalized-contact-injection}
\end{align}
The reduced contact maps obey the same massless relations as the integrals, so the pair block
keeps its previous form,
\begin{equation}
  (\nd\Omega^{\rm J})_{s_0s_1}
  =\ii\bigl(\nd\widehat{\mathcal K}_{v_0,s_0}
            -\nd\widehat{\mathcal K}_{v_1,s_0}\bigr)
       \widehat C^{(v_0v_1)}_{s_0s_1}.
  \label{eq:mixed-contact-block}
\end{equation}
All projection and basis-change matrices are constant. For a massive pinch $F_e$ is
cancelled by $\cN_{s_0}/\cN_{s_1}$; for a massless pinch it is already constant because
$\varrho_e=0$. Both cases therefore preserve the d log form.

In particular, take the pinched propagator itself to be massless and order it as
$(u_e,v_e)=(v_0,v_1)$. Its local column in \eqref{eq:contact-tensor} becomes $-|1\rangle$
under $P_e$; multiplication by $F_e=-4\ii\sigma_e/\pi$ and then by
$(\pi/2)\diag(1,\ii\sigma_e)$ makes it $-2|1\rangle$.
There is no massless pinch factor in $\cN_s$. With the shared slot displayed first,
\begin{equation}
  \widehat C^{(v_0v_1)}_{s_0s_1}
  =-2|1\rangle_e\otimes\bigotimes_{\text{surviving slots }\ell}\Id_2,
  \qquad
  \bigl(\widehat C^{(v_0v_1)}_{s_0s_1}\bigr)_{\bm a\bm c}
  =-2\delta_{a_e,1}\,\delta_{\operatorname{Del}_e\bm a,\bm c}.
  \label{eq:contact-atom-massless}
\end{equation}
Here $a_e$ is the shared bit and $\operatorname{Del}_e$ removes it, keeping all other bits in
order. This matrix has $2^{n_{s_0}^{\rm slot}}$ rows and $2^{n_{s_0}^{\rm slot}-1}$ columns:
one shared bit is removed. Equivalently, let $\operatorname{Ins}_e(\bm c,1)$ insert that bit
back into the source string at its original position. The individual DE matrix element is
\begin{equation}
  \bigl((\nd\Omega^{\rm J})_{s_0s_1}\bigr)_{\bm a\bm c}
  =2\ii\left[
      \nd\widehat{\mathcal K}_{v_1,s_0}
     -\nd\widehat{\mathcal K}_{v_0,s_0}
    \right]_{\bm a,\operatorname{Ins}_e(\bm c,1)}.
  \label{eq:massless-contact-block}
\end{equation}
The endpoint weights are $(-2,+2)$ in the order $(v_0,v_1)$, in agreement with
\eqref{eq:package-massless-basis}. The shared-slot kernels are built from the same $2\times2$
tensor factors as in Sec.~\ref{sec:dlog}, using the Hadamard matrix on shared slots and
$\text{T}$ on massive endpoints.

\paragraph{The complete system.}
\begin{samepage}
Each (anti-)time-ordered propagator supplies its pair block, and each sector supplies its diagonal block.
Order the sectors so that every subsector follows its supersectors. Then
\begin{equation}
  \nd
  \begin{pmatrix}\bm{\text J}_{s_0}\\\bm{\text J}_{s_1}\\\vdots\end{pmatrix}
  =
  \begin{pmatrix}(\nd\Omega^{\rm J})_{s_0s_0}&(\nd\Omega^{\rm J})_{s_0s_1}&\cdots\\
  0&(\nd\Omega^{\rm J})_{s_1s_1}&\cdots\\ \vdots&\vdots&\ddots\end{pmatrix}
  \begin{pmatrix}\bm{\text J}_{s_0}\\\bm{\text J}_{s_1}\\\vdots\end{pmatrix}.
  \label{eq:block-triangular-de}
\end{equation}\par
\end{samepage}
A block vanishes if its subsector cannot be reached by the corresponding contact. All nonzero
blocks are sums of constant matrices times $\nd\log$ of the energy letters, with the additional
$\nd\log q_e$ terms from $\Omega_{\rm ex}$ and the sector normalization. The result is a closed
d log-form system for the complete set of masters, solvable recursively from the lowest sectors.
The package transports this normalized stack as one system (Sec.~\ref{sec:numerics}).

\subsection{Boundary conditions by nested blow-up}
\label{sec:boundary}
The last ingredient is the boundary condition in the large-energy limit. Two mechanisms make
it computable: the damping exponential localizes each time integral near $\tau_v=0$, where
every factor except the thetas reduces to a power of $-\tau$, so each single-time integral
becomes a Gamma function; and the strict energy rank fixes every theta to $0$ or $1$, so no
theta couples two times and every master factorizes into a product of single-time integrals.
We define the \emph{damping energies} by $K_v\equiv\ii p_{0,v}=\ii\varsigma_vk_{0,v}$ and take
$k_{0,v}=-\ii\varsigma_vK_v$ with $K_v>0$. On these damping rays, the exponential of
\eqref{eq:phase-block} becomes $e^{K_v\tau_v}$, with $p_{0,v}$ the component energy
\eqref{eq:component-energy} on a merged component. The package records the substitution as
\texttt{dampingDefinition}
in the vertex coordinates. The energy hierarchy below is understood on these rays,
ordered by the positive damping energies.
The limit $K_v\to\infty$ is one multivariate limit in
which all energies diverge together, and the point it approaches is in general a degenerate
multivariate singularity of the differential equation: several singular divisors meet there, so a
direct series expansion about that point is not allowed. Blowing the point up first makes the
expansion well defined; the nested coordinates introduced below are exactly the blow-up charts
that replace the degenerate point by a chain of one-energy limits. This continues the procedure of
\cite{Chen:2024glu}, where the boundary condition of the multivariate hypergeometric system is
fixed by the same resolution of its degenerate point. In a $c_V$-vertex correlator there are $c_V$
exponential factors with energies $k_{0,v}$; assuming the ordering
\begin{equation}
  k_{\sigma(1)}\gg k_{\sigma(2)}\gg\cdots\gg k_{\sigma(c_V)}\gg1,
  \label{eq:energy-rank}
\end{equation}
with a permutation $\sigma$, we define the nested coordinates
\begin{equation}
  x_j=\frac{k_{\sigma(j+1)}}{k_{\sigma(j)}}\quad(1\leq j<c_V),
  \qquad
  x_{c_V}=\frac1{k_{\sigma(c_V)}},
  \label{eq:nested-blowup-coordinates}
\end{equation}
so that
\begin{equation}
 \frac{1}{k_{\sigma(j)}}=\prod_{r=j}^{c_V}x_r,
 \qquad x_1,\ldots,x_{c_V}\longrightarrow0.
 \label{eq:nested-energy-inverse}
\end{equation}
The exponential localization scales satisfy
\begin{equation}
|\tau_{\sigma(1)}|\ll|\tau_{\sigma(2)}|\ll\cdots
 \ll|\tau_{\sigma(c_V)}|\ll1.
 \label{eq:tau-rank}
\end{equation}
Since all $\tau$ approach zero from the negative side, large $k$ corresponds to small $|\tau|$;
hence $k_u\gg k_v$ implies $\tau_u>\tau_v$, and therefore
\begin{equation}
 \theta(\tau_u-\tau_v)\longrightarrow1,
 \qquad
 \theta(\tau_v-\tau_u)\longrightarrow0.
 \label{eq:theta-rank-rule}
\end{equation}
For $G_{++}$ or $G_{--}$, \eqref{eq:theta-rank-rule} applies to every explicit theta in the
propagator expansion according to the order of its argument. For a bundle of propagators that share
one theta (the \emph{common-theta bundle} of Sec.~\ref{sec:common-theta}) it applies to that single
theta, which is the boundary counterpart of Sec.~\ref{sec:common-theta}: the bundle contributes one
frozen theta value, either $0$ or $1$. If a term requires a cyclic
inequality, for example
\begin{equation}
 \theta(\tau_1-\tau_2)\theta(\tau_2-\tau_3)\theta(\tau_3-\tau_1),
 \label{eq:cyclic-theta}
\end{equation}
then it vanishes in every strict-total-order chart.

After the thetas are frozen, every master integral --- in the top sector as well as in every
subsector --- factorizes into one one-dimensional time integral per vertex $v\in V_s$: each
vertex is integrated over its single time $\tau_v$, and the legs it carries enter that
integral through their mode-function small-$\tau$ coefficients. A massive leg behaves like
\eqref{eq:hankel-tau0} and a massless leg contributes its leading power or a pure exponential, so
each single-time integral has the form
$\int_{-\infty}^0\nd\tau\,(-\tau)^{A}e^{\ii K\tau}=(-\ii)^{A+1}\Gamma(A+1)K^{-A-1}$, which gives a
Gamma function; worked examples of the resulting boundary data and of the blow-up charts are given
in \cite{Chen:2024glu,Chen:2023iix}. The boundary conditions of all master integrals are therefore
products of Gamma functions with some overall coefficients, as used in the numerical evaluation of
Sec.~\ref{sec:numerics}.

Each $h$ factor is expanded at small $\tau$ and integrated term by term. In the convention
\eqref{eq:h-definition} the mode-function two-state block behaves as
\begin{align}
  h^{(1)}(\nu,0;z)&=c_2\bigl(1+O(z^2)\bigr)+c_1z^{2\nu}\bigl(1+O(z^2)\bigr),\nn\\
  h^{(1)}(\nu,1;z)&=O(z)+2\nu\,c_1\,z^{2\nu-1}\bigl(1+O(z^2)\bigr),
  \label{eq:h-two-powers}
\end{align}
with $c_{1,2}$ as in \eqref{eq:hankel-tau0} and $z=-k\tau$; the second Hankel branch has the
same two powers with its own constants. For a vertex with a single massive leg of order
$\nu_1$, the master integral of the undifferentiated state is
\begin{multline}
  \int_{-\infty}^0\nd\tau\,(-\tau)^{\nu_0}e^{\ii K\tau}\,h^{(1)}(\nu_1,0;-k\tau)
  =c_2\,(-\ii)^{\nu_0+1}\Gamma(\nu_0+1)K^{-\nu_0-1}\\
  +c_1\,k^{2\nu_1}(-\ii)^{\nu_0+2\nu_1+1}\Gamma(\nu_0+2\nu_1+1)K^{-\nu_0-2\nu_1-1}
  +\cdots,
  \label{eq:boundary-example}
\end{multline}
where each term of the small-$\tau$ expansion was integrated with the one-dimensional formula
above, $\int_{-\infty}^0\nd\tau\,(-\tau)^{A}e^{\ii K\tau}=(-\ii)^{A+1}\Gamma(A+1)K^{-A-1}$. For
generic nonresonant orders the constant term anchors the undifferentiated state and the
$z^{2\nu-1}$ term the differentiated state; the differentiated master integral is treated
identically and anchors at the exponent $\nu_0+2\nu_1$. The corrections sit one integer power
above the anchor of the other state, the Frobenius separation described after eq.~(3.16) of
\cite{Chen:2024glu}. The two masters of such a vertex therefore anchor at the leading
inverse-energy exponents $\nu_0+1$ and $\nu_0+2\nu_1$. Products
of several $h$ factors are handled the same way: the product is expanded in small $\tau$, and
each term integrates to one Gamma-function coefficient.

The anchoring exponents are the indicial exponents of the differential equation at the endpoint
of a blow-up chart. At this regular singular point the one-variable pullback of the system reads
$\nd\bm{\text J}/\nd\zeta=\bigl(A_0/\zeta+O(1)\bigr)\bm{\text J}$, with $\zeta$ a local
inverse-energy coordinate and $A_0$ the residue matrix, so the leading power $\zeta^\rho$ of
every local solution is a root of the indicial equation
\begin{equation}
  \det\bigl(A_0-\rho\,\Id\bigr)=0.
  \label{eq:indicial-equation}
\end{equation}
Logarithmic terms are absent whenever no two roots of \eqref{eq:indicial-equation} differ by an
integer. At such a resonance a log term still appears only if the resonance is coupled: at root
coincidence, through a nontrivial Jordan block of $A_0$; at separation by a positive integer $m$,
through the first $m$ coefficients of the regular part projecting nontrivially onto the left
eigenspace of the higher root. The small-$\tau$ behavior of every master is fixed by the Hankel asymptotics
\eqref{eq:hankel-tau0}: each master is a sum of products of the two power branches, the
term-by-term integration \eqref{eq:boundary-example} gives every leading coefficient in closed
form, and the indicial roots, $\nu_0+1$ and $\nu_0+2\nu_1$ above, are read off as sums of the
selected per-leg powers. These asymptotics and the one-dimensional integration supply the
analytic boundary data. The DE then generates the higher coefficients of the local series.

The construction requires a damping energy at every vertex to rank it in \eqref{eq:tau-rank}
and localize its time integral. A vertex without external legs has no exponential factor,
so the strict-order chart is undefined there. The package introduces an \emph{auxiliary}
external-leg energy $k_{0,v}^{\rm aux}$ in the same d log-form system. It enters the energy letters
and component signed energies as a physical $k_{0,v}$ does. Boundary data are constructed at
$K_v^{\rm aux}\to\infty$, then transported to the physical target $k_{0,v}^{\rm aux}=0$.
All auxiliary coordinates enter the system and every requested point together, giving a single
transport in the enlarged coordinate space. The package handles these coordinates internally.
The zero target can be singular: if a component's only nonzero signed energy is auxiliary,
its letters vanish at $k_{0,v}^{\rm aux}=0$. At such a finite singular point, the package
returns a finite value or reports an explicit divergence.

\subsection{Several propagators sharing one theta}
\label{sec:common-theta}

\begin{figure}[!htbp]
  \centering
  \includegraphics[width=0.32\textwidth]{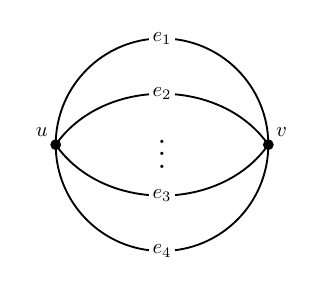}
  \caption{A common-theta bundle of four time-ordered and anti-time-ordered propagators. The vertices $u$ and $v$ carry no external
  legs, and every propagator $e_1,\ldots,e_4$ joins this same pair of vertices, hence the same pair of
  current components at every stage of the reduction. All four propagators therefore carry the identical
  theta argument $\Delta=\tau_u-\tau_v$ and form one bundle $\cB$ with $|\cB|=4$. The two vertices
  sit at the left and the right end, the outer two propagators make up the halves of one circle, and the
  vertical dots between the inner two propagators stand for any number of further propagators joining the same
  pair, so the bundle is not restricted to the four propagators drawn.}
  \label{fig:bundle}
\end{figure}

Up to now every contact has been attached to one (anti-)time-ordered propagator. Two or more time-ordered and anti-time-ordered propagators can,
however, join the same pair of current components, so that all of them carry the same
theta argument $\Delta=\tau_u-\tau_v$; the package groups exactly such propagators into a
\emph{common-theta bundle} $\cB$, with $|\cB|\ge2$. Such diagrams occur in loop integrands, the class read by \texttt{MSInitTimeGraph} of
Sec.~\ref{sec:inputs}. A strict tree never reaches this case:
pinching propagators of a tree leaves a tree, hence never two propagators on one and the same pair of
components. This case is absent from
\cite{Chen:2023iix}, where each contact pinches a single propagator.
Figure~\ref{fig:bundle} shows the minimal diagram of the first kind: two vertices, no external legs, and four
time-ordered and anti-time-ordered propagators that all join the same pair of vertices. The handling is best
summarized in four steps, with the details in App.~\ref{app:theta}.

For each propagator $e$ of the bundle, write $\alpha_e$ for the coefficient of
$\theta(\Delta)$ and $\beta_e$ for the coefficient of $\theta(-\Delta)$.
\begin{enumerate}
  \item \emph{Multiply the thetas out first.} Since inside a bundle all thetas are the same
        function, $\theta(\Delta)^2=\theta(\Delta)$ and $\theta(\Delta)\theta(-\Delta)=0$, the
        product of $|\cB|$ propagators collapses to \emph{two} terms rather than $2^{|\cB|}$,
        \begin{equation}
          \prod_{e\in\cB}\Big(\theta(\Delta)\alpha_e+\theta(-\Delta)\beta_e\Big)
          =\theta(\Delta)\prod_{e\in\cB}\alpha_e
          +\theta(-\Delta)\prod_{e\in\cB}\beta_e .
          \label{eq:bundle-two-terms}
        \end{equation}
  \item \emph{One theta, one delta.} Differentiating the shared theta --- and only the theta, the two
        coefficients being left as spectators --- yields a single
        $\delta(\Delta)$ whose coefficient is the antisymmetric difference under
        $\alpha\leftrightarrow\beta$,
        $\big(\partial_\Delta\theta(\Delta)\big)\prod_{e\in\cB}\alpha_e
        +\big(\partial_\Delta\theta(-\Delta)\big)\prod_{e\in\cB}\beta_e
        =\delta(\Delta)\big(\prod_{e\in\cB}\alpha_e-\prod_{e\in\cB}\beta_e\big)$, so all contact information of
        the bundle is concentrated in one distribution and the two vertices are forced to coincide
        only once. The terms in which the derivative acts on $\alpha_e$ or $\beta_e$ instead are not
        contact terms: they are the single-vertex part of the derivative, already carried by the
        $\text{M}_0$ and $\text{M}_1$ generators of Sec.~\ref{sec:ibp}, and they do not merge any component
        (App.~\ref{app:theta}).
  \item \emph{Split and change basis.} Telescoping the antisymmetric difference and then collecting
        in $\mathsf d_e=\alpha_e-\beta_e$ and $\mathsf m_e=(\alpha_e+\beta_e)/2$ leaves only terms
        with an odd number of $\mathsf d$ factors, each with a numerical coefficient,
        \begin{equation}
          \prod_{e\in\cB}\alpha_e-\prod_{e\in\cB}\beta_e
          =\sum_{\substack{\varnothing\neq S\subseteq\cB\\|S|\ \mathrm{odd}}}
          2^{1-|S|}\prod_{e\in S}\mathsf d_e\prod_{e\notin S}\mathsf m_e .
          \label{eq:bundle-odd-subsets}
        \end{equation}
  \item \emph{Read off the events.} Every non-empty odd subset $S$ is one simultaneous contact
        event: the propagators of $S$ are pinched at once, their two vertices merge once, the component
        count drops by one, and the base power of the merged component follows
        \eqref{eq:base-power} with $|C|-1$ merge terms. The coefficient mapping the subsector
        integral into the supersector remainder generalizes $F_eC^{(uv)}_{s_0s_1}$ in
        \eqref{eq:R-two-vertex}: take the tensor product of the local selectors and raw
        Wronskian factors for the propagators in $S$, multiply by $2^{1-|S|}$, and retain
        $\mathsf m_e$ for each unselected propagator in the bundle.
\end{enumerate}
The package uses the collected form \eqref{eq:bundle-odd-subsets} rather than the telescoping sum term
by term, because enumerating the terms one by one would count the same sector transition
several times; the argument, the parity proof and the treatment of unselected propagators whose two vertices
become coincident are given in App.~\ref{app:theta}. The same symmetric part governs a propagator whose
two vertices have become coincident, its boundary data included: the two branch arrangements are
replaced by their mean, one mean per propagator.

\section{Using the package}
\label{sec:usage}

The package turns the formulas of Sec.~\ref{sec:method} into a sequential workflow: initialize
a family from an input specification, obtain its masters, reduction system and d log-form DE,
then
transport along a kinematic path with the built-in backend and, where the time integrals
diverge, reconstruct the Laurent series in the analytic regulator. This section describes
installation (Sec.~\ref{sec:overview}), the input formats and the functions of each stage
(Sec.~\ref{sec:inputs}), the numerical interface with its built-in backend (Sec.~\ref{sec:numerics}),
the Laurent reconstruction (Sec.~\ref{sec:regulator}) and the options with validation
and common pitfalls (Sec.~\ref{sec:options}). Sec.~\ref{sec:examples-section} then exercises
the workflow on the distributed examples.

\subsection{Installation and loading}
\label{sec:overview}

\paragraph{System requirements.}
\begin{itemize}
  \item A Wolfram Language kernel: Mathematica or the free Wolfram Engine (the package is a
        \texttt{.wl} package; no external Wolfram libraries are needed).
  \item Python 3.10 or newer with \texttt{python-flint>=0.6} and \texttt{sympy>=1.12} installed, for the numerical backend.
        \texttt{MadStree} follows a zero-probe policy: the interpreter is selected only by the explicit
        \texttt{PythonExecutable} option or, when that is \texttt{Automatic} (the default), by the
        \texttt{MADSTREE\_PYTHON} environment variable, falling back to \texttt{"python"}.
\end{itemize}
Install the dependencies in the Python interpreter used for numerical evaluation:
\begin{verbatim}
python -m pip install "python-flint>=0.6" "sympy>=1.12"
\end{verbatim}
Here \texttt{python} should be replaced by the selected interpreter if necessary.
No further backend installation is needed: the numerical backend is the built-in module
\texttt{FlintNDE} under \texttt{MadStree-1.0/Vendor/FlintNDE-dS}, a solver for first-order
matrix differential equations built on the exact ball arithmetic of \texttt{python-flint};
\texttt{MadStree} delegates every numerical task of this package to it.
Its role in the numerical workflow is described in Sec.~\ref{sec:backend}.

\paragraph{Loading the package.}

Download or clone the official release repository
\url{https://github.com/J7C/dSIBP-Packages} and
set \texttt{madStreeDirectory} to the \texttt{MadStree-1.0} directory in the local repository,
add it to \texttt{\$Path}, and load with \texttt{Needs}:
\begin{verbatim}
madStreeDirectory = ".../dSIBP-Packages/MadStree-1.0";
AppendTo[$Path, madStreeDirectory];
Needs["MadStree`"];
\end{verbatim}
Replace the example path with the absolute path to the local \texttt{MadStree-1.0} directory.
This loads \texttt{MadStree.m}, which loads \texttt{Kernel/MadStree.wl} with explicit UTF-8.
Equivalently, load the kernel file directly:
\begin{verbatim}
Get[FileNameJoin[{madStreeDirectory, "Kernel", "MadStree.wl"}],
  CharacterEncoding -> "UTF-8"];
\end{verbatim}
Loading through \texttt{Needs} requires no user encoding option.
\texttt{MadStree} uses its built-in \texttt{FlintNDE} backend without requiring \texttt{Needs["FlintNDE`"]}.
Use a fresh kernel when switching versions, since \texttt{Needs} does not reload
an already loaded package.

\subsection{Input formats and core functions}
\label{sec:inputs}

An initialized context is the root of the whole interface. The input specification fixes the
family, from which the package derives the masters, reduction system and d log-form DE.
We first describe the three initialization formats --- trees, loop
integrands and single-vertex families --- together with the package conventions, then the core
functions and the organization of their outputs.

\paragraph{Tree diagrams: \texttt{MSInitTree}.}

A tree is given by an association of vertices and propagators, with propagators listed under
the input key \texttt{"lines"}. A tree with two vertices and one massless propagator between them:
\begin{verbatim}
spec = <|
  "name" -> "masslessTEdge",
  "vertices" -> {
    <|"id" -> v1, "externalLegEnergy" -> k1,
      "timePower" -> a1, "vertexType" -> "+"|>,
    <|"id" -> v2, "externalLegEnergy" -> k2,
      "timePower" -> a2, "vertexType" -> "+"|>
  },
  "lines" -> {
    <|"type" -> "massless", "endpoints" -> {v1, v2},
      "momentum" -> q, "nu" -> 1/2|>
  }
|>;
context = MSInitTree[spec];
\end{verbatim}
Tables~\ref{tab:vertex-fields} and \ref{tab:propagator-fields} summarize the diagram input fields.
\begin{table}[htbp]
\centering
\begin{tabular}{@{}lp{5.5cm}p{3.8cm}@{}}
\toprule
field & meaning & note\\
\midrule
\texttt{id} & vertex symbol & required\\
\texttt{externalLegEnergy} & external energy parameter (e.g.\ \texttt{k1}) & optional; default $0$\\
\texttt{timePower} & time-power variable $a_v$ (e.g.\ \texttt{a1}) & required\\
\texttt{vertexType} & contour assignment \texttt{"+"} or \texttt{"-"} & required\\
\bottomrule
\end{tabular}
\caption{Vertex fields for tree and time-diagram input.}
\label{tab:vertex-fields}
\end{table}
\begin{table}[htbp]
\centering
\begin{tabular}{@{}p{4.4cm}p{5.3cm}p{3.8cm}@{}}
\toprule
field & meaning & note\\
\midrule
\texttt{type} & \texttt{"massive"} or \texttt{"massless"} & required\\
\texttt{endpoints} & list of one external or two internal vertices & required\\
\texttt{momentum} & momentum variable $k_i$ (e.g.\ \texttt{q}) & required\\
\texttt{nu} & supplied Bessel/Hankel order $\nu$ & required for massive propagators; massless default $1/2$\\
\bottomrule
\end{tabular}
\caption{Propagator fields for tree and time-diagram input.}
\label{tab:propagator-fields}
\end{table}
A massless propagator joining two vertices with equal contour signs is classified internally as
\texttt{masslessT} and uses the shared two-state quotient basis of Sec.~\ref{sec:massless}.
Here the internal suffix T denotes both same-contour components, $G_{++}$ and $G_{--}$,
while W denotes the Wightman components, $G_{+-}$ and $G_{-+}$. The four internal
types are \texttt{massiveT}, \texttt{masslessT}, \texttt{massiveW} and \texttt{masslessW}.
For opposite contour signs it is theta-free and contributes no state slot.
The \texttt{"massless"} input is the pure exponential block at the fixed massless order
\texttt{nu -> 1/2} of Sec.~\ref{sec:conventions}; propagators of other half-integer orders must
first be expanded as described there.
The initialization option \texttt{NuConvention} takes \texttt{"Positive"} (default) or
\texttt{"Negative"} and chooses the prefactor sign of $h(\nu,0;z)=z^{\pm\nu}H_\nu(z)$ introduced in
Sec.~\ref{sec:conventions}; it is fixed for the whole context and cannot be changed per call.

Propagator numbers are assigned from the input order. The (anti-)time-ordered, Wightman or external class and the contour signs are derived from the \texttt{vertexType} of the two vertices the propagator joins; they are not supplied as a type value or as separate sign fields. A \texttt{"+"} vertex has phase $e^{-\ii E\tau}$ and a \texttt{"-"} vertex has phase $e^{+\ii E\tau}$, where $E$ is \texttt{externalLegEnergy}. An omitted or zero external energy is represented internally by an auxiliary variable whose target value is zero. time-ordered and anti-time-ordered propagators that join the same pair of vertices are grouped into one common-theta bundle automatically; the optional top-level field \texttt{"thetaBundles"} pins such a group explicitly as a list of propagator ids, and any group whose members do not join the same pair of vertices is rejected.
\paragraph{Loop integrands: \texttt{MSInitTimeGraph}.}

For a pure time-order diagram the vertex and propagator fields are the same, but the diagram may contain
loops. A triangle loop reads
\begin{verbatim}
cycleContext = MSInitTimeGraph[<|
  "vertices" -> {
    <|"id" -> t1, "externalLegEnergy" -> k1,
      "timePower" -> a1, "vertexType" -> "+"|>,
    <|"id" -> t2, "externalLegEnergy" -> k2,
      "timePower" -> a2, "vertexType" -> "+"|>,
    <|"id" -> t3, "externalLegEnergy" -> k3,
      "timePower" -> a3, "vertexType" -> "+"|>
  },
  "lines" -> {
    <|"type" -> "massless", "endpoints" -> {t1, t2},
      "momentum" -> q12, "nu" -> 1/2|>,
    <|"type" -> "massless", "endpoints" -> {t2, t3},
      "momentum" -> q23, "nu" -> 1/2|>,
    <|"type" -> "massless", "endpoints" -> {t3, t1},
      "momentum" -> q31, "nu" -> 1/2|>
  }
|>];
\end{verbatim}
Here all three propagators join vertices with equal contour signs, so they are classified as
\texttt{masslessT}. Once one of them is pinched, the remaining two connect the same pair of
components and therefore form the common-theta bundle of Sec.~\ref{sec:common-theta}.

\paragraph{Single-vertex integral families: \texttt{MSInitVertexFamily}.}

A single vertex needs no diagram input. The compact input follows the reference
\texttt{ki/nui} convention,
\begin{verbatim}
vertexContext = MSInitVertexFamily[<|
  "ki" -> {k0, k1, k2},
  "nui" -> {a0, nu1, nu2},
  "hankelBranches" -> {1, 2}
|>, NuConvention -> "Positive"];
\end{verbatim}
where \texttt{First[ki]} is the pure vertex phase energy and \texttt{First[nui]} the base power of
$(-\tau)$; the remaining entries give, position by position, the $h$-block momenta and their Hankel
orders $\nu$, which are used exactly as in Sec.~\ref{sec:conventions}. The explicit alternative
\texttt{externalLegEnergy/timePower/hBlocks/exponentialBlocks} is also accepted, where each
$h$-block is an association with the keys \texttt{momentum}, \texttt{nu} and
\texttt{hankelBranch}. In contrast to the tree and time-diagram functions, \texttt{MSInitVertexFamily} rejects any
field outside its own key set.

\paragraph{Package conventions.}
\label{sec:package}

The package implements the $h$ mode-function states of \eqref{eq:h-definition} with $z=-k\tau$ and the
single function name $h$, using
\begin{equation}
  h(\nu,0;z)=z^{\pm\nu}H_\nu(z),
  \label{eq:positive-order-h}
\end{equation}
where the plus sign is the default (\texttt{NuConvention->"Positive"}) and the minus sign gives
the convention of \cite{Chen:2023iix}; the two bases are related by $\nu\to-\nu$
(Sec.~\ref{sec:conventions}). The master integrals are the
$\text{I}_s(\bm 0;\bm a)$ of \eqref{eq:master-vector}, which, for a single vertex, are the one-vertex
integrals $\text{I}^{(0)}_{\tilde{\bm a};\alpha_1\cdots\alpha_n}$ of \eqref{eq:vertex-integral-family}; for
multi-vertex correlators they are the products and
sub-sector integrals of Sec.~\ref{sec:dlog}. The package represents their normalized counterparts
$J_s=\cN_s\widehat{\text I}_s$ by
\texttt{MSIntegral[sectorKey, aShifts, stateBits]}: \texttt{sectorKey} is a binary string over
the root propagators in the input order, with \texttt{"1"} for an unpinched propagator and
\texttt{"0"} for a propagator pinched by a contact; \texttt{aShifts} assigns one integer shift of
the base time power to every vertex component, where a component is a maximal set of vertices
merged by the pinches; \texttt{stateBits} gives the mode-function state of every two-state slot of the
sector, in slot order; the slot order itself is that of \texttt{MSSlotRegistry[context, sector]}.
The bare counterpart $\widehat{\text{I}}_s(\bm n;\bm a)$ in the package slot basis with the same three indices is available as the inert
\texttt{MSBareIntegral[sectorKey, aShifts, stateBits]} for display and audit, and
\texttt{MSIntegralDefinition[integral, context]} returns for any single valid \texttt{MSIntegral}
its bare label, the exact sector normalization $\cN_s$ of \eqref{eq:sector-normalization} --- one
pinch factor $-\sigma_e\frac{4\ii}{\pi}q_e^{\,2\nu_e-1}$ per pinched massive (anti-)time-ordered propagator ---
and the inert identity
$J_s(\bm n;\bm a)=\cN_s\,\widehat{\text{I}}_s(\bm n;\bm a)$.
Their connection is the normalized d log-form system
\eqref{eq:normalized-de}.

We give a few examples of \texttt{MSIntegral[sectorKey, aShifts, stateBits]}. For the three-vertex chain $v_1$--$v_2$--$v_3$ with internal propagators $l_{12}$, $l_{23}$, vertex
energies $k_1,k_2,k_3$ and time powers $a_1,a_2,a_3$, the contact-reachable sectors are the four
strings \texttt{"11"}, \texttt{"01"}, \texttt{"10"}, \texttt{"00"}. Table~\ref{tab:chain-masters}
shows their master indices for three choices of propagators.
\begin{table}[!htbp]
\centering
\emph{(i) both propagators massive: $l_{12}$ with $\nu_{12}$, $l_{23}$ with $\nu_{23}$.}\\[2pt]
\footnotesize\begin{tabular}{@{}lllp{5.8cm}@{}}
\toprule
\texttt{sectorKey} & pinched & components & master integrals\\
\midrule
\texttt{"11"} & --- & $\{v_1\},\{v_2\},\{v_3\}$
  & \texttt{MSIntegral["11",\{0,0,0\},\{a,b,c,d\}]} (16)\\
\texttt{"01"} & $l_{12}$ & $\{v_1,v_2\},\{v_3\}$
  & \texttt{MSIntegral["01",\{0,0\},\{c,d\}]} (4)\\
\texttt{"10"} & $l_{23}$ & $\{v_1\},\{v_2,v_3\}$
  & \texttt{MSIntegral["10",\{0,0\},\{a,b\}]} (4)\\
\texttt{"00"} & $l_{12},l_{23}$ & $\{v_1,v_2,v_3\}$
  & \texttt{MSIntegral["00",\{0\},\{\}]} (1)\\
\bottomrule
\end{tabular}
\par\medskip
\emph{(ii) $l_{12}$ massive with $\nu_{12}$, $l_{23}$ massless.}\\[2pt]
\footnotesize\begin{tabular}{@{}lllp{5.8cm}@{}}
\toprule
\texttt{sectorKey} & pinched & components & master integrals\\
\midrule
\texttt{"11"} & --- & $\{v_1\},\{v_2\},\{v_3\}$
  & \texttt{MSIntegral["11",\{0,0,0\},\{a,b,r\}]} (8)\\
\texttt{"01"} & $l_{12}$ & $\{v_1,v_2\},\{v_3\}$
  & \texttt{MSIntegral["01",\{0,0\},\{r\}]} (2)\\
\texttt{"10"} & $l_{23}$ & $\{v_1\},\{v_2,v_3\}$
  & \texttt{MSIntegral["10",\{0,0\},\{a,b\}]} (4)\\
\texttt{"00"} & $l_{12},l_{23}$ & $\{v_1,v_2,v_3\}$
  & \texttt{MSIntegral["00",\{0\},\{\}]} (1)\\
\bottomrule
\end{tabular}
\par\medskip
\emph{(iii) both propagators massless.}\\[2pt]
\footnotesize\begin{tabular}{@{}lllp{5.8cm}@{}}
\toprule
\texttt{sectorKey} & pinched & components & master integrals\\
\midrule
\texttt{"11"} & --- & $\{v_1\},\{v_2\},\{v_3\}$
  & \texttt{MSIntegral["11",\{0,0,0\},\{r,s\}]} (4)\\
\texttt{"01"} & $l_{12}$ & $\{v_1,v_2\},\{v_3\}$
  & \texttt{MSIntegral["01",\{0,0\},\{r\}]} (2)\\
\texttt{"10"} & $l_{23}$ & $\{v_1\},\{v_2,v_3\}$
  & \texttt{MSIntegral["10",\{0,0\},\{r\}]} (2)\\
\texttt{"00"} & $l_{12},l_{23}$ & $\{v_1,v_2,v_3\}$
  & \texttt{MSIntegral["00",\{0\},\{\}]} (1)\\
\bottomrule
\end{tabular}
\caption{Master indices of an equal-sign three-vertex chain with massive or massless
propagators. Each state index takes the values $0$ and $1$; parentheses give the number
of masters in each sector.}
\label{tab:chain-masters}
\end{table}

The vertex \texttt{vertexType} field (\texttt{"+"} or \texttt{"-"}) determines the sign of the pure
vertex exponential $e^{\ii k_0\tau}$ of \eqref{eq:vertex-integral-family}, and the pair of vertex types determines the SK class of each propagator, from which the default Hankel branches of
\eqref{eq:h-definition} follow; a propagator may override them with its own \texttt{hankelBranches}
field.

Once the context is initialized, the following functions expose the formula products of Sec.~\ref{sec:method}.

\paragraph{Core functions.}
\label{sec:core}

Most functions take the initialized \texttt{context};
\texttt{MSExportEvaluationData} takes a completed evaluation result instead.
Table~\ref{tab:core-functions} lists the principal functions.

\begin{table}[htbp]
\centering
\begin{tabular}{@{}p{7.1cm}p{7.3cm}@{}}
\toprule
function & purpose\\
\midrule
\texttt{MSSectors[context]} & contact-reachable sectors in deterministic order (keys, pinched
propagators, master counts)\\
\texttt{MSMasterIntegrals[context]} & ordered master records; each \texttt{"integral"} field is an
\texttt{MSIntegral[sectorKey, aShifts, stateBits]}\\
\texttt{MSIntegralDefinition[integral, context]} & bare label, exact normalization $\cN_s$ and
inert identity of one valid \texttt{MSIntegral}\\
\texttt{MSFormulaMatrices[context, key]} & per-component $\text{M}_1$, $\text{M}_0$ and energy letters $\kappa_{\bm d}$;
sector diagonalizer \texttt{"U"} and \texttt{"UInverse"}\\
\texttt{MSDLogDE[context]} & block-triangular d log-form potential $\Omega$, letters, certification
status\\
\texttt{MSRecurrenceStep[integral, component, context]} & one recurrence step moving one nonzero
time shift toward zero\\
\texttt{MSReduce[expr, context]} & iterative IBP reduction to the master basis; scalar, list or nested table input\\
\texttt{MSContactMaps[context, key]} & supersector-to-subsector contact maps\\
\texttt{MSWriteFormulaArtifacts[context]} & write formula products beside the caller\\
\texttt{MSFormulaData[context]} & aggregate all-sector masters, recurrence metadata and the d log-form DE\\
\texttt{MSEvaluatePath[context, pointSequence, opts]} & single-stage evaluation: boundary initialization and segment transport in one backend process; optional FlintNDE path planning\\
\texttt{MSReconstructEpSeries[context, ep, pointSequence, opts]} & Laurent series in the common analytic regulator: certified lowest power, automatic sample points, incremental fitting\\
\texttt{MSExportEvaluationData[evaluation, opts]} & export the saved ordinary-point records of an evaluation to CSV/JSON\\
\bottomrule
\end{tabular}
\caption{Principal formula, evaluation and export functions.}
\label{tab:core-functions}
\end{table}

\begin{description}
  \item[\texttt{MSSectors[context]}] returns the sector list. Each sector key is the fixed-length
        bit string over the root propagators in input order (\texttt{"0"} pinched,
        \texttt{"1"} unpinched); the
        sub-sector structure is discussed in Sec.~\ref{sec:dlog}. Keys are
        strings, not integers; leading zeros are part of the identity.
  \item[\texttt{MSMasterIntegrals[context]}] returns the ordered master records.
        Their \texttt{"integral"} fields contain the
        \texttt{MSIntegral[sectorKey, aShifts, stateBits]} objects of Sec.~\ref{sec:package}, i.e.\
        the normalized masters $J_s(\bm 0;\bm a)$ of \eqref{eq:normalized-vector}.
        Use \texttt{Lookup[MSMasterIntegrals[context], "integral"]} to obtain this ordered basis.
        Each record also contains \texttt{"bareIntegral"}, \texttt{"normalization"} and
        \texttt{"definition"}. All matrices, boundary vectors and numerical results follow this order.
  \item[\texttt{MSFormulaMatrices[context, key]}] returns the component records under
        \texttt{"components"}, with fields \texttt{"M1"}, \texttt{"M0"} and
        \texttt{"energyLetters"} for the matrices and letters of Sec.~\ref{sec:global}.
        The sector diagonalizer of $\text{M}_0$ and its inverse are returned as
        \texttt{"U"} and \texttt{"UInverse"}; $\text{M}_1$ is already diagonal in the mode-function state basis.
        Here \texttt{key} is a sector-key string; omitting it, or passing \texttt{All},
        returns an association for all sectors. \texttt{MSContactMaps} accepts the same
        sector selection.
  \item[\texttt{MSDLogDE[context]}] returns \texttt{"omegaPotential"}, the matrix $\Omega$ of
        Sec.~\ref{sec:dlog} whose derivative along a path is the DE matrix, together with the
        letters (the arguments of the logarithms of \eqref{eq:omega-form}),
        \texttt{"letterMatrices"}, \texttt{"dlogResidual"}, \texttt{"constantResidueQ"} and the
        certification \texttt{"dlogStatus"}. The closure test itself reduces the shifted integrals
        generated by the connection with \texttt{MSReduce}, requiring status \texttt{"reduced"}, a
        zero \texttt{"nonMasterResidual"} and no \texttt{"remainingShiftedIntegrals"} on every
        entry; the connection matrix is then read off the reduction coefficient vectors. Only the
        status \texttt{"certifiedByFormulaChecks"}
        certifies a closed d log-form, and only a certified potential may be used numerically.
  \item[\texttt{MSReduce[expr, context, MasterBasis -> Automatic]}] reduces finite linear
        combinations of legal \texttt{MSIntegral}s, or lists and nested tables of such
        expressions, by
        the lowering/raising recurrences $A_\pm$ of Sec.~\ref{sec:iterative} along the
        sub-sector structure of Sec.~\ref{sec:dlog}. The single step is exposed separately as
        \texttt{MSRecurrenceStep[integral, component, context]}, which moves one nonzero time
        shift of one component toward zero; with the component omitted it picks the first nonzero
        shift, and it refuses any step that would move a component into one of the divergent
        small-time layers fixed at initialization. Results of \texttt{MSReduce} include
        \texttt{"result"}, \texttt{"masterBasis"}, \texttt{"coefficientVector"},
        \texttt{"masterRules"},
        \texttt{"nonMasterResidual"}, \texttt{"remainingShiftedIntegrals"} and
        \texttt{"singularLayers"}; a full reduction requires a zero residual and an empty shifted
        list for every scalar expression. For list input, these fields preserve the input
        shape; \texttt{"elementResults"}, \texttt{"failurePositions"} and
        \texttt{"incompletePositions"} provide the element-level reports. The reduction runs
        on the time powers fixed at initialization: a context initialized with specialized or
        numerical \texttt{timePower} values, such as the regulated powers of Example~06,
        reduces with the substitution in place, and purely symbolic results may equivalently
        be substituted afterwards. A worked reduction of a two-term shifted combination is
        displayed in Sec.~\ref{sec:app-regulator}.
  \item[\texttt{MSFormulaData[context]}] returns all-sector masters, the per-sector
        $\text{M}_1/\text{M}_0/\text{T}_n$ recurrence metadata of Sec.~\ref{sec:global} and the complete
        block-triangular d log-form DE in the master order of \texttt{MSDLogDE};
        \texttt{TimePowerRules} may substitute the vertex time powers $a_i$ in the recurrence
        metadata, but the returned \texttt{"dlogDE"} remains analytic and unsubstituted.
  \item[\texttt{MSEvaluatePath[context, pointSequence, opts]}] is the single-stage numerical
        entry for evaluating the master integrals at the user points.
        \texttt{pointSequence} is a coordinate table in the format of Sec.~\ref{sec:numerics}.
        Fixed parameters are supplied once through \texttt{ParameterRules}. Boundary generation,
        transport and result records are described in Sec.~\ref{sec:numerics}; the principal
        options and point tags are summarized in Sec.~\ref{sec:options}.
  \item[\texttt{MSReconstructEpSeries[context, ep, pointSequence, opts]}] reconstructs the Laurent
        series of the master integrals in a common analytic regulator \texttt{ep}, with
        \texttt{MaximumEpPower} the highest returned power (default $0$, i.e.\ the pole terms and
        the finite part). It uses the same \texttt{pointSequence} and \texttt{ParameterRules}
        format as \texttt{MSEvaluatePath}. Only \texttt{ParameterRules}, not the coordinates,
        may depend on \texttt{ep}. By default, no \texttt{ep} values need to be supplied;
        \texttt{EpSamplePoints} and \texttt{EpValidationPoints} may instead specify the production
        points and disjoint validation points. The certificate conditions,
        sampling and fitting procedure are described in Sec.~\ref{sec:regulator}; the corresponding
        options are summarized in Sec.~\ref{sec:options}, and a complete application is
        presented in Sec.~\ref{sec:app-regulator}.
  \item[\texttt{MSExportEvaluationData[evaluation, opts]}] exports the saved ordinary-point
        records of one \texttt{MSEvaluatePath} result, including the coordinates, the
        real/imaginary parts of every master, \texttt{"status"},
        \texttt{"relativeDifferenceInf"} and
        \texttt{"targetRelativeErrorMet"}, to CSV and JSON for downstream plotting tools; transient
        records are excluded. The two precision fields are the corresponding segment reports,
        not separate error estimates computed at every point. \texttt{ExportFormats} selects
        CSV, JSON or both, and \texttt{SignificantDigits} defaults to $16$;
        see Sec.~\ref{sec:outputs}.
\end{description}

\paragraph{Output organization.}
\label{sec:outputs}

\begin{itemize}
  \item Transient request JSON and backend logs:
        \texttt{results\_temp/\allowbreak nde/} beside the caller by default.
        Request and log files are deleted after a successful run and retained for backend
        failures; the package also deletes them for the structured \texttt{"singularPathRefused"}
        and \texttt{"leadingOrderRefused"} outcomes, which are results rather than failures.
        Successful JSON results remain in \texttt{results\_temp/\allowbreak cache/} for reuse.
        \texttt{MSRuntimeDirectory} may specify a different runtime directory.
  \item Saved point records: \texttt{MSEvaluatePath} returns the records under
        \texttt{"pointResults"}. An ordinary untagged coordinate row has
        \texttt{"status" -> "saved"}; \texttt{"saved"} is a status value, not a top-level key.
        A row \texttt{\{\{values...\}, "tmp"\}} gives a transient waypoint, represented by
        \texttt{"status" -> "transient"} and excluded from export.
        Singular-point classifications are returned under \texttt{"singularityClassifications"}
        by the same backend evaluation run.
  \item Formula artifacts: initialization itself writes no files.
        \texttt{MSWriteFormulaArtifacts[context, opts]} --- with its own options
        \texttt{TimePowerRules}, which specializes the recurrence metadata only, and
        \texttt{MSOutputDirectory}, which redirects the artifact directory --- writes
        \texttt{masters.wl},
        \texttt{recurrence\_metadata.wl} (the per-sector $\text{M}_1$, $\text{M}_0$, diagonalizer
        and energy letters that feed the recurrence),
        \texttt{dlog\_de.wl} and \texttt{manifest.wl} (version, master digest, sector order,
        certification status and actual paths) to
        \texttt{results/\allowbreak madstree\_formula/\allowbreak run-UUID/} by default.
        \texttt{MSEvaluatePath} and \texttt{MSReconstructEpSeries} automatically write or reuse
        these analytic files before numerical transport and return their paths under
        \texttt{"formulaArtifacts"}; the write is refused unless the d log-form DE is certified.
  \item Evaluation export: \texttt{MSExportEvaluationData[result, MSOutputDirectory -> ...]} writes
        \texttt{evaluation\_data.csv} and \texttt{evaluation\_data.json} under
        the caller-chosen output directory (default
        \texttt{results/\allowbreak madstree\_evaluation/\allowbreak run-UUID/}), for downstream
        plotting tools.
  \item Cache identity: a successful-result cache key includes the request and the SHA-256
        identities of the backend adapter, the built-in backend's \texttt{pyproject.toml}
        and sorted \texttt{flintnde/*.py}, so
        in-place source fixes invalidate stale cache entries automatically.
\end{itemize}

\subsection{Numerical evaluation}
\label{sec:numerics}

With an initialized context, numerical evaluation proceeds in two steps: boundary generation and
transport. The workflow is single-stage: \texttt{MSEvaluatePath} partitions the user point
sequence into maximal consecutive complex-affine one-variable segments, pulls the d log-form
connection back once per segment, and sends all segments, the exact boundary conditions and the
user options to \texttt{FlintNDE} in one UTF-8 JSON request. With the default
\texttt{FlintNDEPathPlanning -> True}, \texttt{FlintNDE} plans the transport nodes inside each segment and
evaluates user points by fast multipoint evaluation; with \texttt{False} it uses every supplied
point as a node in input order. The mechanisms
of the built-in \texttt{FlintNDE} backend are described in Sec.~\ref{sec:backend}. \texttt{MSBoundaryData} and \texttt{MSEvaluatePath}
default to 200 decimal working digits; an explicit \texttt{WorkingPrecision} overrides the
default.

\paragraph{Boundary generation.}

\begin{verbatim}
targetRules = {k1 -> 9 I, k2 -> 3 I, q -> 1, a1 -> 1, a2 -> 1};

certificate = MSBoundaryChartCertificate[context, targetRules,
  RankOrder -> All];

boundary = MSBoundaryData[context, targetRules,
  BoundaryScale -> 4, WorkingPrecision -> 200];
\end{verbatim}
\texttt{MSBoundaryChartCertificate} verifies the nested blow-up chart of Sec.~\ref{sec:boundary}
(normal crossing, theta fixing, quotient checks) for the chosen strict energy rank;
\texttt{RankOrder -> All} checks every strict chart. \texttt{MSBoundaryData} consumes the
certificate and returns the leading Frobenius data at the $K_v\to\infty$ boundary. For every master
whose leading weight does not vanish, it returns one branch record with the Frobenius exponent, the
overall coefficient and the normalized leading vector, together with the anchor rules, the rank order
and the certificate itself. The anchor rules normalize each leading vector: the entry of the
anchoring master is fixed to one, entries of masters not reached by that solution are zero, and the
remaining entries follow from a linear system that propagates leading subsector data into the
masters of higher sectors. A master with a vanishing leading coefficient produces no record, and an
unsolvable system is reported as a failure.
If the chart is not certified, it returns \texttt{BoundaryChartNotCertified} and does not proceed
to numerics. The energies in the example are positive imaginary so that the vertex factors
$e^{-\ii E\tau}$ damp at $\tau\to-\infty$; the rank order of the boundary is derived from the
supplied target point, largest damping energy first, unless \texttt{RankOrder} fixes it.

Boundary values from any independent representation of the masters --- a hypergeometric
evaluation, for example --- can replace the automatic construction through the low-level
entry \texttt{MSNumericalSystem[de, spec]}, where \texttt{de} is the association returned
by \texttt{MSDLogDE}. The association \texttt{spec} fixes the path parameter, the path
rules that pull the letters back to one variable, the numerical substitutions, the
interval, and the key \texttt{"boundaryValues"} with one value per master in master order:
\begin{verbatim}
dlogDE = MSDLogDE[context];

system = MSNumericalSystem[dlogDE, <|
   "pathParameter" -> t,
   "pathRules" -> {k1 -> 9 I t, k2 -> 3 I},
   "substitutions" -> {q -> 1, a1 -> 1, a2 -> 1},
   "interval" -> {0, 1},
   "boundaryPoint" -> 0,
   "boundaryValues" -> {1, 1, 1}|>];

system["status"]
system["initialConditions"]
\end{verbatim}
\texttt{MSNumericalSystem} validates the substitutions and the boundary vector and returns
the pulled-back differential equations together with one initial condition per master,
fixing the corresponding unknown function at the boundary point; in this example the
status is \texttt{"ready"} and the three initial conditions fix the three unknown master
functions to $1$ at $t=0$. Without \texttt{"boundaryValues"} the returned status is
\texttt{"boundaryDataRequired"} and no initial conditions are constructed. The single-stage
entry \texttt{MSEvaluatePath} does not accept boundary values and always constructs the
automatic boundary above.

\paragraph{Transport.}

\begin{verbatim}
pointSequence = {{k1, k2}, {9 I, 3 I}};
parameterRules = {q -> 1, a1 -> 1, a2 -> 1};

result = MSEvaluatePath[context, pointSequence,
  ParameterRules -> parameterRules,
  BoundaryScale -> 4, RankOrder -> {v1, v2},
  WorkingPrecision -> 200, TransportOrder -> 80,
  ReferenceTransportOrder -> 104,
  TargetRelativeError -> "1e-20",
  MessageLanguage -> "EN"];

result["pointResults"]
\end{verbatim}
The second argument is the \emph{point sequence}: its first row lists the movable coordinate
symbols, and every further row supplies one numerical value for each of them, in the same order.
The remaining symbols of the target point, here \texttt{q}, \texttt{a1} and \texttt{a2}, are fixed
once for the whole run through \texttt{ParameterRules}. For several target points, append one row per
point,
\begin{verbatim}
pointSequence = {{k1, k2}, {8 I, 2 I}, {7 I, 4 I}, {6 I, 5 I}};
\end{verbatim}
and \texttt{MadStree} groups the
points into maximal consecutive complex-affine segments and pulls each segment back once, while
\texttt{FlintNDE} evaluates the requested points in one process from the local series associated
with the planned transport nodes. For a single-vertex integral family, run the same
single-stage entry on the vertex context. The principal transport options and the point tags are
summarized in Sec.~\ref{sec:options}. If every d log-form letter is constant on a segment, every
pulled-back derivative vanishes; the segment is then a zero connection with no finite poles and
transport proceeds as for any ordinary segment.

\paragraph{Reading the result.}

\texttt{result} is an association whose \texttt{"pointResults"} entry is the list of point records.
Every saved ordinary point carries
\texttt{"coordinate"}, \texttt{"value"} (the master vector in master order), \texttt{"status"}
(equal to
\texttt{"saved"} for ordinary points) and \texttt{"userIndex"}; transient \texttt{"tmp"} points
are used only as intermediate points of the path. The current version does not provide leading-order
point labels: every tag other than \texttt{"tmp"} is rejected. With
\texttt{FlintNDEPathPlanning -> False}, a user chain that
crosses a pole or leaves a local convergence disk fails closed; the two options that decide this,
\texttt{SingularityMode} and \texttt{FlintNDEPathPlanning}, are described in
Sec.~\ref{sec:options}. The exported per-point
\texttt{"relativeDifferenceInf"} and \texttt{"targetRelativeErrorMet"} are obtained from the
truncation certification of Sec.~\ref{sec:backend}.
To read the first saved value, use
\begin{verbatim}
savedPoints = Select[result["pointResults"], #["status"] === "saved" &];
firstPoint = First[savedPoints];
firstPoint["value"]
\end{verbatim}

\subsection{The built-in \texttt{FlintNDE} backend}
\label{sec:backend}

\texttt{FlintNDE} solves first-order matrix differential equations with exact
$Q(\ii)(s)$-coefficient systems; \texttt{MadStree} includes it as a built-in module and
feeds it single-variable pullbacks of the d log-form connection. The entries
\texttt{MSEvaluatePath} and \texttt{MSReconstructEpSeries} call the built-in backend,
as described in Secs.~\ref{sec:numerics} and~\ref{sec:regulator}. Four mechanisms are responsible
for the current performance:
\begin{itemize}
  \item \emph{Structured tensor letters.} The state labels of the d log-form masters
        factorize over propagator slots, and each letter matrix is fixed by tensor
        products of two-state local bases on these slots, with the identity on
        unaffected slots. Along a segment every certified letter is affine,
        $\alpha+\beta s$, and is serialized as this factor together with structured
        per-sector contributions --- sparse vectors, letter indicators and scalar
        post-factors --- while the sector records carry the slot basis labels and the
        sparse quotient maps. The backend rebuilds every residue exactly in $Q(\ii)$
        from this factorized form.
  \item \emph{Pole--residue transport.} An ordinary segment is serialized as its pole--residue
        representation $A(x)=P(x)+\sum_jR_j/(x-p_j)$, with $P$ of arbitrary degree and the residues
        $R_j$ of coincident letters merged. The backend generates the solution Taylor coefficients
        by a recurrence that carries one auxiliary state vector per pole, avoiding the Cauchy sampling
        and discrete-Fourier coefficient reconstruction of a generic analytic-matrix backend.
  \item \emph{Path planning and fast multipoint evaluation.} With the default
        \texttt{FlintNDEPathPlanning -> True}, \texttt{FlintNDE} places its own transport nodes
        inside each segment according to the local convergence geometry;
        user points covered by one expansion node form an
        evaluation bucket and are evaluated from the cached vector series. Buckets with at least
        eight points use the subproduct/remainder tree of fast multipoint evaluation, and small
        buckets use a simple iterative algorithm to avoid the overhead of the tree. Within each
        complex-affine segment $x(s)=x_0+s\,v$, points in the same convergence disk share
        one set of local series coefficients. With
        \texttt{FlintNDEPathPlanning -> False}, every supplied point is used as a node in input
        order.
  \item \emph{Embedded truncation certification.} The backend is always called in its embedded
        certification mode, which runs one chain at \texttt{ReferenceTransportOrder} and takes the
        primary result from the coefficient prefix truncated at \texttt{TransportOrder}, re-evaluated
        by Horner's rule. The per-segment truncation difference is returned as a local error estimate,
        so the \texttt{"relativeDifferenceInf"} and \texttt{"targetRelativeErrorMet"} fields remain
        available without the cost of a second chain. The call is upgraded to independent primary and
        reference chains whenever the prefix cannot carry the certification: on a multi-segment path,
        on a path with singular local data, for a run that starts from Frobenius boundary data, and
        for a request with saved points.
\end{itemize}
Singularity handling is explicit: the default \texttt{SingularityMode -> "Automatic"} maps to the
backend jump mode, so a segment that meets a supported singularity is crossed by an explicit
local-basis jump. The three modes and their interaction with
\texttt{FlintNDEPathPlanning} are listed in Sec.~\ref{sec:options}. Exact Lee--Moser balances,
strictly decoupled exponential times, and power-log recurrences are handled, while resonances
whose local bases cannot be constructed, and any ramification or general Stokes requirement, fail
closed. The internal working bits are $\lceil\mathrm{WP}\log_2 10\rceil+32$, where WP is the value of
\texttt{WorkingPrecision} in decimal digits. Matrix arithmetic and local series calculations
use Arb/Acb precision, while some geometric comparisons use double precision.\footnote{In real-variable coordinates,
$k=-\ii\kappa$ and $-k^2=+\kappa^2$.
The code variable \texttt{ik} denotes $\kappa$; \texttt{ik\^{}2} denotes minus $k^2$.}

\subsection{Laurent-series reconstruction}
\label{sec:regulator}

When all vertex time powers share a common analytic regulator $\varepsilon$, e.g.\
$a_v=1+\varepsilon$, the master integrals develop a Laurent series
\begin{equation}
  I(\varepsilon)=\sum_{p=n_{\min}}^{m}c_p\,\varepsilon^p
  +O(\varepsilon^{m+1}),
  \label{eq:regulator-laurent}
\end{equation}
and \texttt{MSReconstructEpSeries[context, ep, pointSequence, MaximumEpPower -> m]}
reconstructs the coefficients $c_{n_{\min}},\ldots,c_m$. The user supplies no
$\varepsilon$ values; the lowest power is not a numerical fit parameter.

This procedure follows the one used by AMFlow \cite{Liu:2022chg,Huang:2026rjb}. The transport
is numerical, while the regulator dependence of the result has to stay analytic, and carrying that
analytic dependence through every matrix multiplication of a long path would dominate the cost.
Each sample point is therefore evaluated numerically at a fixed value of $\varepsilon$, and the
coefficients of \eqref{eq:regulator-laurent} are recovered afterwards from the final values of all
runs, which is what keeps a multi-order expansion affordable.

\paragraph{Certification of the lowest power.}
Before any numerical solve at nonzero $\varepsilon$, \texttt{MadStree} constructs a support
certificate from the symbolic boundary conditions and the d log-form DE of the production
context. It requires:
\begin{enumerate}
  \item regulator-independent path coordinates and no negative Laurent power at
        $\varepsilon=0$ in any d log-form matrix, pulled-back connection, or residue;
  \item finite-Laurent boundary branches, where branches with the same limiting Frobenius
        exponent and log degree are first combined with their physical weights and leading
        vectors, and only then are the component valuations of the record field
        \texttt{"componentValuations"} extracted;
  \item at least one provably nonzero coefficient at the lowest power of the combined
        boundary; a noninteger power, non-Laurent structure, undecidable coefficient, or
        support entirely above the requested user range fails closed.
\end{enumerate}
For $n_{\min}<0$ the certificate describes a finite meromorphic Laurent boundary, not an
analytic boundary at $\varepsilon=0$; analyticity is recorded only when $n_{\min}\geq0$.
Along the current finite path an $\varepsilon$-analytic invertible fundamental matrix
preserves the valuation of nonzero formal boundary conditions, so the certified $n_{\min}$ is
passed directly to the sampling plan. A rank drop of an individual Frobenius recurrence
operator at $\varepsilon=0$ is retained as a resonance diagnostic only; it is not by itself
interpreted as a $1/\varepsilon$ pole of the already combined physical boundary.

\paragraph{Sampling, fitting and validation.}
Given the highest requested power $m$ and the certified lowest power $n_{\min}$, the plan sets
\begin{equation}
  P=\max(0,-n_{\min}),\qquad
  \Delta=m-n_{\min},\qquad
  N_{\rm fit}=\max\!\left(\left\lceil\tfrac52\Delta+P\right\rceil,\Delta+1+f\right),
  \label{eq:regulator-fit-count}
\end{equation}
where $f=\texttt{EpFitExtraOrder}$, default $2$, is the number of powers the fit keeps above the
user maximum: the second term in the maximum in \eqref{eq:regulator-fit-count} would be $\Delta+1$ if
only $c_{n_{\min}},\ldots,c_m$ were fitted, and the extra $f$ is what makes the interpolation reach
past $c_m$ in every run. A user-supplied pool \texttt{EpSamplePoints} and the refit rounds below may
raise $N_{\rm fit}$ further, but $N_{\rm fit}$ is never below $\Delta+1$.
The plan then chooses the $\varepsilon$ scale and precisions
\begin{equation}
  \alpha_\varepsilon=\frac{P}{4}+\frac{g}{\Delta+1},\qquad
  \varepsilon_0=10^{-\alpha_\varepsilon},\qquad
  p_0=\max\!\left(\left\lceil(N_{\rm fit}+P)\alpha_\varepsilon\right\rceil,30\right),
  \label{eq:regulator-scale-precision}
\end{equation}
where $g=\texttt{EpGoalDigits}$ sets the validation tolerance $10^{-g}$. The working precision is
$\max(200,\lceil2(1+e)p_0\rceil)$ with $e$ the extra-precision fraction; the primary
transport order is $4p_0$ and the reference chain adds
$\lceil\max(50,p_0/5)\rceil$ orders. All $N_{\rm fit}$ production points enter one square
Acb Vandermonde system with powers up to $\varepsilon^{n_{\min}+N_{\rm fit}-1}$; only
$c_{n_{\min}},\ldots,c_m$ are returned.
Independent validation points lie at a fixed fraction of the production scale and are held out
from the fit. At each point, the fitted series includes all $N_{\rm fit}$ internal powers through
$n_{\min}+N_{\rm fit}-1$, including the buffer powers above the requested maximum $m$.
This sum must reproduce the directly transported value to the relative tolerance $10^{-g}$. If the target
is not met after the allowed refit rounds, the result carries a warning status and
\texttt{"precisionTargetMet" -> False}. The residual tests the fitted series at the validation
points; it does not establish the number of correct digits of each returned coefficient
$c_{n_{\min}},\ldots,c_m$.

\paragraph{Incremental expansion and parallelism.}
The initial fit includes at least two powers above the user maximum
(\texttt{EpFitExtraOrder -> 2}). If validation fails, each subsequent round adds
\texttt{EpFitOrderIncrement -> 2} powers, solves only the new production points, and reuses
every previous production and validation value, for at most \texttt{EpFitMaximumRounds -> 3}
rounds, at the prescribed tolerance. Independent
$\varepsilon$ tasks run in an outer process pool bounded by \texttt{ParallelTaskCount -> 12},
whose effective concurrency is the smaller of the limit and the number of tasks; this is
separate from the in-process \texttt{ctx.threads} of \texttt{python-flint}.
The application in Sec.~\ref{sec:app-regulator} runs this procedure on a divergent
configuration.

\subsection{Options, validation and common pitfalls}
\label{sec:options}

The principal options of the numerical workflow are summarized in Table~\ref{tab:numerical-options}.
\begin{table}[!htbp]
\centering
\begin{tabular}{@{}p{6.2cm}p{8.1cm}@{}}
\toprule
option & meaning\\
\midrule
\texttt{FlintNDEPathPlanning} & \texttt{True} (default): \texttt{FlintNDE} plans nodes and uses fast
multipoint evaluation; \texttt{False}: user points as nodes in input order\\
\texttt{SingularityMode} & \texttt{"Automatic"} (default), \texttt{"Avoid"} or \texttt{"SingularityJump"}\\
\texttt{MessageLanguage} & \texttt{"EN"} (default) or \texttt{"CN"}, case-sensitive\\
\texttt{BoundaryScale} & scale setting the damping energies at the finite boundary anchor ($>1$; default 8)\\
\texttt{BoundarySeriesOrder} & boundary-order metadata and regulator recurrence diagnostics; not the numerical Frobenius start order in the present generic workflow\\
\texttt{RankOrder} & vertex order by damping energy $K_v$ (default \texttt{Automatic}: from the first target point)\\
\texttt{WorkingPrecision} & integer working precision in decimal digits, at least 20; default 200\\
\texttt{TransportOrder} / \texttt{ReferenceTransportOrder} & primary and reference local-series orders, including the numerical Frobenius start; defaults 48 and 64, with the reference order strictly larger\\
\texttt{TargetRelativeError} & target relative error of the primary/reference check; default \texttt{"1e-25"}\\
\texttt{MaximumEpPower} & highest returned regulator power of \texttt{MSReconstructEpSeries}
(default 0)\\
\texttt{EpGoalDigits} & decimal tolerance exponent for validation of the reconstructed regulator series
(default 20)\\
\texttt{EpFitExtraOrder} / \texttt{EpFitOrderIncrement} / \texttt{EpFitMaximumRounds} / \texttt{EpInitialInternalMaximumPower} & internal
buffer powers, per-round increment, maximum refit rounds and first-round internal highest power of
the regulator fit (defaults 2, 2, 3, \texttt{Automatic})\\
\texttt{ParallelTaskCount} & maximum concurrent \texttt{ep} backend processes in \texttt{MSReconstructEpSeries} (default 12)\\
\texttt{PythonExecutable} & Python command for the \texttt{FlintNDE} adapter (default \texttt{Automatic})\\
\texttt{MSRuntimeDirectory} & temporary-file root (default: \texttt{results\_temp} in the calling script or notebook directory)\\
\bottomrule
\end{tabular}
\caption{Principal numerical options and their defaults.}
\label{tab:numerical-options}
\end{table}

The transport orders and \texttt{TargetRelativeError} in Table~\ref{tab:numerical-options} apply to
\texttt{MSEvaluatePath}. For \texttt{MSReconstructEpSeries}, these three options
are not accepted; the numerical settings are planned from \texttt{EpGoalDigits}.
Its \texttt{BoundarySeriesOrder} defaults to \texttt{Automatic}, whereas
\texttt{MSEvaluatePath} uses 24.

The saved/transient distinction is not a transport option: it is a tag attached to the coordinate
rows below the symbol header, which are bare value lists or \texttt{\{\{values...\},"tmp"\}}.
With \texttt{SingularityMode -> "Avoid"}, interior singularity crossings are refused, and the
refusal names the interior singular points of the path. The default \texttt{"Automatic"} mode
allows supported singularity jumps when \texttt{FlintNDEPathPlanning -> True};
an explicit \texttt{"SingularityJump"} also requires this setting.
When planning is disabled, \texttt{"Automatic"} behaves as \texttt{"Avoid"}.
An explicitly requested
terminal singular point is handled by local classification and evaluation,
rather than by ordinary Taylor transport through the singularity.

Validity checks and result status:
\begin{itemize}
  \item The package evaluates time integrals, including loop integrands, and obtains
        IBP reductions from the closed recurrence formulas.
  \item \texttt{"dlogStatus"} must be \texttt{"certifiedByFormulaChecks"}; otherwise the DE is not
        consumed.
  \item Numerical evaluation requires a certified boundary chart and a supported local basis
        at each singular point; unsupported cases return structured refusals.
  \item An incomplete \texttt{MSReduce} call may return an association with
        \texttt{"status" -> "partiallyReduced"} or \texttt{"status" -> "failed"},
        rather than a top-level \texttt{Failure}. Inspect
        \texttt{"remainingShiftedIntegrals"} and \texttt{"nonMasterResidual"}.
  \item For \texttt{MSEvaluatePath}, check
        \texttt{result["flintNDE"]["targetRelativeErrorMet"]}, not only
        \texttt{result["status"]}. For \texttt{MSReconstructEpSeries}, check
        \texttt{"precisionTargetMet"}; coefficients may be returned with
        \texttt{"status" -> "computed\_with\_warning"} if the target is not met.
\end{itemize}

\paragraph{Common pitfalls.}
\label{sec:pitfalls}

\begin{enumerate}
  \item \textbf{Convention}: \texttt{NuConvention} is chosen at initialization only
        (\texttt{"Positive"} by default). To compare with the formulas of
        \cite{Chen:2024glu}, initialize with \texttt{NuConvention -> "Negative"}, which selects
        the minus sign in \eqref{eq:positive-order-h}.
  \item \textbf{Sector keys}: fixed-length strings; leading zeros and the string type are part of
        the identity (Sec.~\ref{sec:dlog}).
  \item \textbf{Single-stage workflow}: call \texttt{MSEvaluatePath} with a coordinate table
        and fixed \texttt{ParameterRules}. The call generates the boundary data once and
        transports the solution along the full point sequence.
  \item \textbf{Singularity handling}: the default is \texttt{SingularityMode -> "Automatic"},
        where \texttt{"Automatic"} is a string, not the symbol \texttt{Automatic}.
        Use \texttt{"Avoid"} to refuse interior crossings. Supported jumps require
        \texttt{FlintNDEPathPlanning -> True}, and the user must check the multivalued branch.
  \item \textbf{Point tags}: after the coordinate-symbol header, only bare value lists and
        \texttt{\{\{values...\},"tmp"\}} are accepted; every other tag, including
        \texttt{"lo"}, is rejected.
  \item \textbf{Backend environment}: \texttt{MadStree} follows the zero-probe policy of
        Sec.~\ref{sec:overview}, so the interpreter is the one selected there and must be able to
        import the embedded \texttt{flintnde}. Backend failures report captured output through
        messages such as \texttt{backendLaunchFailed} and \texttt{backendRunFailed}; retained runtime
        logs contain the details. After a successful call, and after a structured refusal of a
        singular path, the package removes the temporary input and log files.
  \item \textbf{Message language}: \texttt{MessageLanguage} accepts only \texttt{"EN"} and
        \texttt{"CN"} (case-sensitive), with \texttt{"EN"} the default.
  \item \textbf{Encoding}: loading through \texttt{Needs} uses explicit UTF-8 internally and needs
        no user encoding option; if loading the kernel file directly fails with syntax errors, set
        \texttt{\$CharacterEncoding = "UTF-8"} before loading.
\end{enumerate}

\section{Examples}
\label{sec:examples-section}

This section lists the examples distributed with the package and presents four applications
computed with them, using \texttt{FlintNDE} for numerical transport; all timings measure
elapsed time on a personal computer.
Sec.~\ref{sec:examples} describes the seven distributed examples, which exercise
recurrence reduction, single-vertex families, time-order charts, massless and massive trees,
zero external-leg energy, and the regulator reconstruction of Sec.~\ref{sec:regulator}.
The first application (Sec.~\ref{sec:app-massive-tree}) is the tree with three vertices and
two massive exchanges. It compares all 25 $(+++)$ masters with an independent hypergeometric
series; the second (Sec.~\ref{sec:app-bubble}) is the bubble loop
integrand; and the third
(Sec.~\ref{sec:app-regulator}) is the massless three-vertex chain evaluated in the regulator
mode, which demonstrates
the certified Laurent reconstruction of Sec.~\ref{sec:regulator} on a divergent configuration.
The fourth (Sec.~\ref{sec:app-five-vertex}) is a massless chain with five vertices and 81 masters.
The comparisons underlying the verification tasks of Sec.~\ref{sec:verification} are not repeated
here. They cover the single-vertex families and the massive two-vertex correlator against
\cite{Chen:2024glu}, and the IBP relations and differential equations against the
\texttt{dSIBP} package~\cite{Chen:2026dSIBP}.

\subsection{List of examples}
\label{sec:examples}
Each example is a Wolfram Language script; together they exercise the complete workflow:
\begin{itemize}
  \item \texttt{01\_massless\_full\_edge.wl}: the massless shared mode-function two-state simplification of
        Sec.~\ref{sec:massless}, masters, recurrence, d log-form DE and the automatic
        boundary/numerical entry; it also demonstrates the
        single-stage multipoint workflow, with CSV/JSON export;
  \item \texttt{02\_single\_vertex\_family.wl}: single-vertex integral family input, local tensor
        inverses (Sec.~\ref{sec:global}), and finite-combination reduction;
  \item \texttt{03\_time\_only\_cycle\_chart.wl}: time-order loop input, common-theta contacts
        (Sec.~\ref{sec:common-theta} and App.~\ref{app:theta}), and
        all strict-rank chart certificates; contacts are discussed in Sec.~\ref{sec:dlog};
  \item \texttt{04\_three\_vertex\_tree.wl}: a three-vertex, two-propagator massless tree
        ($+++$ vertex structure), following the same flow as example 01, with a single-stage
        multipoint evaluation and CSV/JSON export;
  \item \texttt{05\_massive\_three\_vertex\_tree.wl}: a three-vertex, two-propagator massive tree
        ($+++$ vertex structure) with non-half-integer \texttt{nu12=3/4}, \texttt{nu23=1/3},
        followed by a single-stage multipoint evaluation and CSV/JSON export;
  \item \texttt{06\_massless\_three\_vertex\_ep\_regularization.wl}: a three-vertex massless tree
        with the analytic regulator $a_1=-1+\varepsilon$, $a_2=a_3=1+\varepsilon$, whose
        $\varepsilon=0$ time integrals genuinely diverge; before the numerical stage it reduces
        the sum of two integrals whose time powers lie one unit below the master powers
        (displayed in Sec.~\ref{sec:app-regulator}); the user requests
        only the highest regulator power (\texttt{MaximumEpPower -> 0}), and the package
        certifies the lowest power $n_{\min}$ from the symbolic boundary and DE before any
        numerical solve (text below); its numerical
        reconstruction is presented in Sec.~\ref{sec:app-regulator};
  \item \texttt{07\_zero\_external\_leg\_energy.wl}: zero external-leg energy, represented through an auxiliary energy with zero target value.
\end{itemize}

In Example~06, the vertex-1 unit $u_1=a_1+1$ equals $\varepsilon$, and every
factorized boundary weight that isolates this unit as a standalone factor carries a
$\Gamma(\varepsilon)$ simple pole, while the nine-dimensional DE has no negative $\varepsilon$
power. The certificate therefore gives $n_{\min}=-1$, decided by the combined
physical boundary and not by any individual resonance.

Each example ends with a failure gate: it prints \texttt{Example PASSED} and exits $0$ only when
every checked result is a non-\texttt{Failure} object, and exits $1$ with the reason otherwise.
Run them in a Mathematica front end section by section, or with
\begin{verbatim}
wolframscript -file .../Examples/01_massless_time_ordered_edge.wl
\end{verbatim}

\subsection{Example: tree with three vertices and two massive scalar exchanges}
\label{sec:app-massive-tree}

\begin{figure}[!htbp]
  \centering
  \includegraphics[width=0.72\textwidth]{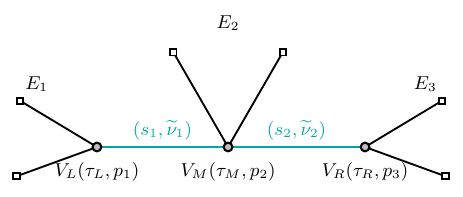}
  \caption{Tree with three vertices and two massive scalar propagators, drawn with the
  in-in Feynman rules of Sec.~\ref{sec:conventions} for the all-plus assignment
  $(+++)$: the time vertices are gray-filled circles on a horizontal axis,
  each carrying its conformal time $\tau_L$, $\tau_M$
  or $\tau_R$ and time power $p_1,p_2,p_3$ in the label $V_v(\tau_v,p_v)$ below it, together
  with a schematic fan of two external legs ending in open squares and labeled by the
  summed external energy $E_v=\sum_j k_{v,j}$. The internal propagators are drawn in teal
  and carry their momentum magnitudes and mass labels
  $(s_1,\widetilde\nu_1)$, $(s_2,\widetilde\nu_2)$ above the propagator.}
  \label{fig:tree-diagram}
\end{figure}

The first example is the chain with three vertices in figure~\ref{fig:tree-diagram}.

The vertices $L,M,R$ have conformal times $\tau_L,\tau_M,\tau_R$.
The two internal propagators connect $L$ to $M$ and $M$ to $R$, respectively.
Each vertex carries a fan of external legs; the momentum magnitudes $k_{v,j}$
of the legs at vertex $v$ sum to its external energy, $E_v=\sum_j k_{v,j}$.
The internal momentum magnitudes are $s_1,s_2$, and the Hankel orders are
$\nu_1,\nu_2$. For the masses in the principal series used here,
$\nu_i=\ii\widetilde\nu_i$, with real $\widetilde\nu_i$.
Each vertex has an independent contour sign $\sigma_v=\pm1$;
figure~\ref{fig:tree-diagram} is drawn for the all-plus assignment
$\sigma_L=\sigma_M=\sigma_R=+1$, and the masters, normalizations and
integrands below are worked out for this assignment as the example.

This application uses \texttt{NuConvention -> "Negative"} for the $z^{-\nu}$ prefactor
of \cite{Chen:2024glu}, while retaining the common Hankel order of the package.
The replacement $\nu\to-\nu$ in the common-order formulas of Sec.~\ref{sec:method} applies
throughout, including the effective vertex time powers below.
From figure~\ref{fig:tree-diagram}, the function family of this example is built from one time
integration per vertex, carrying the time power $(-\tau_v)^{\nu_{0v}}$ and the external exponential
$e^{\ii\sigma_vE_v\tau_v}$, and from one propagator kernel per internal propagator, each kernel
carrying one mode-function state at each of its two vertices. We first make these mode-function
states explicit:
\begin{equation}
 \begin{aligned}
 h^{(r)}(\nu,0;z)&=z^{-\nu}H^{(r)}_\nu(z),\\
 h^{(r)}(\nu,1;z)&=\partial_z h^{(r)}(\nu,0;z)
                  =-z^{-\nu}H^{(r)}_{\nu+1}(z),
 \qquad r=1,2.
 \end{aligned}
 \label{eq:application-endpoint-states}
\end{equation}
Here $r$ selects the Hankel function and the binary index selects the
undifferentiated or differentiated mode-function state. We use the same Hankel order
at both vertices of each propagator; constant factors from the physical mode
normalization are restored when forming the correlator.
To display the time ordering explicitly, define
\begin{equation}
 \begin{aligned}
 W_{>,\nu;ab}(s;\tau,\tau')
 &=h^{(1)}(\nu,a;-s\tau)h^{(2)}(\nu,b;-s\tau'),\\
 W_{<,\nu;ab}(s;\tau,\tau')
 &=h^{(2)}(\nu,a;-s\tau)h^{(1)}(\nu,b;-s\tau'),
 \qquad a,b\in\{0,1\}.
 \end{aligned}
 \label{eq:application-line-products}
\end{equation}
These two products are $G_>$ and $G_<$ of \eqref{eq:bulk-bulk-propagators} with the mode prefactors
removed and the two mode-function states made explicit. The propagator kernel of contour type
$\sigma\sigma'$, written $\mathcal G_{\sigma\sigma',\nu;ab}(s;\tau,\tau')$ below, is then the
contour-dependent combination of the two products exactly as \eqref{eq:sk-propagators} combines $G_>$
and $G_<$: $\mathcal G_{++}$ and $\mathcal G_{--}$ take $\theta(\tau-\tau')W_{>}$ plus
$\theta(\tau'-\tau)W_{<}$ and $\theta(\tau-\tau')W_{<}$ plus $\theta(\tau'-\tau)W_{>}$, while
$\mathcal G_{+-}=W_{<,\nu;ab}$ and $\mathcal G_{-+}=W_{>,\nu;ab}$ carry no theta function. The binary
indices act only on the Hankel factors, never on the step functions, and every kernel has the
arguments $(s;\tau,\tau')$.

For $\boldsymbol\sigma=(\sigma_L,\sigma_M,\sigma_R)$, the Top-sector master
integrals are
\begin{equation}
 \begin{aligned}
 \text{I}^{\boldsymbol\sigma}_{abcd}
 ={}&\int_{-\infty}^{0}d\tau_L\,d\tau_M\,d\tau_R\,
 (-\tau_L)^{\nu_{0L}}(-\tau_M)^{\nu_{0M}}(-\tau_R)^{\nu_{0R}}\\
 &\times e^{\ii(\sigma_LE_1\tau_L+\sigma_ME_2\tau_M+\sigma_RE_3\tau_R)}\\
 &\times\mathcal G_{\sigma_L\sigma_M,\nu_1;ab}(s_1;\tau_L,\tau_M)
 \mathcal G_{\sigma_M\sigma_R,\nu_2;cd}(s_2;\tau_M,\tau_R).
 \end{aligned}
 \label{eq:tree-indexed-masters}
\end{equation}
Here $a,b,c,d\in\{0,1\}$. Each time integral runs from $-\infty$ to $0$ with the Bunch--Davies
prescription. Each massive propagator carries one mode-function state at each of its two vertices.
The four mode-function state indices $a,b,c,d$ are ordered $(L_1,M_1,M_2,R_2)$: $a,b$ are the
states of propagator $s_1$ at $\tau_L,\tau_M$ and $c,d$ those of $s_2$ at $\tau_M,\tau_R$. Thus the
Top sector consists of the sixteen masters
\begin{equation}
 \begin{aligned}
 \bm{\text I}^{\boldsymbol\sigma}_{\mathrm{Top}}
 =\bigl(&\text{I}^{\boldsymbol\sigma}_{0000},
        \text{I}^{\boldsymbol\sigma}_{0001},
        \text{I}^{\boldsymbol\sigma}_{0010},
        \text{I}^{\boldsymbol\sigma}_{0011},
        \text{I}^{\boldsymbol\sigma}_{0100},
        \text{I}^{\boldsymbol\sigma}_{0101},
        \text{I}^{\boldsymbol\sigma}_{0110},
        \text{I}^{\boldsymbol\sigma}_{0111},\\
       &\text{I}^{\boldsymbol\sigma}_{1000},
        \text{I}^{\boldsymbol\sigma}_{1001},
        \text{I}^{\boldsymbol\sigma}_{1010},
        \text{I}^{\boldsymbol\sigma}_{1011},
        \text{I}^{\boldsymbol\sigma}_{1100},
        \text{I}^{\boldsymbol\sigma}_{1101},
        \text{I}^{\boldsymbol\sigma}_{1110},
        \text{I}^{\boldsymbol\sigma}_{1111}\bigr)^{\mathsf T},
 \end{aligned}
 \label{eq:tree-top-vector}
\end{equation}
one for each of the sixteen values of $(a,b,c,d)\in\{0,1\}^4$.
The entry $\text{I}^{\boldsymbol\sigma}_{0000}$ is the one with no
derivative acting on the mode-function states and no additional
integer shifts in the time powers. Equation \eqref{eq:tree-indexed-masters} is the sector
$\mathrm{Top}=(\xi_1\xi_2)=(11)$ of \eqref{eq:sector-key}, with $c_{\mathrm{Top}}=c_V=3$ components
by \eqref{eq:component-count} and $n^{\rm slot}=4$ slots, and
\begin{equation}
 \text{I}^{\boldsymbol\sigma}_{abcd}
 =\text{I}_{\mathrm{Top}}\bigl(\bm 0;(a,b,c,d)\bigr),
 \label{eq:tree-top-identification}
\end{equation}
the bit order of \eqref{eq:binary-label} being the one used by $\bm{\text I}^{\boldsymbol\sigma}_{\mathrm{Top}}$.

The parameters $\nu_{0L},\nu_{0M},\nu_{0R}$ in the integrand are the
effective vertex time powers. If $p_1,p_2,p_3$ are the powers before
absorbing the massive mode prefactors, then
\begin{equation}
 \begin{aligned}
 \nu_{0L}&=p_1+\frac32+\nu_1,\qquad
 \nu_{0R}=p_3+\frac32+\nu_2,\\
 \nu_{0M}&=p_2+3+\nu_1+\nu_2.
 \end{aligned}
 \label{eq:tree-time-power-map}
\end{equation}
The vertex factors $\ii\sigma_v$ and the external and mode normalizations
are not included in \eqref{eq:tree-indexed-masters}; they are restored
when constructing the physical correlator.

\begin{figure}[!htbp]
  \centering
  \includegraphics[width=0.88\textwidth]{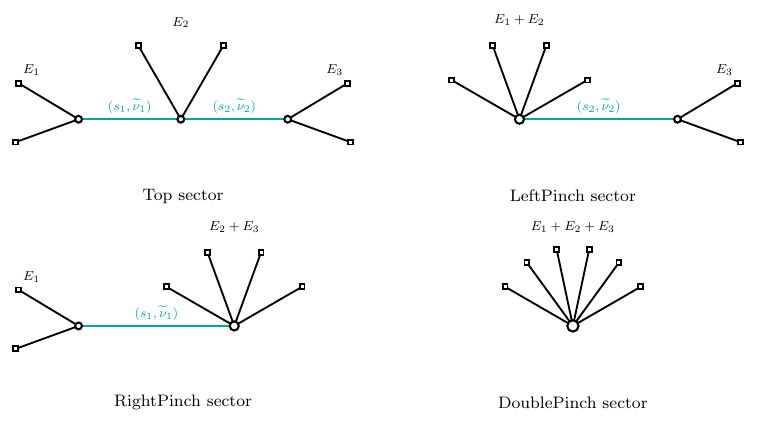}
  \caption{The four sectors of the three-vertex chain of figure~\ref{fig:tree-diagram},
  drawn in the same visual language with white vertices marking
  the equal-sign assignments ($+++$ and $---$), for
  which both internal propagators carry a theta and can be pinched.
  Pinching propagator~1 merges $L$ and $M$ (LeftPinch), pinching propagator~2 merges
  $M$ and $R$ (RightPinch), and both pinches merge all three vertices (DoublePinch).
  A merged component is drawn as one larger circle --- the pinch is visible through
  the absent propagator and the summed external-leg fans --- carrying the fans of
  all vertices it contains;
  the surviving propagator keeps its mode-function two-state (massive chain) or shared
  (massless chain) slot and its teal label. The
  massless chain of the regulator example (figure~\ref{fig:ep-tree}) carries the same
  four sectors.}
  \label{fig:tree-sectors}
\end{figure}

In the IBP reduction, derivatives of the functions that impose time ordering generate
contact terms that merge adjacent vertices. These terms define the
LeftPinch, RightPinch and DoublePinch sectors.
The left pinch merges $L$ and $M$ and leaves the two indices of propagator 2, so its four masters
are the subsector integrals of the contact term \eqref{eq:R-two-vertex}, written out as
\begin{equation}
 \begin{aligned}
 \text{I}^{\boldsymbol\sigma,{\rm L}}_{cd}
 ={}&\int_{-\infty}^{0}\!\!\dd\tau_{LM}\!\int_{-\infty}^{0}\!\!\dd\tau_R\,
 (-\tau_{LM})^{\nu_{0L}+\nu_{0M}-2\nu_1}(-\tau_R)^{\nu_{0R}}\\
 &\times e^{\ii(\sigma_LE_1+\sigma_ME_2)\tau_{LM}+\ii\sigma_RE_3\tau_R}\,
 \mathcal G_{\sigma_M\sigma_R,\nu_2;cd}(s_2;\tau_{LM},\tau_R),
 \qquad c,d\in\{0,1\},
 \end{aligned}
 \label{eq:application-child-master}
\end{equation}
where $\tau_{LM}$ is the single time of the merged component and the merged power is
$\nu_{0,C}$ of \eqref{eq:base-power}: the raw contact power of the pinched propagator is
$\varrho_1=-2\nu_1-1$, and the component count $|C|=2$ adds the remaining $+1$, so
that \eqref{eq:tree-time-power-map} gives $\nu_{0L}+\nu_{0M}-2\nu_1=p_1+p_2+\tfrac92+\nu_2$.
The right pinch merges $M$ and $R$ and keeps the two indices of propagator 1; its four masters
are obtained from \eqref{eq:application-child-master} by the exchange
$(L,\nu_{0L},\sigma_LE_1,s_1,\nu_1)\leftrightarrow(R,\nu_{0R},\sigma_RE_3,s_2,\nu_2)$,
\begin{equation}
 \begin{aligned}
 \text{I}^{\boldsymbol\sigma,{\rm R}}_{ab}
 ={}&\int_{-\infty}^{0}\!\!\dd\tau_L\!\int_{-\infty}^{0}\!\!\dd\tau_{MR}\,
 (-\tau_L)^{\nu_{0L}}(-\tau_{MR})^{\nu_{0M}+\nu_{0R}-2\nu_2}\\
 &\times e^{\ii\sigma_LE_1\tau_L+\ii(\sigma_ME_2+\sigma_RE_3)\tau_{MR}}\,
 \mathcal G_{\sigma_L\sigma_M,\nu_1;ab}(s_1;\tau_L,\tau_{MR}),
 \qquad a,b\in\{0,1\},
 \end{aligned}
 \label{eq:application-right-master}
\end{equation}
where $\tau_{MR}$ is the time of the merged $M,R$ component and
$\nu_{0M}+\nu_{0R}-2\nu_2=p_2+p_3+\tfrac92+\nu_1$.
The double pinch merges all three vertices and leaves no massive propagator, so its single
master is the one-time integral
\begin{equation}
 \text{I}^{\boldsymbol\sigma,{\rm D}}
 =\int_{-\infty}^{0}\dd\tau\,
 (-\tau)^{\nu_{0L}+\nu_{0M}+\nu_{0R}-2\nu_1-2\nu_2}
 e^{\ii(\sigma_LE_1+\sigma_ME_2+\sigma_RE_3)\tau},
 \label{eq:application-double-master}
\end{equation}
whose power is $p_1+p_2+p_3+6$.
In the index convention \eqref{eq:graph-index} the four sectors are
$\mathrm{Top}=(\xi_1\xi_2)=11$ with three components and four slots,
$(\mathrm{LeftPinch},\mathrm{RightPinch})=(01,10)$ with two components and two slots each, and
$\mathrm{DoublePinch}=00$ with one component and no slot, exactly as \eqref{eq:component-count}
requires of a diagram with $c_V=3$ vertices.
When both internal propagators join vertices with equal contour signs, the full
hierarchy is
\begin{equation}
  \mathrm{Top}(16)\oplus\mathrm{LeftPinch}(4)\oplus
  \mathrm{RightPinch}(4)\oplus\mathrm{DoublePinch}(1),
  \qquad \dim=25,
  \label{eq:tree-hierarchy}
\end{equation}
and the four sectors are shown in figure~\ref{fig:tree-sectors}.
For other contour assignments, a propagator whose two
vertices carry opposite signs is theta-free and cannot be pinched, so only the allowed
pinched sectors appear.

We write $\sigma_{LM}$ and $\sigma_{MR}$ for the contour signs of the two internal propagators,
$+1$ for $G_{++}$ and $-1$ for $G_{--}$ as in \eqref{eq:massless-relations}. For the IBP at $M$,
the contact that pinches the $LM$ propagator is \eqref{eq:R-two-vertex} at zero source shift:
\begin{equation}
  \bigl(\cR^{(ML)}_{\rm Top}(\bm0)\bigr)_{(a,b,c,d)}
  =-\sigma_{LM}\,\delta_{a,1-b}\,(-1)^{a+1}\,\frac{4\ii}{\pi}\,s_1^{-2\nu_1-1}\,
   \text{I}_{\rm L}\bigl(-\bm e_{LM};(c,d)\bigr).
  \label{eq:tree-contact-term}
\end{equation}
Here $(a,b,c,d)$ is the full Top state string: $a,b$ belong to the $L,M$ ends of the
pinched propagator, and $c,d$ survive in the LeftPinch sector. The ordered pair $(ML)$
specifies that the derivative acts at $M$; the contact in the IBP at $L$ has the opposite sign.
The subsector shift is $-1$ on the merged component $LM$ and $0$ on $R$, so its merged
time power is one below \eqref{eq:application-child-master}. The total remainder at $M$
also contains the analogous $(MR)$ term.
The four sectors then carry the hidden constants
\begin{equation}
  \begin{aligned}
  \cN_{\rm Top}&=1, & \cN_{\rm L}&=-\sigma_{LM}\,\frac{4\ii}{\pi}\,s_1^{-2\nu_1-1},\\
  \cN_{\rm R}&=-\sigma_{MR}\,\frac{4\ii}{\pi}\,s_2^{-2\nu_2-1}, &
  \cN_{\rm D}&=-\sigma_{LM}\sigma_{MR}\,\frac{16}{\pi^2}\,s_1^{-2\nu_1-1}s_2^{-2\nu_2-1},
  \end{aligned}
  \label{eq:tree-sector-normalizations}
\end{equation}
which is the product rule \eqref{eq:sector-normalization} with $\cN_s^{(0)}=1$. The object the
package propagates in sector $s$ is $J_s(\bm n;\bm a)=\cN_s\,\text{I}_s(\bm n;\bm a)$ ---
concretely $J^{\boldsymbol\sigma}_{abcd}=\text{I}^{\boldsymbol\sigma}_{abcd}$,
$J^{\boldsymbol\sigma,{\rm L}}_{cd}=\cN_{\rm L}\text{I}^{\boldsymbol\sigma,{\rm L}}_{cd}$,
$J^{\boldsymbol\sigma,{\rm R}}_{ab}=\cN_{\rm R}\text{I}^{\boldsymbol\sigma,{\rm R}}_{ab}$ and
$J^{\boldsymbol\sigma,{\rm D}}=\cN_{\rm D}\text{I}^{\boldsymbol\sigma,{\rm D}}$ --- and for the
$(+++)$ assignment \eqref{eq:tree-sector-normalizations} is the value of the
\texttt{"normalization"} field of each master record.

For the timing comparison, we evaluate only the $(+++)$ assignment and its complete
25-dimensional hierarchy in \eqref{eq:tree-hierarchy}. Both methods return all 25 master
integrals at every point. The independent hypergeometric evaluation is constructed directly
from the integral definitions above: complete leaf integrals are expressed through Gamma
and Gauss $\,{}_2F_1$ functions, the finite leaf integrals are expanded in powers, and the
remaining central integration is performed term by term using Gamma integrals.

We set $p_1=p_2=p_3=0$, $(\nu_1,\nu_2)=(\ii,2\ii)$ and $s_1=s_2=1/10$,
and fix the largest external energy at $E_2=1$. The two remaining energies vary independently
over $E_1,E_3\in[1/5,1/4]$. For an $n\times n$ grid with $n>1$, each axis contains
$1/5+j/[20(n-1)]$, $j=0,\ldots,n-1$; the single point is the centre
$(E_1,E_2,E_3)=(9/40,1,9/40)$. We use $n=1,5,10,25,100,1000$,
giving $N=1,25,100,625,10^4,10^6$ points, and traverse each grid along a snake path.
This small rectangle is chosen for a convenient comparison using the same hypergeometric
series expression throughout its convergence domain; it satisfies the sufficient condition
$E_2>E_1+E_3+2s_1+2s_2$. The DE method is not restricted to this region and can
continue the chosen branch beyond it along paths that avoid singularities.

For a direct comparison, both methods run in one process with one CPU thread on the same
personal computer. Both use 80 decimal digits. DE transport uses local Taylor series of
order 80, while the hypergeometric calculation uses total series degree 160.
Separate order-96 transport and degree-208 series calculations check the smaller grids.
The DE primary time includes numerical preparation, the native boundary startup,
path planning, transport and numerical output conversion. Timings exclude symbolic DE
construction, interpreter startup, reference checks and file writing.
The $N\leq625$ results are medians of three sequential runs, with error bars spanning
the minimum and maximum; the two larger grids are each measured once.

The DE times for the six grids are $23.77$, $23.83$, $25.16$, $26.86$, $37.45$ and
$966.94\,\mathrm{s}$. The hypergeometric times for the first five grids are
$0.29$, $9.11$, $44.66$, $138.65$ and $2900.00\,\mathrm{s}$; the million-point
hypergeometric calculation is not performed. The single-point DE call is slower than
the direct series evaluation, while the $10^4$-point grid gives a speedup of $77.43$.
The million-point DE call includes one native boundary startup of $20.59\,\mathrm{s}$.
Its average cost is $0.967\,\mathrm{ms}$ per point for the full vector, or
$38.68\,\mu\mathrm{s}$ per master integral per point.

\begin{figure}[!htbp]
  \centering
  \includegraphics[width=0.99\textwidth]{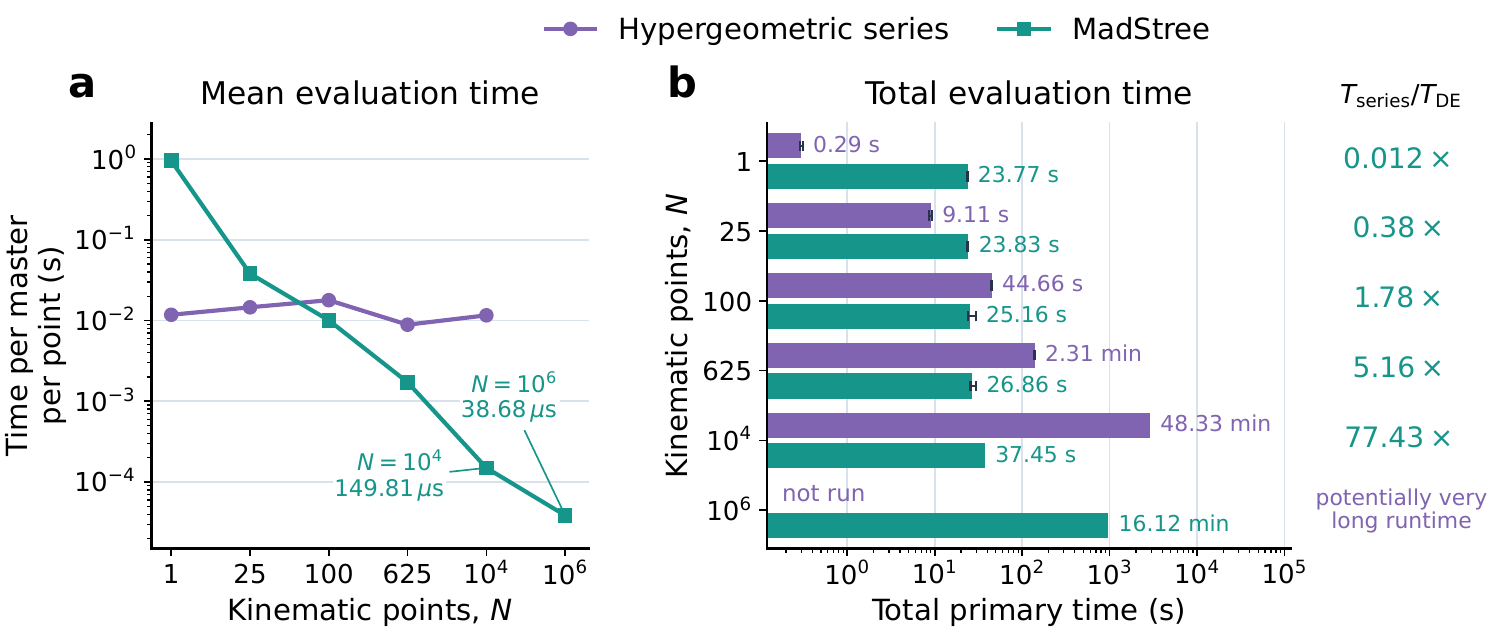}
  \caption{Single-thread comparison for the $(+++)$ branch and all 25 master integrals.
  \textbf{a}, Average primary time per master integral per point, $T/(25N)$.
  \textbf{b}, Total primary times for the six grid sizes.}
  \label{fig:tree-timing}
\end{figure}

For the first five grid sizes in figure~\ref{fig:tree-timing}\textbf{b}, the ratios
$T_{\rm series}/T_{\rm DE}$ are $0.012$, $0.38$, $1.78$, $5.16$ and $77.43$, respectively.
All outputs through $10^4$ points are compared with the independent
series. For the million-point grid, 100 points with existing degree-208 reference values
are checked in all 25 components. The maximum componentwise relative difference over
the compared points is $2.65\times10^{-32}$.

Three properties explain the shape of the efficiency curve in figure~\ref{fig:tree-timing}:
\begin{enumerate}
  \item \emph{Shared transport and local expansions.} Boundary initialization and local
        series coefficients are shared across the scan. The backend plans transport nodes
        from the local convergence geometry, and each local Taylor expansion evaluates
        the target points within its convergence disk. This reuse reduces the average
        cost per output point as the grid becomes denser.
  \item \emph{Fixed boundary cost.} The single-point call is dominated by the automatic
        Frobenius boundary construction of Sec.~\ref{sec:numerics} and transport from
        the boundary anchor to the target point. This fixed cost
        amortizes over larger batches; at one point the direct series evaluation is
        faster. When an independent representation such as the hypergeometric series
        evaluates the masters at a given point faster, those values can serve as the
        boundary vector through the \texttt{MSNumericalSystem} entry of
        Sec.~\ref{sec:numerics}.
  \item \emph{Scaling of the hypergeometric representation.} For more complicated graphs
        the representation involves hypergeometric functions of several variables:
        expanding to total degree $m$ in $n$ variables requires summing on the order of
        $m^n$ terms, and the evaluation becomes laborious near the boundary of the
        convergence domain. DE transport
        constructs successive local power-series solutions around ordinary points along the path.
\end{enumerate}

\begin{figure}[!htbp]
  \centering
  \includegraphics[width=0.99\textwidth]{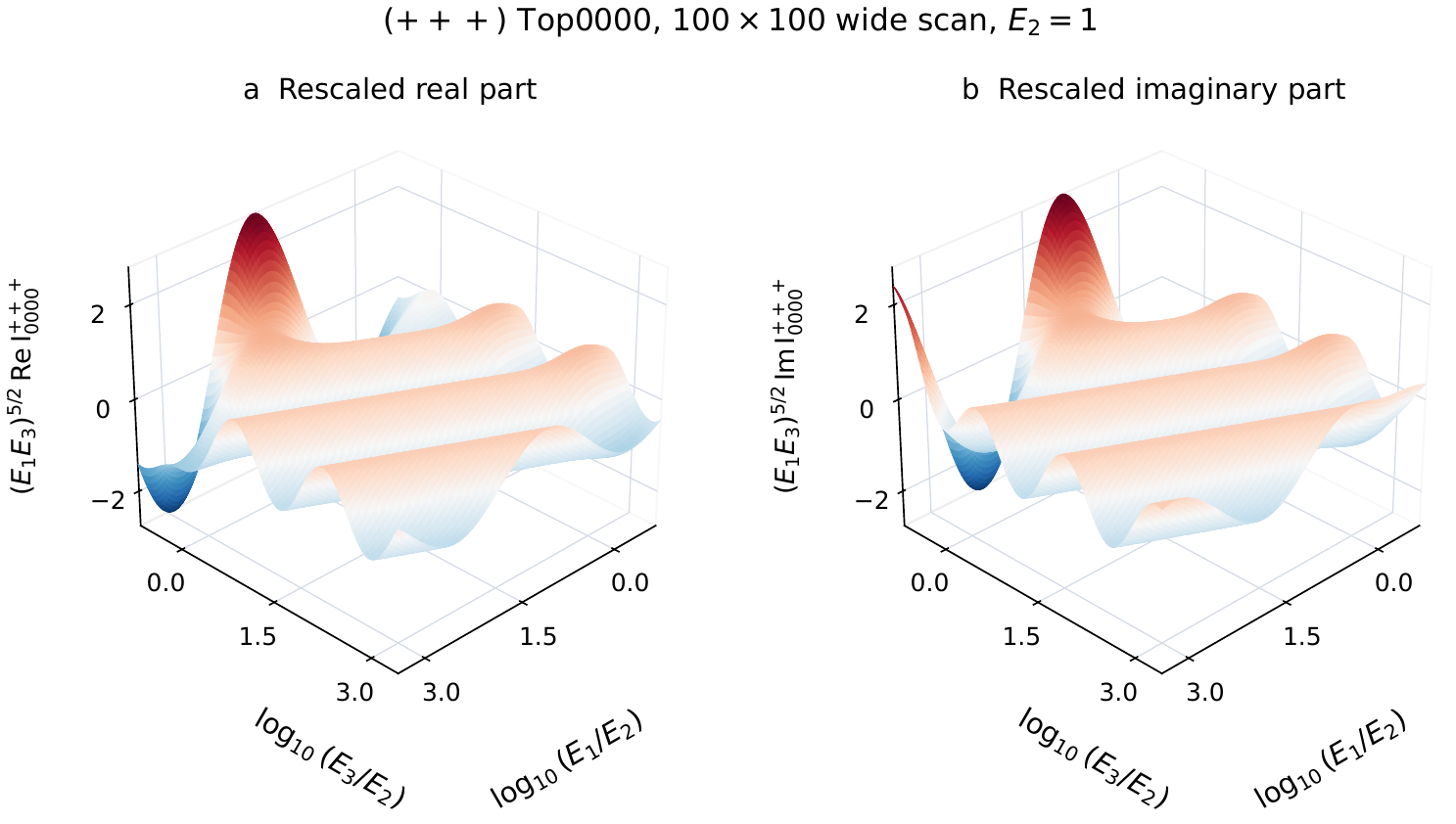}
  \caption{Wide-range scan of the $(+++)$ Top0000 master at $10^4$ points.
  \textbf{a}, Real part. \textbf{b}, Imaginary part.
  Both are multiplied by $(E_1E_3)^{5/2}$, in units of $E_2=1$.
  Height and color represent the indicated quantity, with the same scale in both panels.}
  \label{fig:tree-top0000-magnitude}
\end{figure}

For a separate scan over a wider range, we evaluate the $100\times100$ geometric grid
\[
  (E_1,E_2,E_3)=E_2\left(\frac15\left(\frac{11}{10}\right)^j,\,1,\,
  \frac15\left(\frac{11}{10}\right)^k\right),\qquad j,k=0,\ldots,99.
\]
Both energy ratios range from $0.2$ to about $2505.6$, with the remaining parameters
unchanged and all quantities expressed in units of $E_2=1$.
Figure~\ref{fig:tree-top0000-magnitude} shows the real and imaginary parts of the
rescaled $(+++)$ Top0000 master $(E_1E_3)^{5/2}\text{I}^{+++}_{\mathrm{Top},0000}$.
This factor compensates the leading power decay at large $E_1$ and $E_3$ and makes the oscillations visible.
This scan uses 80 working digits and order 96, with a maximum complex relative difference of
$1.26\times10^{-38}$ for Top0000 against a separate order-112 calculation.
The primary calculation takes $266.84\,\mathrm{s}$, including $42.00\,\mathrm{s}$ for boundary startup.

\subsection{Example: the bubble loop integrand}
\label{sec:app-bubble}

\begin{figure}[!htbp]
  \centering
  \includegraphics[width=0.55\textwidth]{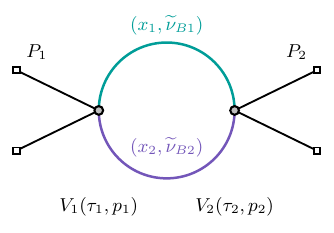}
  \caption{Time integral in the bubble, drawn with the in-in Feynman rules of
  Sec.~\ref{sec:conventions} for the assignment $pp$: the vertices are gray-filled circles
  and carry the times $\tau_1,\tau_2$ and powers $p_1,p_2$ in the labels
  $V_1(\tau_1,p_1)$, $V_2(\tau_2,p_2)$ below the loop; each vertex carries a schematic fan
  of two external legs ending in open squares and labeled by the energy sums
  $P_1,P_2$. The two massive propagators join the same pair of vertices and are drawn as
  the two halves of one circular loop --- propagator~1 with momentum magnitude $x_1$ and
  mass label $\widetilde\nu_{B1}$ on the upper half, propagator~2 with $x_2$ and
  $\widetilde\nu_{B2}$ on the lower half.}
  \label{fig:bubble-diagram}
\end{figure}

The second example is the loop integrand of the bubble: two time vertices joined by two
massive propagators (figure~\ref{fig:bubble-diagram}).

\begin{figure}[!htbp]
  \centering
  \includegraphics[width=0.85\textwidth]{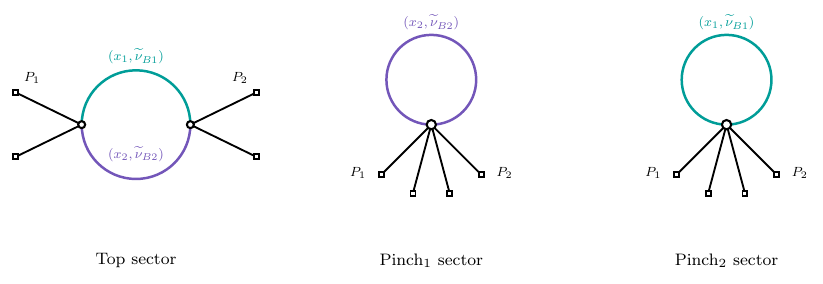}
  \caption{The three sectors of the time integral in the bubble of
  figure~\ref{fig:bubble-diagram}, drawn for the equal-sign assignments ($pp$ and $mm$)
  with the same propagator colors as in figure~\ref{fig:bubble-diagram} and white
  vertices marking the equal-sign assignments, for which both propagators carry a theta
  and can be pinched. The Top sector
  keeps both propagators of the circular loop;
  Pinch$_1$ pinches propagator~1 (upper half, $x_1$) and leaves propagator~2 as a
  circular tadpole on the single merged component, drawn as one larger circle carrying
  both external-leg fans; Pinch$_2$ pinches
  propagator~2 and leaves propagator~1. The opposite-sign assignments $pm$ and $mp$ have
  only the Top sector.}
  \label{fig:bubble-sectors}
\end{figure}

From figure~\ref{fig:bubble-diagram}, the function family of this example has two time
integrals, one per vertex, and two propagator kernels, both joining the same pair of vertices; as
in the first example each kernel carries one mode-function state at each of its two vertices.
Let $\tau_1,\tau_2$ be the vertex times, $P_1,P_2$ the external energy sums,
and $x_1,x_2$ the internal momentum magnitudes.
The Hankel orders of the two internal propagators are $\nu_{B1},\nu_{B2}$.
This example uses \texttt{NuConvention -> "Negative"}, as in the tree example,
with the Hankel orders unchanged:
\begin{equation}
 \begin{aligned}
 h^{(r)}(\nu,0;z)&=z^{-\nu}H^{(r)}_\nu(z),\\
 h^{(r)}(\nu,1;z)&=\partial_z h^{(r)}(\nu,0;z)
                   =-z^{-\nu}H^{(r)}_{\nu+1}(z),\qquad r=1,2.
 \end{aligned}
 \label{eq:bubble-endpoint-states}
\end{equation}
Using these states in the branch products of \eqref{eq:application-line-products} and the
SK combinations of \eqref{eq:sk-propagators} defines the propagator kernels
$\mathcal G_{\sigma\sigma',\nu;ab}$. The Top masters are
\begin{equation}
 \begin{aligned}
 \text{I}^{\sigma_1\sigma_2}_{abcd}
 ={}&\int_{-\infty}^{0}d\tau_1\,d\tau_2\,
 (-\tau_1)^{p_1}(-\tau_2)^{p_2}
 e^{\ii(\sigma_1P_1\tau_1+\sigma_2P_2\tau_2)}\\
 &\times\mathcal G_{\sigma_1\sigma_2,\nu_{B1};ab}(x_1;\tau_1,\tau_2)
 \mathcal G_{\sigma_1\sigma_2,\nu_{B2};cd}(x_2;\tau_1,\tau_2).
 \end{aligned}
 \label{eq:bubble-indexed-masters}
\end{equation}
Here $a,b,c,d\in\{0,1\}$. As in the first example, $a,b$ label the mode-function states of
propagator~1 at vertices~1 and~2, and $c,d$ those of propagator~2 in the same vertex order.
Here $p_1,p_2$ are the effective time powers after absorbing the mode
prefactors, unlike the bare powers used in \eqref{eq:tree-time-power-map}.
The contour signs are $\sigma_1,\sigma_2=\pm1$.
We denote positive and negative signs by $p$ and $m$, respectively;
these labels are unrelated to the time powers $p_1,p_2$.
The four contour assignments are $pp$, $pm$, $mp$ and $mm$.
Figure~\ref{fig:bubble-diagram} is drawn for the assignment $pp$ as the example;
the masters, integrands and hidden constants below carry the signs $\sigma_1,\sigma_2$
and apply to every assignment.
An assignment with equal contour signs has 22 masters after the
exact mode-function state projection; an assignment with opposite contour signs has only the
Top sector, with 16 masters.
For each assignment, the projection onto the Top sector is
\begin{equation}
 \begin{aligned}
 \bm{\text I}^{\sigma_1\sigma_2}_{\rm Top}
 =\bigl(&\text{I}^{\sigma_1\sigma_2}_{0000},
        \text{I}^{\sigma_1\sigma_2}_{0001},
        \text{I}^{\sigma_1\sigma_2}_{0010},
        \text{I}^{\sigma_1\sigma_2}_{0011},
        \text{I}^{\sigma_1\sigma_2}_{0100},
        \text{I}^{\sigma_1\sigma_2}_{0101},
        \text{I}^{\sigma_1\sigma_2}_{0110},
        \text{I}^{\sigma_1\sigma_2}_{0111},\\
       &\text{I}^{\sigma_1\sigma_2}_{1000},
        \text{I}^{\sigma_1\sigma_2}_{1001},
        \text{I}^{\sigma_1\sigma_2}_{1010},
        \text{I}^{\sigma_1\sigma_2}_{1011},
        \text{I}^{\sigma_1\sigma_2}_{1100},
        \text{I}^{\sigma_1\sigma_2}_{1101},
        \text{I}^{\sigma_1\sigma_2}_{1110},
        \text{I}^{\sigma_1\sigma_2}_{1111}\bigr)^{\mathsf T},
 \end{aligned}
 \label{eq:bubble-top-vector}
\end{equation}
one for each of the sixteen values of $(a,b,c,d)\in\{0,1\}^4$.
The two pinched sectors of an equal-sign assignment arise from pinching one of the two
propagators. Both vertices then merge into a single component carrying one time $\tau$, and the
surviving propagator joins this component to itself. When its two vertices coincide, the theta pair of
the (anti-)time-ordered propagator collapses and its factor becomes the mean of the two chamber products, the
spectator factor $\mathsf m_e$ of \eqref{eq:bundle-odd-subsets},
\begin{equation}
 T_{cd}(x;\tau)
 =\tfrac12\Big[h^{(1)}(\nu,c;-x\tau)\,h^{(2)}(\nu,d;-x\tau)
 +h^{(2)}(\nu,c;-x\tau)\,h^{(1)}(\nu,d;-x\tau)\Big],
 \label{eq:bubble-tadpole-kernel}
\end{equation}
which is symmetric under $(c,d)\leftrightarrow(d,c)$, so the coincidence quotient identifies
$(1,0)$ with $(0,1)$ and keeps the three combinations $(0,0),(0,1),(1,1)$ of the two mode-function
states --- this is the exact mode-function state
projection quoted above. The pinched masters are therefore, with propagator~1 pinched,
\begin{equation}
 \text{I}^{\sigma_1\sigma_2,{\rm P1}}_{cd}
 =\int_{-\infty}^{0}\dd\tau\,
 (-\tau)^{p_1+p_2-2\nu_{B1}}
 e^{\ii(\sigma_1P_1+\sigma_2P_2)\tau}\,
 T_{cd}(x_2;\tau),
 \qquad (c,d)\in\{(0,0),(0,1),(1,1)\},
 \label{eq:bubble-pinch1-masters}
\end{equation}
where the merged power follows from \eqref{eq:base-power} with the raw contact power
$\varrho_1=-2\nu_{B1}-1$ and one vertex merge, and, with propagator~2 pinched,
\begin{equation}
 \text{I}^{\sigma_1\sigma_2,{\rm P2}}_{ab}
 =\int_{-\infty}^{0}\dd\tau\,
 (-\tau)^{p_1+p_2-2\nu_{B2}}
 e^{\ii(\sigma_1P_1+\sigma_2P_2)\tau}\,
 T_{ab}(x_1;\tau),
 \qquad (a,b)\in\{(0,0),(0,1),(1,1)\}.
 \label{eq:bubble-pinch2-masters}
\end{equation}
The equal-sign hierarchy is
$\mathrm{Top}(16)\oplus\mathrm{Pinch}_1(3)\oplus\mathrm{Pinch}_2(3)$ with $\dim=22$
(figure~\ref{fig:bubble-sectors}).
The vertex factors $(\ii\sigma_1)(\ii\sigma_2)$ give
\begin{equation}
  \bm K_{\mathrm{SK}}=-\bm{\text I}^{pp}_{\rm Top}+\bm{\text I}^{pm}_{\rm Top}
  +\bm{\text I}^{mp}_{\rm Top}-\bm{\text I}^{mm}_{\rm Top}.
  \label{eq:bubble-sk}
\end{equation}
Only the equal-sign assignments reach a pinched sector, and there both propagators are of one
branch, so they share the contour sign $\sigma_e=\sigma_1=\sigma_2$ of
\eqref{eq:sector-normalization}. The hidden constants of the three sectors are
\begin{equation}
  \cN_{\rm Top}=1,\qquad
  \cN_{\rm P1}=-\sigma_e\,\frac{4\ii}{\pi}\,x_1^{-2\nu_{B1}-1},\qquad
  \cN_{\rm P2}=-\sigma_e\,\frac{4\ii}{\pi}\,x_2^{-2\nu_{B2}-1},
  \label{eq:bubble-sector-normalizations}
\end{equation}
with $\cN_s^{(0)}=1$; the propagated objects are
$J^{\sigma_1\sigma_2}_{abcd}=\text{I}^{\sigma_1\sigma_2}_{abcd}$,
$J^{\sigma_1\sigma_2,{\rm P1}}_{cd}=\cN_{\rm P1}\text{I}^{\sigma_1\sigma_2,{\rm P1}}_{cd}$ and
$J^{\sigma_1\sigma_2,{\rm P2}}_{ab}=\cN_{\rm P2}\text{I}^{\sigma_1\sigma_2,{\rm P2}}_{ab}$.
As in the tree example, \eqref{eq:sector-normalization} is used with the prefactor
exponent $-\nu_e$, so the momentum power is $-2\nu_e-1$.
The code passes $-P_i$ as \texttt{externalLegEnergy} to reproduce the vertex phases
in \eqref{eq:bubble-indexed-masters}.
The computation covers the full family of contour signs: the scan
below evaluates all four assignments $pp$, $pm$, $mp$ and $mm$, which enter the combination
\eqref{eq:bubble-sk}, the equal-sign ones with their pinched sectors included in the reduced
system.
We use $k_s=x_1$ as the momentum unit and vary the dimensionless energy
$\rho=P_1/k_s=P_2/k_s$. The parameter values in
\eqref{eq:bubble-indexed-masters} are
\begin{equation}
 \begin{aligned}
 (k_s,x_1,x_2)&=(1,1,\sqrt2),\qquad P_1=P_2=\rho,\\
 (\nu_{B1},\nu_{B2})&=(\ii/4,\ii/4),\qquad
 p_1=p_2=1+\ii/2.
 \end{aligned}
 \label{eq:bubble-parameters}
\end{equation}
The normalized $0000$ entry of $\bm K_{\mathrm{SK}}$, written $\widehat K_{\mathrm{SK}}(\rho)$, is
\begin{equation}
  \widehat K_{\mathrm{SK}}(\rho)
  =\frac{[\bm K_{\mathrm{SK}}(\rho)]_{0000}}
  {[\bm K_{\mathrm{SK}}(10)]_{0000}}.
  \label{eq:bubble-normalized}
\end{equation}
The four zeros select the four undifferentiated mode-function states of the two massive propagators.
Panel \textbf{b} instead uses $[\bm K_{\mathrm{SK}}(\rho)]_{0000}/
\operatorname{Re}[\bm K_{\mathrm{SK}}(10)]_{0000}$; division by this real scale
preserves the complex phase.

The scan contains 201 uniformly spaced points,
$\rho_j=10-9j/250$, $j=0,\ldots,200$.
The calculation uses 80 decimal working digits, with local series order 96
for the results and order 112 for a separate reference check.
The numerical time for all four contour assignments is $83.01$~s, including
numerical initialization, boundary startup and path transport, but excluding
symbolic construction and the reference calculation.
The batch average is $0.413$~s per output point.
The maximum reported relative difference between the branch vectors at orders
96 and 112 is $9.97\times10^{-34}$, using the backend's relative infinity-norm check.

\begin{figure}[!htbp]
  \centering
  \includegraphics[width=0.99\textwidth]{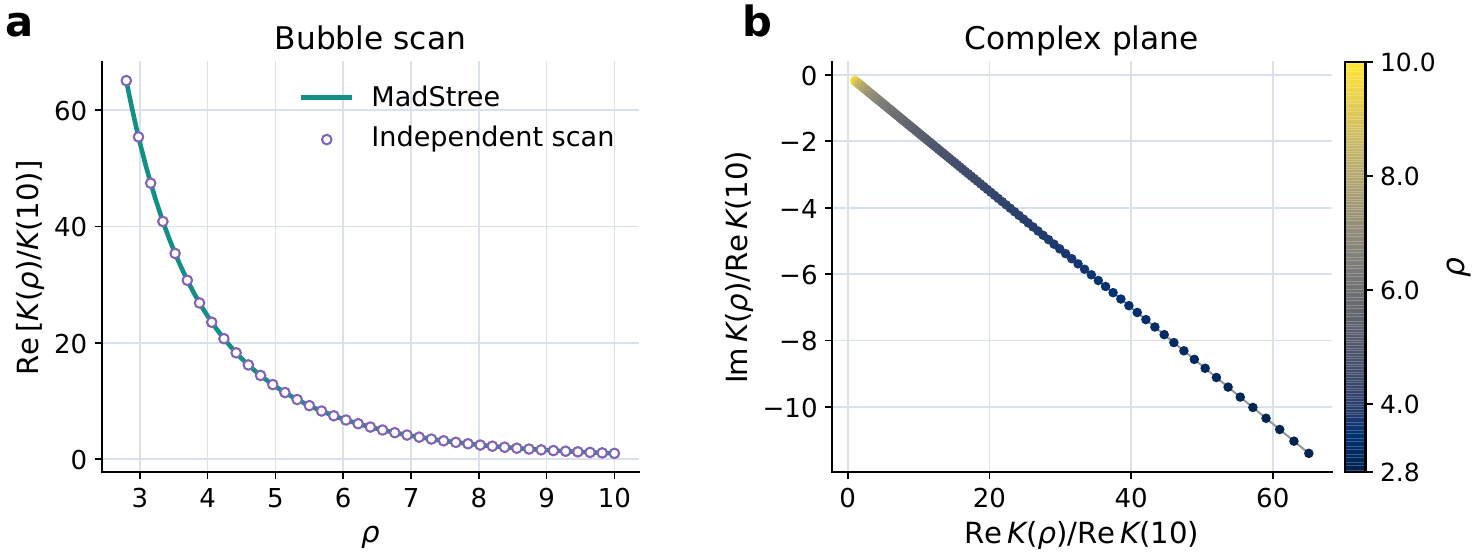}
  \caption{Time integral in the bubble evaluated with \texttt{MadStree}.
  \textbf{a}, The real part of the normalized entry \eqref{eq:bubble-normalized} and the
  independent reference.
  \textbf{b}, The normalized entry in the complex plane, with color indicating $\rho$.}
  \label{fig:bubble-scan}
\end{figure}

Both panels of figure~\ref{fig:bubble-scan} use the same 201 points. Panel \textbf{a} compares the
real part of the normalized entry \eqref{eq:bubble-normalized} with the same quantity computed
independently by a program for this bubble diagram; the maximum relative
difference between the normalized complex entries over the scan is $3.71\times10^{-14}$. The nearly radial
distribution in panel \textbf{b} indicates an approximately constant phase.

\subsection{Example: massless three-vertex tree in the regulator mode}
\label{sec:app-regulator}

\begin{figure}[!htbp]
  \centering
  \includegraphics[width=0.72\textwidth]{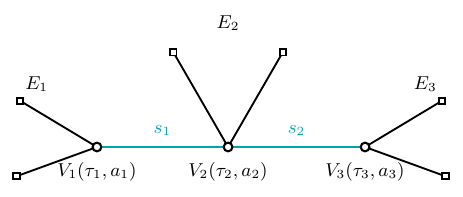}
  \caption{Massless three-vertex chain in the regulator mode, drawn as in
  figure~\ref{fig:tree-diagram} with the massless counterparts of its parameters: the same
  external-leg fans with summed energies $E_v=\sum_j k_{v,j}$ and propagator momentum
  magnitudes $s_1,s_2$, with no mass labels since the massless propagators carry the
  fixed order $\nu=1/2$. The vertex labels $V_v(\tau_v,a_v)$ below the vertices carry the
  regularized time powers $a_1=-1+\varepsilon$ and $a_2=a_3=1+\varepsilon$, i.e. the time
  factors $(-\tau_1)^{-1+\varepsilon}$, $(-\tau_2)^{1+\varepsilon}$ and
  $(-\tau_3)^{1+\varepsilon}$. The fans are schematic: in the input of Example~06 each
  vertex carries a single external leg, $E_v=k_v$, $s_1=q_{12}$ and $s_2=q_{23}$.
  The white vertices indicate the fixed all-plus assignment $(+++)$.}
  \label{fig:ep-tree}
\end{figure}

The third example returns to the chain topology of the first example with both exchanges
massless, and lowers the time power of vertex~1 by two units before shifting all vertex time
powers by the common analytic regulator (figure~\ref{fig:ep-tree}): $a_1=-1+\varepsilon$,
$a_2=a_3=1+\varepsilon$. At $\varepsilon=0$ the time integrals diverge.
The regulator mode of Sec.~\ref{sec:regulator} evaluates them at the regularized powers;
requesting the return level
$\varepsilon^0$ delivers both the $\varepsilon^{-1}$ pole coefficient and the finite part. The
momenta are fixed by
\texttt{ParameterRules} and the target point is $\{k_1,k_2,k_3\}=\{9\ii,3\ii,5\ii\}$.
\Needspace{8\baselineskip}
The input is
\begin{verbatim}
treeSpec = <|
  "vertices" -> {
    <|"id" -> 1, "externalLegEnergy" -> k1,
      "timePower" -> a1, "vertexType" -> "+"|>,
    <|"id" -> 2, "externalLegEnergy" -> k2,
      "timePower" -> a2, "vertexType" -> "+"|>,
    <|"id" -> 3, "externalLegEnergy" -> k3,
      "timePower" -> a3, "vertexType" -> "+"|>},
  "lines" -> {
    <|"type" -> "massless", "endpoints" -> {1, 2},
      "momentum" -> q12, "nu" -> 1/2|>,
    <|"type" -> "massless", "endpoints" -> {2, 3},
      "momentum" -> q23, "nu" -> 1/2|>}|>;
context = MSInitTree[treeSpec];
epSeries = MSReconstructEpSeries[context, ep,
  {{k1, k2, k3}, {9 I, 3 I, 5 I}},
  ParameterRules -> {q12 -> 1, q23 -> 2,
    a1 -> -1 + ep, a2 -> 1 + ep, a3 -> 1 + ep},
  MaximumEpPower -> 0, EpGoalDigits -> 12,
  EpSampleAngleRange -> {-Pi/3, Pi/3},
  RankOrder -> {1, 2, 3}]
\end{verbatim}
The example additionally sets \texttt{BoundaryScale -> 4}, \texttt{BoundarySeriesOrder -> 24}
and the default pool of \texttt{ParallelTaskCount -> 12} workers.

From figure~\ref{fig:ep-tree}, the family has three time integrals and two propagator kernels;
because both propagators are massless time-ordered and anti-time-ordered propagators of the fixed order $\nu=1/2$, each
contributes one shared mode-function two-state slot (Sec.~\ref{sec:massless}) rather than two
independent mode-function state slots.
We write $E_v=k_v$ for the input external energy and $s_1=q_{12}$, $s_2=q_{23}$ for
the internal momentum magnitudes. The all-plus input has the vertex phase
$e^{-\ii E_v\tau_v}$, which damps at the positive-imaginary target energies.
At $\nu=1/2$ the positive-prefactor branches are pure exponentials, so the input powers
$a_v$ are the effective integrand powers. We define the integrals in this example directly
in the unit-amplitude package basis of \eqref{eq:package-massless-basis}. For an (anti-)time-ordered
propagator with contour sign $\sigma=\pm1$ and shared bit $b\in\{0,1\}$, its kernel is
\begin{equation}
 \mathcal G^{\rm m}_{\sigma,b}(s;\tau_u,\tau_v)
 =(-1)^b\theta_{uv}e^{-\ii\sigma s(\tau_u-\tau_v)}
  +\theta_{vu}e^{+\ii\sigma s(\tau_u-\tau_v)}.
 \label{eq:ep-massless-kernel}
\end{equation}
The two entries are the chamber sum and the second chamber minus the first; the factors
$\pi/2$ and $\ii\sigma$ relating them to the raw Hankel states are already included.
Here $\boldsymbol\sigma=(+1,+1,+1)$ and both propagators have $\sigma=+1$.
With one shared slot per propagator, the Top sector has the four masters
\begin{equation}
 \begin{aligned}
 \text{I}^{\boldsymbol\sigma}_{b_1b_2}
 ={}&\int_{-\infty}^{0}\dd\tau_1\,\dd\tau_2\,\dd\tau_3\,
 (-\tau_1)^{a_1}(-\tau_2)^{a_2}(-\tau_3)^{a_3}
 e^{-\ii(E_1\tau_1+E_2\tau_2+E_3\tau_3)}\\
 &\times\mathcal G^{\rm m}_{+1,b_1}(s_1;\tau_1,\tau_2)\,
 \mathcal G^{\rm m}_{+1,b_2}(s_2;\tau_2,\tau_3),
 \qquad b_1,b_2\in\{0,1\},
 \end{aligned}
 \label{eq:ep-top-masters}
\end{equation}
with the bits taken in propagator order, and the Top vector is
$(\text{I}^{\boldsymbol\sigma}_{00},\text{I}^{\boldsymbol\sigma}_{01},
\text{I}^{\boldsymbol\sigma}_{10},\text{I}^{\boldsymbol\sigma}_{11})^{\mathsf T}$.
A pinched massless propagator has the raw contact power $\varrho=0$ of \eqref{eq:base-power},
and its pure exponentials multiply into the merged exponential. The four sectors, shown in
figure~\ref{fig:tree-sectors}, are
\begin{equation}
 \begin{aligned}
 \text{I}^{\boldsymbol\sigma,{\rm L}}_{b}
 ={}&\int_{-\infty}^{0}\!\!\dd\tau_{12}\!\int_{-\infty}^{0}\!\!\dd\tau_3\,
 (-\tau_{12})^{a_1+a_2+1}(-\tau_3)^{a_3}
 e^{-\ii(E_1+E_2)\tau_{12}-\ii E_3\tau_3}\\
 &\times\mathcal G^{\rm m}_{+1,b}(s_2;\tau_{12},\tau_3),
 \qquad b\in\{0,1\},
 \end{aligned}
 \label{eq:ep-left-masters}
\end{equation}
where propagator~1 is pinched and $\tau_{12}$ is the time of the merged component,
\begin{equation}
 \begin{aligned}
 \text{I}^{\boldsymbol\sigma,{\rm R}}_{b}
 ={}&\int_{-\infty}^{0}\!\!\dd\tau_1\!\int_{-\infty}^{0}\!\!\dd\tau_{23}\,
 (-\tau_1)^{a_1}(-\tau_{23})^{a_2+a_3+1}
 e^{-\ii E_1\tau_1-\ii(E_2+E_3)\tau_{23}}\\
 &\times\mathcal G^{\rm m}_{+1,b}(s_1;\tau_1,\tau_{23}),
 \qquad b\in\{0,1\},
 \end{aligned}
 \label{eq:ep-right-masters}
\end{equation}
with propagator~2 pinched, and the single one-time master of the DoublePinch sector,
\begin{equation}
 \text{I}^{\boldsymbol\sigma,{\rm D}}
 =\int_{-\infty}^{0}\dd\tau\,
 (-\tau)^{a_1+a_2+a_3+2}
 e^{-\ii(E_1+E_2+E_3)\tau}.
 \label{eq:ep-double-master}
\end{equation}
The hierarchy is therefore
$\mathrm{Top}(4)\oplus\mathrm{LeftPinch}(2)\oplus\mathrm{RightPinch}(2)\oplus
\mathrm{DoublePinch}(1)$ with $\dim=9$, the nine-dimensional DE recorded in
Sec.~\ref{sec:examples}; the massive chain of \eqref{eq:tree-hierarchy} carries four mode-function state slots
and $25$ masters. No pinched propagator contributes a pinch factor in
\eqref{eq:sector-normalization}, so every sector normalization is $1$. Since the integrals
in this example are already defined in the package basis, $J_s=\text{I}_s$ here, and the
propagated vector is $\bm{\text I}$ itself, without vertex contour factors.

Before the regulator enters, the initialized context is already a closed reduction
system, and \texttt{MSReduce} accepts shifted combinations of the integrals just defined.
Taking the two integrals whose time powers lie one unit below the master powers, at
vertex~1 and at vertex~2, and reducing their sum in one call,
\begin{verbatim}
combination = MSIntegral["11", {-1, 0, 0}, {0, 0}] +
               MSIntegral["11", {0, -1, 0}, {1, 0}];
reduction = MSReduce[combination, context];
\end{verbatim}
returns \texttt{"status" -> "reduced"} with a zero \texttt{"nonMasterResidual"} and an
empty \texttt{"remainingShiftedIntegrals"}. Writing
$\text{I}^{\boldsymbol\sigma}_{b_1b_2}\big|_{a_v\to a_v-1}$ for the integral with the
time power $a_v$ of \eqref{eq:ep-top-masters} lowered by one unit,
\begin{equation}
 \begin{aligned}
 \text{I}^{\boldsymbol\sigma}_{00}\big|_{a_1\to a_1-1}
 +\text{I}^{\boldsymbol\sigma}_{10}\big|_{a_2\to a_2-1}
 ={}&-\ii\,\frac{a_2E_1+a_1s_1}{a_1a_2}\,\text{I}^{\boldsymbol\sigma}_{00}
 +\ii\!\left(\frac{s_1}{a_1}-\frac{E_2}{a_2}\right)\text{I}^{\boldsymbol\sigma}_{10}
 +\ii\,\frac{s_2}{a_2}\,\text{I}^{\boldsymbol\sigma}_{11}\\
 &-\frac{2\ii\,(E_1+E_2)}{a_2(1+a_1+a_2)}\,\text{I}^{\boldsymbol\sigma,\rm L}_{0}
 +\frac{2\ii\,s_2}{a_2(1+a_1+a_2)}\,\text{I}^{\boldsymbol\sigma,\rm L}_{1}.
 \end{aligned}
 \label{eq:ep-reduction-demo}
\end{equation}
The lowering directions invert $\text{M}_1$ eigenvalues, the contact of the first lowering
injects into the LeftPinch sector, whose merged component is lowered once more, and the
returned \texttt{"singularLayers"} names exactly the inverted denominators $\{a_1\}$,
$\{a_2\}$ and $\{1+a_1+a_2\}$. The reduction uses the time powers fixed at
initialization, so substituted values are entered by initializing with them: with the
regularized powers $a_1=-1+\varepsilon$, $a_2=a_3=1+\varepsilon$ in the
\texttt{timePower} fields of the same topology, the same call returns the LeftPinch
coefficient $-2\ii(E_1+E_2)/[(1+\varepsilon)(1+2\varepsilon)]$ and the singular surfaces
$\{-1+\varepsilon\}$, $\{1+\varepsilon\}$, $\{1+2\varepsilon\}$.
These denominators are nonzero at $\varepsilon=0$; the pole reconstructed below
comes from the boundary factor $\Gamma(\varepsilon)$.

The certificate of Sec.~\ref{sec:regulator} runs before any numerical solve and returns
$n_{\min}=-1$: the d log-form DE is regular at $\varepsilon=0$, but the vertex-1 unit
$u_1=a_1+1=\varepsilon$ is isolated as a standalone factor in part of the factorized boundary
weights, and every such branch carries a single $\Gamma(\varepsilon)$ simple pole, so the
combined physical boundary starts at $\varepsilon^{-1}$ with component valuations $-1$ or
$0$ only. Three Frobenius recurrence operators do drop rank at $\varepsilon=0$, with rank
changes $9\to5$, $9\to8$ and $9\to8$ at the exponents $6(1+\varepsilon)$ (twice) and
$8+7\varepsilon$; per
Sec.~\ref{sec:regulator} these are retained as resonance diagnostics and produce no additional
pole, since the combined physical boundary decides the support. With $m=0$, $n_{\min}=-1$ and
$g=12$, \eqref{eq:regulator-fit-count} and \eqref{eq:regulator-scale-precision} give
$P=1$, $\Delta=1$, $N_{\rm fit}=4$, $\alpha_\varepsilon=25/4$, $\varepsilon_0=10^{-25/4}$,
$p_0=32$, a working
precision of $200$ digits and a primary transport order of $128$ with $50$ reference orders.
The four production points are exact Gaussian rational approximations to points with moduli
$\varepsilon_0(1+j/100)$, $j=1,\ldots,4$, cycling through the rays $-\pi/6$, $0$ and $+\pi/6$ of the
open window $(-\pi/3,\pi/3)$; the two validation points similarly approximate points at half the first two moduli, on the
rays $-\pi/6$ and $0$, and are held out from the fit.
All six solves are dispatched to one bounded pool
($12$ requested, $6$ effective); the whole reconstruction, including the certificate, both
validation solves and the fit, takes $170$~s of wall time.

The returned Laurent data are the $\varepsilon^{-1}$ pole coefficient and the
$\varepsilon^0$ coefficient, both $9$-entry master vectors. Entries below $10^{-20}$ in
absolute value are numerical noise at the achieved accuracy and are reported as zero. For scale,
the relative validation residual is of order $10^{-25}$, while the finite DoublePinch entry
listed below is of order $10^{-5}$. With this convention three entries of the pole coefficient vanish,
among them the DoublePinch entry, because the fully merged boundary branch carries no
isolated $\Gamma(\varepsilon)$ factor. Two representative entries are
\begin{equation}
  \begin{aligned}
  [c_{-1}]_{\mathrm{Top},00}&=\bigl(9.49200055589-25.86650516658\,\ii\bigr)\times10^{-4},\\
  [c_0]_{\mathrm{D}}&=\frac{\Gamma(4)}{17^4}=7.183822032782174\times10^{-5}.
  \end{aligned}
  \label{eq:ep-finite-entries}
\end{equation}
The first is the Top-sector entry with both shared slots in their symmetric state, $b_1=b_2=0$,
and the second is the DoublePinch master, a single one-time integral. At the two independent
validation points the fitted series is summed over all four internal powers
$\varepsilon^{-1},\ldots,\varepsilon^{2}$, which include the two buffer powers above the requested
maximum $m=0$. The maximum relative residual against direct transport is
$7.43\times10^{-25}$ for the requested tolerance $10^{-12}$.
The coefficients are computed at a working precision of $200$ digits and returned with about $199$ significant digits.

The tree example illustrates efficient multipoint evaluation by DE transport.
The bubble example applies the same framework to a loop integrand.
The regulator example certifies the Laurent support symbolically on a divergent
massless chain and delivers the $\varepsilon\to0$ pole coefficient and finite part from six
numerical solves.

\subsection{Example: massless tree with five vertices}
\label{sec:app-five-vertex}

\begin{figure}[!htbp]
  \centering
  \includegraphics[width=0.99\textwidth]{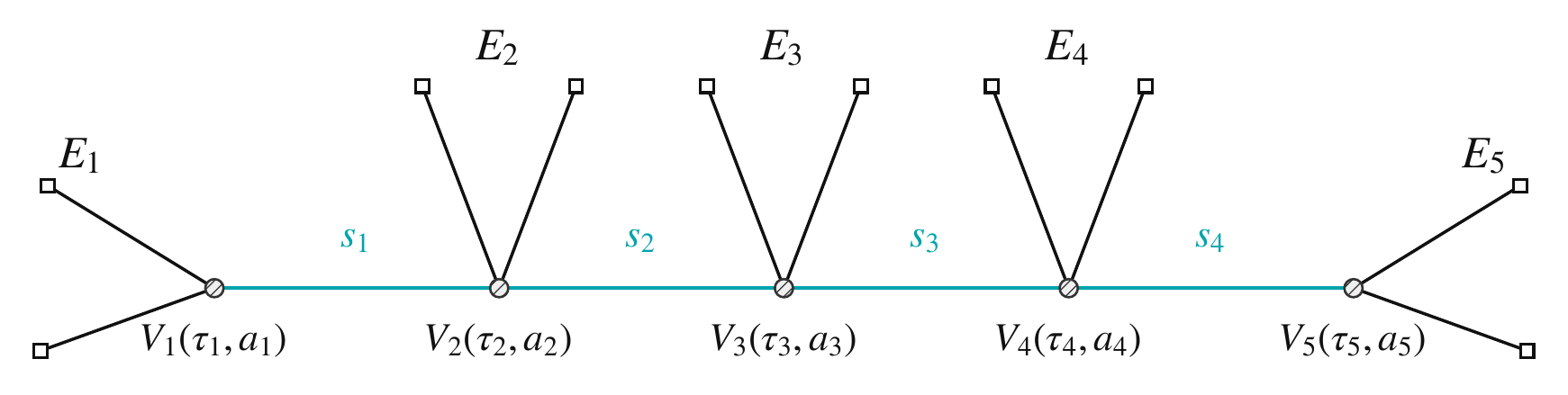}
  \caption{Chain with five vertices and four massless internal propagators.}
  \label{fig:five-vertex-tree}
\end{figure}

Figure~\ref{fig:five-vertex-tree} shows a chain with five vertices.
All four internal propagators have $\nu=1/2$, and the vertex time powers
are $a_v=2$, corresponding to the weights $(-\tau_v)^2$.
For the $(+++++)$ branch and its pinched sectors, the package
automatically constructs 16 sectors and 81 master integrals.
We fix the external energies $E_1,\ldots,E_4$ and internal momenta
$s_1,\ldots,s_4$, and sample 100 values of $E_5$:
\begin{equation}
 \begin{gathered}
 (E_1,E_2,E_3,E_4)=(10,9,8,7),\qquad
 (s_1,s_2,s_3,s_4)=\left(1,\frac65,\frac75,\frac85\right),\\
 E_5^{(j)}=4+\frac{2j}{99},\qquad j=0,\ldots,99.
 \end{gathered}
 \label{eq:five-vertex-kinematics}
\end{equation}

The calculation uses 80 decimal working digits and local series order 80,
with a single thread on a personal computer.
The numerical stage takes $493.43$~s for 100 points, or $4.93$~s per point.
Including symbolic preparation and process overhead, the total is $555.93$~s,
or $5.56$~s per point.
The points share the boundary initialization and local expansions along the path.
A separate order-96 check gives a maximum componentwise relative difference
of $6.81\times10^{-32}$ over all values.

For example, at $E_5=6$, the Top master with zero time shifts and shared
state indices $0000$ is
\begin{equation}
 \left.\text{I}^{+++++}_{\mathrm{Top},0000}\right|_{E_5=6}
 \simeq -5.69426134208370276618914087738\times10^{-12}\,\ii .
 \label{eq:five-vertex-value}
\end{equation}

\section{Verifications}
\label{sec:verification}

The package was checked against independent references in the following verification tasks.

\begin{itemize}
  \item \textbf{Single-vertex families and the massive two-vertex correlator.} The exact
        differential equations, the boundary coefficients and the numerically transported results
        are compared with the known results of \cite{Chen:2024glu}, which are computed by an
        independent multivariate hypergeometric construction.
  \item \textbf{The massive three-vertex tree diagram.} The package results are compared with the
        published eight-branch result of \cite{Xianyu:2023ytd} and an independent three-vertex
        implementation generated by an agent from the hand-written two-vertex code. This reference
        implementation was first checked against the published result, then used to validate
        \texttt{MadStree}.
  \item \textbf{IBP relations and differential equations.} These are compared with the output of the
        \texttt{dSIBP} package \cite{Chen:2026dSIBP}, which is released in the same repository
        as this one, \url{https://github.com/J7C/dSIBP-Packages}. \texttt{dSIBP} samples naive IBP
        seeds and solves the resulting linear reduction system, whereas \texttt{MadStree} evaluates
        the closed formulas derived in Sec.~\ref{sec:method}. Their agreement tests these formulas.
\end{itemize}

\section{Summary and outlook}
\label{sec:summary}
In this paper we derived the IBP relations and the differential equations in the general case, for massive as well as massless propagators, and hence for tree correlators of arbitrary topology. For a massless (anti-)time-ordered propagator the extra relations reduce the four endpoint states to a single shared two-state factor, and the recurrences and the d log-form differential equations are formulated directly on this reduced space. We implemented these derivations in the Mathematica package \texttt{MadStree}, which evaluates dS tree correlators numerically. The package handles arbitrary tree diagrams, loop integrands, and single-vertex integral families.

The loop-level IBP and d log-form construction of \cite{Chen:2026dqp} provides a promising direction for extending the present workflow. A further step is to incorporate the loop-momentum integrals and the additional sectors they generate into the automated numerical evaluation, which would extend our package from tree level to loop level.

\section*{Acknowledgments}

Jiaqi Chen is supported by the National Natural Science Foundation of China (NSFC) under
Grant No.~12505094 and by the Science Foundation of China University of Petroleum, Beijing
(No.~2462025YJRC019). Yi-Xiao Tao is supported by the National Natural Science Foundation of China (NSFC) (Grant No.~124B2094). 

\medskip
\noindent\textbf{Statement on AI Use.}
The authors developed the scientific ideas and methodology and designed the function interfaces, commands, and invocation patterns of the software package.
Several agent programs driven by different large language models were used in this work. The authors have reviewed and verified all the results.

\appendix

\section{Common-theta bundles and the odd-subset expansion}
\label{app:theta}

\paragraph{Bundles and their theta argument.}
Fix a sector $s$ and write $\operatorname{comp}_s(v)$ for the current component that contains the
vertex $v$; a component is a maximal set of vertices already merged by contacts, and a lone vertex
is one component. An unpinched (anti-)time-ordered propagator $e=(u_e,v_e)$ carries the theta
$\theta(\tau_{u_e}-\tau_{v_e})$ of its two component times, so two propagators $e,e'$ carry the
\emph{same} theta if and only if their component pairs coincide as ordered pairs,
\begin{equation}
  \begin{aligned}
  \bigl(\operatorname{comp}_s(u_e),\operatorname{comp}_s(v_e)\bigr)
  &=\bigl(\operatorname{comp}_s(u_{e'}),\operatorname{comp}_s(v_{e'})\bigr)=(U,V),\\
  \text{so}\quad \Delta_e&=\Delta_{e'}\equiv\Delta=\tau_U-\tau_V .
  \end{aligned}
  \label{eq:bundle-criterion}
\end{equation}
A maximal set of propagators with this property is the common-theta bundle $\cB$ of the pair $(U,V)$, with
$|\cB|\ge2$. The steps below also cover $|\cB|=1$, which is the ordinary single-propagator contact
treated in Sec.~\ref{sec:remaining}. Different component pairs define different bundles.
A time derivative acts on their theta functions by the product rule, so each contact term
is assigned to the bundle whose theta was differentiated. The vertex exponentials of Block (i) are
theta-free and take no part in this bookkeeping, so only the mode-function two-state factors of \eqref{eq:sk-propagators} enter the steps below.

\paragraph{Step 1: multiplying the thetas out.}
For each propagator of one bundle, write the SK pair of products as in \eqref{eq:sk-propagators},
$\alpha_e$ for the coefficient of $\theta(\Delta)$ and $\beta_e$ for that of $\theta(-\Delta)$. At
the level of the $h$ blocks the two are the two orderings of the Hankel branches,
\begin{equation}
  \alpha_e = h^{(1)}_{u_e}\,h^{(2)}_{v_e},
  \qquad
  \beta_e = h^{(2)}_{u_e}\,h^{(1)}_{v_e},
  \label{eq:bundle-alphabeta}
\end{equation}
where the orders, mode-function states and arguments are those of \eqref{eq:h-definition}; $G_{--}$ interchanges
the two names, and this is the only place the sign $\sigma_e$ of the bundle enters. The product of
the bundle is a product of $|\cB|$ binomials in the two functions $\theta(\Delta)$ and
$\theta(-\Delta)$. Expanding naively gives $2^{|\cB|}$ terms, but inside one bundle $\theta(\Delta)$ is
the \emph{same} function for every factor, so idempotency $\theta(\Delta)^2=\theta(\Delta)$ makes
all powers of one argument collapse to that argument, and orthogonality
$\theta(\Delta)\theta(-\Delta)=0$ annihilates every term containing both arguments. Exactly the two
terms of \eqref{eq:bundle-two-terms} survive. The factor $2^{|\cB|}$ thus never appears within a bundle:
it would count the events of $|\cB|$ propagators with pairwise different theta arguments, that is
$|\cB|$ bundles of one propagator each, and those are enumerated bundle by bundle.

\paragraph{Step 2: one theta gives one delta.}
The contact part of the $\Delta$ derivative of the two-term form comes from differentiating only
 the step functions, leaving the coefficients of \eqref{eq:bundle-alphabeta} unchanged:
\begin{equation}
  \big(\partial_\Delta\theta(\Delta)\big)\prod_{e\in\cB}\alpha_e
  +\big(\partial_\Delta\theta(-\Delta)\big)\prod_{e\in\cB}\beta_e
  =\delta(\Delta)\left(\prod_{e\in\cB}\alpha_e-\prod_{e\in\cB}\beta_e\right),
  \label{eq:bundle-delta}
\end{equation}
because $\partial_\Delta\theta(\Delta)=\delta(\Delta)$ and
$\partial_\Delta\theta(-\Delta)=-\delta(\Delta)$. The coefficients $\alpha_e$ and $\beta_e$ are
products of mode functions of the two
component times and so depend on $\Delta$ as well; the complete derivative therefore adds the two terms
$\theta(\Delta)\,\partial_\Delta\prod_e\alpha_e$ and $\theta(-\Delta)\,\partial_\Delta\prod_e\beta_e$,
and those are the single-vertex derivative terms carried by the $\text{M}_0$ and $\text{M}_1$
generators of Sec.~\ref{sec:ibp}; they force no two component times to coincide. Only the contact
part \eqref{eq:bundle-delta} enters the bookkeeping of this appendix.
Three consequences are worth stating explicitly.
First, the bundle produces \emph{one} distribution, not $|\cB|$ of them: the whole contact information
of the bundle is concentrated in the single coefficient
$\prod_e\alpha_e-\prod_e\beta_e$, which is antisymmetric under
$\alpha\leftrightarrow\beta$. Second, that one $\delta(\Delta)$ forces the two component times to
coincide \emph{once}, so the components $U$ and $V$ merge once, independently of how many factors
$\mathsf d_e$ a term of the coefficient contains; this is what fixes the merge count in
\eqref{eq:base-power}. Third, for $G_{--}$ the interchange of $\alpha$ and $\beta$ turns
\eqref{eq:bundle-delta} into its negative, which reproduces the single sign $\sigma_e$ of
\eqref{eq:contact-weight} for a whole bundle.

\paragraph{Step 3: telescoping the antisymmetric difference.}
Order the propagators of the bundle as $1,\dots,|\cB|$ and abbreviate $\alpha_e=A_e$, $\beta_e=B_e$. Inserting
and subtracting successive partial products gives the telescoping identity
\begin{equation}
\begin{split}
  \prod_{e=1}^{|\cB|}A_e-\prod_{e=1}^{|\cB|}B_e
  &=\sum_{k=1}^{|\cB|}\left(\prod_{e\leq k}A_e\prod_{e>k}B_e
  -\prod_{e< k}A_e\prod_{e\geq k}B_e\right)\\
  &=\sum_{k=1}^{|\cB|}D_k\prod_{e<k}A_e\prod_{e>k}B_e ,
\end{split}
  \label{eq:bundle-telescope}
\end{equation}
whose intermediate sum collapses because the second summand of the $k$-th term equals the first
summand of the $(k{+}1)$-th term, while the two ends, $k=1$ and $k=|\cB|$, supply the two
products of \eqref{eq:bundle-delta}. Each term of \eqref{eq:bundle-telescope} contains exactly one
difference factor $D_e=A_e-B_e$.

\paragraph{Step 4: collecting by subsets and the parity.}
Introduce, as in \eqref{eq:bundle-odd-subsets}, the difference and the mean of the two
products, $\mathsf d_e=D_e=A_e-B_e$ and $\mathsf m_e=(A_e+B_e)/2$, equivalently
$A_e=\mathsf m_e+\mathsf d_e/2$ and $B_e=\mathsf m_e-\mathsf d_e/2$, and expand the spectators of
\eqref{eq:bundle-telescope} in this basis. In the $k$-th term the propagator $k$ supplies the explicit
$\mathsf d_k$; a propagator $e\in S$ with $e<k$ contributes $+\mathsf d_e/2$, a propagator $e\in S$ with $e>k$
contributes $-\mathsf d_e/2$, and every propagator outside $S$ contributes $\mathsf m_e$. So the
coefficient of the monomial $\prod_{e\in S}\mathsf d_e\prod_{e\notin S}\mathsf m_e$ is
\begin{equation}
  c_S = 2^{1-|S|}\sum_{k\in S}(-1)^{\#\{e\in S:\,e>k\}} .
  \label{eq:bundle-coefficient}
\end{equation}
Writing $S=\{i_1<\dots<i_m\}$ with $m=|S|$, the exponent for $k=i_j$ is $m-j$, hence
\begin{equation}
  \sum_{j=1}^{m}(-1)^{m-j}=\sum_{r=0}^{m-1}(-1)^{r}
  =\begin{cases}1,&m\ \text{odd},\\ 0,&m\ \text{even},\end{cases}
  \qquad
  c_S=\bigl(1-(-1)^{|S|}\bigr)2^{-|S|},
  \label{eq:bundle-parity}
\end{equation}
which is \eqref{eq:bundle-odd-subsets}: only odd subsets survive, each with the pure numerical
weight $2^{1-|S|}$. The empty set cannot occur, since every telescoping term carries at least the
explicit $\mathsf d_k$, in agreement with $c_\varnothing=0$. The same parity follows from symmetry
as the left-hand side of \eqref{eq:bundle-telescope} changes sign under
$\alpha\leftrightarrow\beta$, in which $\mathsf d_e$ is odd and $\mathsf m_e$ even, so a monomial
with an even number of $\mathsf d$ factors must cancel; the explicit computation
\eqref{eq:bundle-coefficient} only fixes the remaining coefficients.

\paragraph{Events, weights and the bookkeeping.}
Each non-empty odd subset $S$ of one bundle is one simultaneous contact event. Let $s_0$ be
its supersector, $s_1$ the sector obtained by pinching all propagators in $S$, and $(u,v)$
the ordered pair of merged vertices, with the derivative acting at $u$. The coefficient of
the subsector integral generalizes $F_eC^{(uv)}_{s_0s_1}$ in \eqref{eq:R-two-vertex}.
Before massless reduction, it is the tensor product of the local complementary-bit selectors
of \eqref{eq:contact-atom-massive} and the raw Wronskian factors $F_e$ for $e\in S$, multiplied
by $2^{1-|S|}$ and by the spectator factors $\mathsf m_e$ of the unselected bundle propagators.
All spectator bits outside the bundle are preserved. Massless propagators are then reduced
and normalized by the maps of Sec.~\ref{sec:remaining}; for a single selected massless
propagator this gives the shared-slot column \eqref{eq:contact-atom-massless}.
The collected form
\eqref{eq:bundle-odd-subsets}, not the term-by-term form \eqref{eq:bundle-telescope}, is what the
package enumerates: the $k$-th telescoping term stands for ``pinch only the $k$-th propagator,'' but
after the spectators are expanded the same subset $S$ is reached from several $k$, so an
enumeration over $k$ would generate the same sector transition more than once. After collecting,
each event appears exactly once with a numerical weight, and the deduplication key of the reduction
is the triple (supersector, subsector, set of selected propagator identifiers). There are only
finitely many such events, because every event strictly reduces the number of current components.

\paragraph{Subsector base power and coincident unselected propagators.}
For a component $C$ produced by such events the base power of time is \eqref{eq:base-power},
$\nu_{0,C}=\sum_{u\in C}\nu_{0,u}+\sum_{e\in E_s(C)}\varrho_e+|C|-1$.
The vertex powers add because the merged integrand is a product.
Each pinched propagator contributes its raw contact power $\varrho_e$ through the factor
$(-\tau)^{\varrho_e}$ accompanying the corresponding $\delta$.
The last term counts component merges: $|C|-1$ for $|C|$ vertices, regardless of how many
bundle propagators are pinched at once. Since \eqref{eq:base-power} depends only on the final component and on
the pinched set, and not on the order or the clustering of the events, different legal event
sequences that reach the same sector give the same subsector base power; the reduction is confluent at
the level of the powers.

An unselected propagator whose two vertices fall into one component after a bundle event must also be
reduced with the equal-time identifications. For a massless (anti-)time-ordered propagator the shared slot then
contributes only its symmetric mode-function state,
because the two Wightman combinations are equal at equal times, so the antisymmetric shared
mode-function state of the
propagator gives zero; for a massive (anti-)time-ordered propagator the two mode-function states carrying a single derivative become
identified, $10\sim01$, which is the equal-time canonical quotient. These are additional
relations in the binary space of massive endpoint slots and massless shared slots. Write
$P_s^{\rm eq}$ for the projection from this space onto the independent equal-time masters
and $S_s^{\rm eq}$ for the embedding that reconstructs the redundant states. For a matrix
$M_s^{\rm bin}$ built on the binary space, the corresponding matrix on the independent masters is
\begin{equation}
  M_s^{\rm eq}=P_s^{\rm eq}M_s^{\rm bin}S_s^{\rm eq},
  \qquad P_s^{\rm eq}S_s^{\rm eq}=\Id.
  \label{eq:equal-time-retraction}
\end{equation}
These constant maps implement the equal-time relations after the four-to-two massless
reduction $P_s,S_s$ of \eqref{eq:sector-massless-maps}.
On the boundary side the same equal-time configuration makes the two
branch arrangements of a coincident massive (anti-)time-ordered propagator enter with the same power, and the product
of its two mode-function asymptotic coefficients is averaged over these two Hankel-branch arrangements,
one mean per coincident propagator, as stated in
Sec.~\ref{sec:common-theta}.

\paragraph{Relation to the single-contact case.}
For $|\cB|=1$ the two-term form \eqref{eq:bundle-two-terms} reduces to the ordinary SK pair of one
propagator, \eqref{eq:bundle-telescope} to the single difference $\mathsf d_e$, and $c_{\{e\}}=1$, so the
bundle machinery reproduces the single-propagator contacts of Sec.~\ref{sec:remaining} and of
\cite{Chen:2023iix}. The new ingredients are the
weight $2^{1-|S|}$, the parity selection, and the identification of the merge count with $|C|-1$
rather than with the number of pinched propagators.

\bibliographystyle{JHEP}
\bibliography{references}

\end{document}